\documentclass[a4paper, 11pt]{article}
\usepackage{graphicx}
\usepackage{amssymb}

\newcommand\Slash[1]{{\ooalign{\hfil/\hfil\crcr$#1$}}}

\begin{document}

\begin{titlepage}

\vspace*{-10ex}
\begin{flushright}
OU-HET 603\\
{\tt arXiv:0804.3506}\\
April 2008
\end{flushright}
\bigskip

\begin{center}
{\LARGE\bf  All orders analysis of three dimensional $CP^{N-1}$ model in $1/N$-expansion}
\vspace{10ex}

\setcounter{footnote}{0}

{\renewcommand{\thefootnote}{\fnsymbol{footnote}}}
{\large\bf Kiyoshi Higashijima$^a$\footnote{E-mail: {\tt
 higashij@het.phys.sci.osaka-u.ac.jp}} and Takahiro
 Nishinaka$^a$\footnote{E-mail: {\tt nishinaka@het.phys.sci.osaka-u.ac.jp}}}

\vspace{2ex}

{\sl
$^a$Department of Physics, Graduate School of Science, Osaka
 University,\\
Toyonaka, Osaka 560-0043, Japan\\
}
\end{center}

\bigskip

\begin{abstract}
 The renormalizability of the three dimensional supersymmetric
 $CP^{N-1}$ model is discussed in the $1/N$-expansion method, to all
 orders of $1/N$. The model has $N$ copies of the dynamical field and the
 amplitudes are expanded in powers of $1/N$. In order to see the effects
 of supersymmetry explicitly, Feynman rules for superfields are
 used. All divergences in amplitudes can be eliminated by the renormalizations of the coupling constant and the wavefunction of the dynamical field to all orders of $1/N$.

The beta function of the coupling constant is also calculated to all
 orders of $1/N$.
 It is shown that this model has a non-trivial
 ultraviolet fixed point. The beta function is shown to have no higher
 order correction in the $1/N$-expansion.

\end{abstract}

\end{titlepage}

\pagestyle{plain}


\section{Introduction}

Three dimensional non-linear sigma models are perturbatively
non-renormalizable according to the power counting. The reason for this
is that non-linear sigma models have an infinite number of interaction
terms in the action. In three dimensions, a scalar field has mass dimension
$1/2$. Therefore, coupling constants of interaction terms involving more than six scalar
fields has negative mass dimensions. This implies
non-renormalizability in perturbation theory.

Some supersymmetric nonlinear sigma models were, however, argued to be renormalizable in three dimensions by the
renormalization group method \cite{paper2}. The supersymmetric
$CP^{N-1}$ model is one of the candidates. The $CP^{N-1}$ model is a non-linear sigma model on the complex projective
manifold $CP^{N-1}$, which was first introduced by Eichenherr
\cite{cpn}.
 The supersymmetric version of the model was formulated by introducing an auxiliary
gauge field \cite{cp1,cp2,cp3}.

The renormalization group method is one of the powerful methods which can reveal the non-perturbative property of the theory.
The renormalizability in the renormalization group method is
equivalent to the existence of a non-trivial ultraviolet (UV) fixed
point of the theory. In the renormalization group analysis in
\cite{paper2}, however, the effective action is
 expanded in powers of derivetives on spacetime and approximated by
 truncating at the second order of derivaties. Although this approximation is
 valid in the low energy scale region, it is not obvious in high
 energy scale whether
 the approximation is valid or not. 

The existence of the UV fixed point of the three
dimensional supersymmetric $CP^{N-1}$ model is also shown by the
$1/N$-expansion method up to the next-to-leading order
\cite{paper3, paper1}. In the $1/N$-expansion, we expand amplitudes in powers of
$1/N$ instead of coupling constant, where
$N$ is the number of fields involved in the theory. In general, each
term of $1/N$ expansion corresponds to a sum of infinite number of
Feynman diagrams in perturbation theory. Therefore,
the $1/N$-expansion is an another powerful non-perturbative method. Indeed, it was argued that the three dimensional nonlinear sigma
models and its supersymmetric versions are renormalizable, order by order
in the $1/N$ expansion \cite{Prof.Nissimov, Prof.Nissimov2}. In
the reference
\cite{paper3, paper1}, the beta function of the coupling constant was explicitly evaluated
by using Feynman rules in the component field formalism. It was shown
that there is no next-to-leading order contribution of $1/N$. Because of
supersymmetry, contributions of bosons and
fermions cancel each other in the next-to-leading order of
$1/N$. There might be, however, contributions of higher orders of
$1/N$.

In this paper, we study the three dimensional ${\mathcal N} =2$
supersymmetric $CP^{N-1}$ model and show that in the $1/N$-expansion there is no higher order correction to the
beta function in this model. We also show explicitly that all divergences can be
eliminated by the renormalizations of the coupling constant and the wave
function of the dynamical field $\Phi$, namely the renormalizability in the
method of $1/N$-expansion. In order to keep the manifest supersymmetry, we use Feynman rules for superfields,
which we call ``super Feynman rules''. For example, a chiral superfield
$\Phi\!\left(x,\theta,\bar{\theta}\right)$ can be expanded in terms of
component fields as
\begin{eqnarray*}
\Phi\!\left(x,\theta,\bar{\theta}\right) = \phi\!\left(y\right) +
 \theta\psi\!\left(y\right) + \frac{1}{2}\theta^2F\!\left(y\right),
\end{eqnarray*}
where $y^\mu:=x^\mu+\frac{i}{2}\bar{\theta}\gamma^\mu\theta$. Therefore,
if we know the two point functions of component fields, we can costruct
the two point function of superfield
$\Phi\!\left(x,\theta,\bar{\theta}\right)$ from them. Using this
propagator of superfield, we can explicitly see the cancellation due to supersymmetry.

In section 2, we review a part of the argument given in
\cite{paper1}. For the $1/N$-expansion, it is useful to introduce an
auxiliary field in the action. Although the auxiliary field has no kinetic
term in the classical action, it acquires quadratic terms in the
effective action induced by the quatum fluctuations of the dynamical
field. When the auxiliary field is introduced, the path integration over
the dynamical field becomes a gaussian integral and can be performed
easily. After performing the integration over the dynamical field, we
obtain the action with respect to the auxiliary field, which is
proportional to $N$. Therefore, the $1/N$-expansion turns out to be the
loop expansion of the auxiliary field. This is the reason for
introducing the auxiliary field.

 We
evaluate the effective action to the leading order of $1/N$ to study
the vacuum structure of the model. The model turns out to have two
phases, ``symmetric phase'' and ``broken phase''. The global $SU(N)$
symmetry is spontaneously broken in the broken phase, while it is
unbroken in the symmetric phase. The supersymmetry is unbroken in both
the symmetric phase and the broken phase.
In the leading order of $1/N$, the effective action has a linear divergence which can be
eliminated by the renormalization of the coupling constant. 

In section 3, we evaluate the propagator of the chiral superfield
$\Phi$ in the symmetric phase, which can be obtained by combining the
propagators of component fields. We call this propagator of superfield
``superpropagator''.
After we modify the chiral superfield $\Phi$ by some similarity
transformation, the superpropagator can be written by using differential operators on superspace. These
differential operators can be obtained by modifing ordinary
supercovariant derivatives $D_\alpha. \bar{D}_\alpha$. We call these
differential operators ``twisted covariant derivatives''.

In section 4, we first evaluate one-loop diagrams of the dynamical field,
which induce the inverse propagator of the auxiliary field in the
effective action. Using the superpropagator of the dynamical field, we
can easily calculate the one-loop diagrams by a partial integration over
grassmann coordinates. From the inverse propagator of the auxiliary
field, we secondly evaluate the propagator of the auxiliary field $V$,
which can be written in terms of ordinary covariant derivatives $D_\alpha, \bar{D}_\alpha$.

In section 5, we study divergent diagrams and the renormalization. We first
evaluate the superficial degree of divergence and find that divergent
diagrams can be classified into two types. We show that all divergences
can be eliminated in each order of $1/N$ by renormalizations of the coupling
constant and the wave function of the dynamical field.
In the last subsection, we evaluate the beta function of the coupling
constant.  In the $1/N$-expansion, there is no contribution to the beta
function except at the leading order. We find that this model has a
non-trivial UV fixed point.

Throughout this paper we work in three dimensions with metric
$\eta_{\mu\nu} = {\rm diag}\left(+,-,-\right)$.

\section{$CP^{N-1}$ model}

\subsection{Action of the $CP^{N-1}$ model with the auxiliary field}
The action of the $CP^{N-1}$ model involves a set of $N$ chiral
superfields $\Phi^j$ ($j = 1\sim N$) and one vector superfield $V$:
\begin{eqnarray}
 S &=& \int \! d^3x \hspace{.25em}d^4\theta \left( \Phi^{j\dagger} e^{-V}
					    \Phi^{j}  + cV\right) \label{eq12}
\end{eqnarray}
where $c$ is a coupling constant and we define
\begin{eqnarray*}
 \int \! d^4\theta := \int \! d^2\theta d^2\bar{\theta} \quad
  ,\quad \int \! d^2\theta \hspace{.25em} \theta^2 = \int \! d^2
  \bar{\theta} \hspace{.25em} \bar{\theta}^2 = 2.
\end{eqnarray*}
This action has ${\mathcal N} = 2$ supersymmetry, $U\!\left(1\right)$
local gauge symmetry, and a global $SU(N)$ symmetry. In appendix A, ${\mathcal N} = 2$ supersymmetry in three dimensions is
reviewed.

The local gauge transformation is
\begin{eqnarray*}
 \Phi^j \to e^{i\Lambda}\Phi^j \quad, \quad \Phi^{j\dagger} \to
  e^{-i\Lambda^\dagger}\Phi^{j\dagger} \quad ,\quad V \to V + \Lambda +\Lambda^\dagger,
\end{eqnarray*}
where $\Lambda$ is any chiral
superfield. Although $V$ itself is not invariant under this
transformation, the following term
\begin{eqnarray*}
 \int \! d^2\theta \hspace{.25em}d^2\bar{\theta} \hspace{.5em} V
\end{eqnarray*}
is gauge invariant.

The equation of motion for $V$
\begin{eqnarray*}
 \Phi^{j\dagger}e^{-V}\Phi^j = c
\end{eqnarray*}
is solved for the auxiliary field $V$:
\begin{eqnarray*}
 V = \log\left(\Phi^{j\dagger}\Phi^j\right) - \log c.
\end{eqnarray*}
Therefore we can eliminate $V$ from the action:
\begin{eqnarray*}
 S = c \int \! d^3\!x \hspace{.125em} d^4\theta \hspace{.25em}\log\left(\Phi^{j\dagger}\Phi^j\right),
\end{eqnarray*}
which reduces to the action with the Fubini-Study metric if we fix the
gauge symmetry by $\Phi^N=1$.
Note that $\int \! d^4\theta \hspace{.5mm} \log c = 0$.

For the $1/N$
expansion, the action
(\ref{eq12}) is more convinient than this action.

In terms of component fields, $\Phi$ can be written as
\begin{eqnarray*}
\Phi^j\!\!\left(x,\theta,\bar{\theta}\right)  &=& \phi\!\left(x\right) + \theta\psi\!\left(x\right)
  + \frac{1}{2}\theta^2 F\!\left(x\right)
  + \frac{i}{2}\left(\bar{\theta}\Slash{\partial}\theta\right)\!\phi\!\left(x\right)
  -
  \frac{i}{4}\theta^2\!\!\left[\hspace{.125em}\bar{\theta}\Slash{\partial}\psi\!\left(x\right)\right]
 -
 \frac{1}{16}\theta^2\bar{\theta}^2\partial^2\!\phi\!\left(x\right)
\end{eqnarray*}
and if we choose the Wess-Zumino gauge, $V$ can be written as
\begin{eqnarray}
V\!\left(x,\theta,\bar{\theta}\right) = \bar{\theta}\Slash{v}\!\left(x\right)\theta\hspace{.125em} 
+ \hspace{.125em}M\!\left(x\right)\bar{\theta}\theta +
  \frac{1}{2}\left[\theta^2\hspace{.1em}\bar{\theta}\hspace{.1em}\lambda\!\left(x\right) +
	      \bar{\theta}^2\theta\bar{\lambda}\!\left(x\right)\right] +
  \frac{1}{4}\theta^2\bar{\theta}^2 \! D\!\left(x\right). \label{eq500}
\end{eqnarray}
Then the action (\ref{eq12}) becomes
\begin{eqnarray*}
 S &=& \int \! d^3x \left\{\partial_\mu\phi^{j*}\partial^\mu\phi^j +
		   i\bar{\psi}^j\!\Slash{\partial}\hspace{.1em}\psi^j +
		   F^{j*}\!F^j - \left[i\left(\phi^{j*}\partial_\mu\phi^j -
					\phi^j\partial_\mu\phi^{j*}\right) + \bar{\psi}^j\!\gamma^\mu\psi^j\right]v^\mu\right.\\
 && \left.+ v^\mu v_\mu\phi^{j*}\!  \phi^j - M^2\phi^{j*}\!\phi^j - M\bar{\psi}^j\!\psi^j - D\phi^{j*}\!\phi^j + cD + \left(\phi^j\hspace{.125em}\bar{\psi}^j\!\!\lambda + \phi^{j*}\bar{\lambda}\psi^j\right)\right\}.
\end{eqnarray*}

\subsection{Vacuum structure of the $CP^{N-1}$ model}

To investigate the vacuum structure we have to calculate the effective
potential. We divide the dynamical
fields into the vacuum expectation values and the quantum fluctuations:
\begin{eqnarray*}
 \phi^j=\phi^j_c + \phi^j_q \quad, \quad \psi^j = \psi^j_q
  \quad, \quad F^j = F^j_c +  F^j_q
\end{eqnarray*}
where
\begin{eqnarray*}
 \phi^j_c = \left<\phi^j\right> \quad , \quad F^j_c = \left<F^j\right>
\end{eqnarray*}
are constant modes independent of space-time and $\left<\psi^j\right> = 0$ because we assume the
translation and Lorentz invariance of the vacuum. Qunatum fluctuations
satisfy $\int\phi_q^id^3x=\int \psi_q^i d^3x = \int F_q^i d^3x = 0$.
Then we perform the path integration over
$\phi^j_q, \psi^j_q, F^j_q$.

We can first perform the gaussian integral over $F^j_q$ and find that the
effective potential for $F^j$ is $F^{j*}_c\!F^j_c$. Therefore $F^j$ does
not have the vacuum expectation value:
\begin{eqnarray*}
 F^j_c = 0.
\end{eqnarray*}

Then we integrate out $\phi^j_q$ and $\psi^j_q$. Notice that the
Lagrangian can be written as
\begin{eqnarray*}
{\mathcal L} &=& \phi^{j*}_q\left\{-\left(\partial_\mu +
				    iv_\mu\right)\left(\partial^\mu +i v^\mu\right) - M^2
	   -D\right\}\phi^j_q + \bar{\psi}^j_q\left(i\Slash{\partial} -
	   \Slash{v} - M\right)\psi^j_q\\
&&\hspace{1.25em} +
	   \phi^j_q\hspace{.125em}\bar{\psi}^j_q\!\lambda +
	   \phi^{j*}_q\bar{\lambda}\psi^j_q\\[1mm]
&&\hspace{2.5em} + cD -
	   \phi^{j*}_c\!\left\{i\left(\partial_\mu v^\mu\right)-v^\mu
		       v_\mu +M^2 +D\right\}\phi^j_c
\end{eqnarray*}
where surface terms are ignored. We shift the integration variables
$\psi^j_q, \bar{\psi}^j_q$:
\begin{eqnarray*}
 \psi'^j_q := \psi^j_q + \left(i\Slash{\partial} - \Slash{v} -
			  M\right)^{-1}\phi^j_q\lambda \quad, \quad
 \bar{\psi}'^j_q := \bar{\psi}^j_q +
 \phi^{j*}\bar{\lambda}\left(i\Slash{\partial}-\Slash{v} - M\right)^{-1},
\end{eqnarray*}
then we find
\begin{eqnarray*}
 {\mathcal L} &=& \phi^{j*}_q\left(\nabla_B -
 \bar{\lambda}\nabla_F^{-1}\lambda\right)\hspace{.12em}\phi^j_q +
  \bar{\psi}'^j_q\nabla_F \hspace{.12em}\psi'^j_q \\
&& + cD -
	   \phi^{j*}_c\!\left\{i\left(\partial_\mu v^\mu\right)-v^\mu
		       v_\mu +M^2 +D\right\}\phi^j_c
\end{eqnarray*}
where
\begin{eqnarray*}
 \nabla_B &:=& -\left(\partial_\mu +
				    iv_\mu\right)\left(\partial^\mu +i v^\mu\right)- M^2 -D\\
\nabla_F &:=& i\Slash{\partial} - \Slash{v} - M.
\end{eqnarray*}
We perform the gaussian integration over $\phi^j_q, \psi^j_q$
and obtain the effective action for the dynamical fields, where the
auxiliary fields are treated as the external background fields:
\begin{eqnarray*}
 S_{\rm eff}\!\left(\phi_c\hspace{.1em};\hspace{.125em}v,M,\lambda,D\right) &=& iN\hspace{.125em}{\rm Tr\hspace{.125em} ln}\left(\nabla_B +
			     \bar{\lambda}\nabla_F^{-1}\lambda\right)
 - iN\hspace{.125em}{\rm Tr \hspace{.125em} ln}\nabla_F\\
 && \hspace{1.25em} + \int \! d^3x \left[ cD -
	   \phi^{j*}_c\!\left\{i\left(\partial_\mu v^\mu\right)-v^\mu
		       v_\mu +M^2 +D\right\}\phi^j_c\right]
\end{eqnarray*}

To obtain the exact effective potential for both the dynamical fields
and the auxiliary fields, we have to perform the
path integration over the fluctuations of the auxiliary fields. In this
section, we calculate the effective potential in the leading order of
the $1/N$ expansion, and we take $c=N/g^2$ in order to make the
Lagrangian of order $N$.

If we take the limit of $N\to\infty$, the path integration
over the auxiliary fields can be performed by the saddle point method
since the $S_{\rm eff}$ is of order $N$. In the leading order of
$1/N$ expansion, the effective potential is given by the value of
$S_{\rm eff}$ at the saddle point.

We take the vacuum expectation values
of the auxiliary fields as follows:
\begin{eqnarray*}
 \left<v^\mu\right>=\left<\lambda\right>= 0 \quad, \quad \left<M\right> = M_c \quad, \quad
  \left<D\right> = D_c,
\end{eqnarray*}
where $M_c, D_c$ is constant fields. Then we find
\begin{eqnarray}
 S_{\rm eff} &=& - \int\! d^3x \hspace{.25em}V_{\rm eff} \nonumber \\
 \frac{V_{\rm eff}}{N} &=& -i\int^\Lambda
  \!\!\!\frac{d^3k}{\left(2\pi\right)^3}\ln\left(-k^2+M_c^2+D_c^2\right) +i
  \int^\Lambda\!\!\!\frac{d^3k}{\left(2\pi\right)^3}\hspace{.25em}{\rm
  tr}\ln\left(\Slash{k}-M_c\right) \nonumber \\
 && \hspace{.5em} +
  \frac{1}{N}\phi^{j*}_c\left(M_c^2+D_c^2\right)\phi^{j}_c-\frac{1}{g^2}D_c\nonumber \\
&=& -\frac{1}{6\pi}\left|M_c^2+D_c\right|^{\frac{3}{2}} +
 \frac{1}{6}\left|M_c\right|^3 + \frac{1}{N}\left(M_c^2 +
	     D_c\right)\phi^{j*}_c\!\phi^j_c +
 \left(\frac{\Lambda}{2\pi^2}-\frac{1}{g^2}\right)D_c \label{eq15000}
\end{eqnarray}
where $\Lambda$ is an ultraviolet cutoff and the last equality is shown in
appendix B. Then we define a
renormalized coupling constant $g_R$ to absorb the linear divergence:
\begin{eqnarray*}
 \frac{\mu}{g_R^2} := \frac{1}{g^2} - \frac{\Lambda}{2\pi^2} + \frac{\mu}{2\pi^2}
\end{eqnarray*}
where $\mu$ is a renormalization scale and $g_R$ is demensionless. Then
we define $m$ as follow:
\begin{eqnarray*}
\frac{m}{4\pi} :=  \mu\left(\frac{1}{2\pi^2} -
					     \frac{1}{g_R^2}\right) =
\frac{\Lambda}{2\pi^2} - \frac{1}{g^2}.
\end{eqnarray*}
which is independent of the renormalization scale $\mu$.

With these definitions, the effective potential can be written as
\begin{eqnarray}
 \frac{V_{\rm eff}}{N} &=& -\frac{1}{6\pi}\left|M_c^2+D_c\right|^{\frac{3}{2}} +
 \frac{1}{6\pi}\left|M_c\right|^3 + \frac{1}{N}\left(M_c^2 +
	     D_c\right)\phi^{j*}_c\!\phi^j_c + \frac{m}{4\pi}D_c. \label{eq10}
\end{eqnarray}
The saddle point condition of $S_{\rm eff}$ is
\begin{eqnarray}
 \frac{1}{N}\frac{\partial V_{\rm eff}}{\partial M_c} &=&
  -2M_c\left(\frac{\epsilon}{4\pi}\left|M_c^2+D_c\right|^{\frac{1}{2}} -
       \frac{1}{4}\left|M_c\right| -
       \frac{1}{N}\phi^{j*}_c\!\phi^j_c\right) = 0 \label{eq2}\\[1mm]
 \frac{1}{N}\frac{\partial V_{\rm eff}}{\partial D_c} &=&
  -\frac{\epsilon}{4\pi}\left|M_c^2 + D_c\right|^{\frac{1}{2}} +
  \frac{1}{N}\phi^{j*}_c\!\phi^j_c + \frac{m}{4\pi} = 0 \label{eq3}\\[2mm]
\frac{1}{N}\frac{\partial V_{\rm eff}}{\partial \phi_c^{*i}} &=&
 \frac{1}{N}\left(M_c^2+D_c\right)\phi_c^{i} = 0 \nonumber
\end{eqnarray}
where $\epsilon = {\rm sgn}\!\left(M_c^2 + D_c\right)$. The first two
conditions fix the value of $M_c$ at the saddle point
\begin{eqnarray*}
\left|M_c\right| = m \quad {\rm or} \quad 0.
\end{eqnarray*}
So there are two candidates for the vacuum configuration. We will evaluate
the values of $V_{\rm eff}$ at these two configurations.

\vspace{5ex}

\noindent \underline{\bf $\left|M_c\right| = m$ case:} \hspace{.5em}Notice that this case is
possible only when $m \geq 0$. The equation (\ref{eq3}) can be solved
for $\frac{1}{N}\phi^{j*}_c\phi^j_c$ as follow
\begin{eqnarray}
  \frac{1}{N}\phi^{j*}_c\!\phi^j_c =
  \frac{\epsilon}{4\pi}\left|M_c^2+D_c\right|^{\frac{1}{2}}-
  \frac{m}{4\pi}. \label{eq11}
\end{eqnarray}
Substituting this and $\left|M_c\right|=m$ to (\ref{eq10}), we find
\begin{eqnarray*}
 \frac{V_{\rm eff}}{N} &=& \frac{1}{12\pi}\left(\left|M_c^2 +
					   D_c\right|^{\frac{3}{2}} - m^3\right).
\end{eqnarray*}
And we can also solve the constraint (\ref{eq3}) for $\left|M_c^2 +
D_c\right|^{\frac{1}{2}}$:
\begin{eqnarray*}
 \left|M_c^2 + D_c\right|^{\frac{1}{2}} = \left|m + \frac{4\pi}{N}\phi^{j*}_c\phi^j_c\right|.
\end{eqnarray*}
Then we obtain the vacuum energy when $\phi^j_c$ is kept fixed
\begin{eqnarray*}
 V_{\rm eff}\!\left(\phi_c\right) =
  \frac{N}{12\pi}\left(\left|\frac{4\pi}{N}\phi^{j*}_c\!\phi^{j}_c+m\right|^3-m^3\right).
\end{eqnarray*}
 Assuming $m>0$, the minimum of this vacuum energy is located at
$\phi^j_c=0$. Then (\ref{eq3}) implies $D_c=0$. Since $\phi_c$ and $D_c$
are the order parameter of SU($N$) and supersymmetry respectively,
both SU($N$) and supersymmetry are unbroken in this case. The fact that
the minimum vacuum energy is exactly zero also implies supersymmetry is not broken.

\vspace{5ex}

\noindent \underline{\bf $M_c = 0$ case:} \hspace{.5em} Substituting (\ref{eq11})
and $M_c=0$ to the effective potential (\ref{eq10}), we find
\begin{eqnarray*}
 V_{\rm eff}\!\left(\phi_c\right) &=& \frac{1}{12\pi}\left|D_c\right|^{\frac{3}{2}}.
\end{eqnarray*}
And by solving the constraint (\ref{eq3}) for
$\left|D_c\right|^{\frac{1}{2}}$ and substituting it to this equation,
we obtain the vacuum energy
\begin{eqnarray*}
 V_{\rm eff}\!\left(\phi_c\right) &=& \frac{N}{12\pi}\left|\frac{4\pi}{N}\phi^{j*}_c\!\phi^j_c+m\right|^3.
\end{eqnarray*}
If $m>0$, the minimum of this vacuum energy is located at $\phi^j_c=0$
and larger than zero, and
therefore the true vacuum is located at $M_c=m$. On the other hand, if
$m<0$, the minimum is located at
\begin{eqnarray*}
 \phi^{j*}_c\!\phi^j_c = \frac{N}{4\pi}\left|m\right|,
\end{eqnarray*}
then (\ref{eq3}) implies $D_c=0$. Therefore supersymmetry is not broken
while SU($N$) symmetry is spontaneousky broken in this case. The minimum
vacuum energy is again exactly zero.

\vspace{5ex}

In summary, in the case of $m\geq 0$ which we call the ``symmetric phase'', both supersymmetry and SU($N$) are unbroken,
and $\phi^j_c=D_c=0,\hspace{1.5mm} \left|M_c\right|=m$ at the
vacuum. On the other hand, in the case of $m<0$ which we call the ``broken phase'', supersymmetry is not broken while
SU($N$) is spontaneously broken, and
$\phi^j_c=\frac{4\pi}{N}\left|m\right|,\hspace{1.5mm}M_c=D_c=0$ at the vacuum.

\section{Propagator of the dynamical field}

In this section, we will evaluate the propagator of the dynamical
field $\Phi^j$ in the symmetric phase. We first evaluate the propagators of the component fields
$\phi^j, \psi^j, F^j$, and then
we construct the propagator of the superfield $\Phi^j$.

\subsection{Propagator of the component fields}

In the symmetric phase, we redefine $M$ as follows
\begin{eqnarray*}
 M \quad \longrightarrow \quad M + m
\end{eqnarray*}
so that $\left<M\right> = 0$.
Then in the Lagrangian, the kinetic term for the dynamical field becomes
\begin{eqnarray*}
 {\mathcal L}_{\rm kin} =
  \phi^{j*}\!\left(-\partial^2-m^2\right)\phi^j +
  \bar{\psi}^j\!\left(i\Slash{\partial}-m\right)\psi^j + F^{j*}\!F^j.
\end{eqnarray*}
Notice that although the dynamical field obtained the mass $m$, neither supersymmetry nor SU($N$) symmetry is broken in this phase. 

Defining the Green's function
\begin{eqnarray*}
\Delta_F\!\left(x-x'\right):=\left(-\partial^2-m^2 + i\epsilon\right)^{-1}\!\delta\left(x-x'\right),
\end{eqnarray*}
the propagators of $\phi^j, \psi^j$ and $F^j$ can be written as
\begin{eqnarray*}
 \left<\phi^{j}\!\left(x\right)\phi^{k*}\!\left(x'\right)\right>_0 &=&
  i\delta^{jk}\Delta_F\!\left(x-x'\right)\\[2mm]
 \left<\psi^{j\alpha}\!\left(x\right)\bar{\psi}^k_\beta\!\left(x'\right)\right>_0
  &=&
  i\delta^{jk}\left[\left(i\Slash{\partial}-m\right)^{-1}\right]^{\alpha}_{\hspace{.3em}\beta}\delta\left(x-x'\right)\\
 &=& i\delta^{jk}\left(i\Slash{\partial}+m\right)^{\alpha}_{\hspace{.3em}\beta}\Delta_F\!\left(x-x'\right)\\[3mm]
 \left<F^j\!\left(x\right)F^{k*}\!\left(x'\right)\right>_0 &=&
 i\delta^{jk}\delta\left(x-x'\right)\\
 &=& i\delta^{jk}\left(-\partial^2-m^2\right)\Delta_F\!\left(x-x'\right)
\end{eqnarray*}
All other two-point functions vanish.

\subsection{Superpropagator of the dynamical field}

Using these component propagators, we can construct the propagator of the
superfield $\Phi^j$ which we call the ``superpropagator''.

Since the superfield $\Phi^j$ can be written in terms of components as
\begin{eqnarray*}
 \Phi^j\!\left(x,\theta,\bar{\theta}\right) &=& \phi^j\!\left(y\right)
  + \theta\psi^j\!\left(y\right) + \frac{1}{2}\theta^2
  F^j\!\left(y\right)
\end{eqnarray*}
where $y^\mu:= x^\mu + \frac{i}{2}\bar{\theta}\gamma^\mu\theta$, the
free-field two-point function of $\Phi^j$ becomes as follows: 
\begin{eqnarray*}
 \left<\Phi^j\!(x,\theta,\bar{\theta})\Phi^{\dagger k}\!(x',\theta',\bar{\theta}')\right>_0
  &=& \left<\phi^j\!\left(y\right)\phi^{k*}\!\left(y'^\dagger\right)\right>_0 +
  \left<\theta\psi^j\!\left(y\right)\hspace{.25em}\bar{\theta}'\bar{\psi}^{k}\!\left(y'^\dagger\right)\right>_0
  + \frac{1}{4}\theta^2\bar{\theta}'^2\left<F^j\!\left(y\right)F^{k*}\!\left(y'^\dagger\right)\right>_0.
\end{eqnarray*}
If we note
\begin{eqnarray*}
\left<\theta\psi^{j}\!\left(y\right) \hspace{.25em}
 \bar{\theta}'\bar{\psi}^{k}\!\left(y'^\dagger\right)\right>_0 &=&
 \theta_\alpha\left<\psi^{j\alpha}\!\!\left(y\right)\hspace{.25em}\bar{\psi}^k_\beta\!\left(y'\right)\right>_0\bar{\theta}'^\beta,
\end{eqnarray*}
then we find
\begin{eqnarray*}
  \left<\Phi^j\!(x,\theta,\bar{\theta})\Phi^{\dagger k}\!(x',\theta',\bar{\theta}')\right>_0
  &=&
  i\delta^{jk}\left[1+\theta\left(i\Slash{\partial}+m\right)\bar{\theta}'
	      +
	      \frac{1}{4}\theta^2\bar{\theta}'^2\!\left(-\partial^2-m^2\right)\right]\Delta_F\!\left(y-y'^\dagger\right)\\
 &=& i\delta^{jk} e^{\theta\left(i\Slash{\partial}+m\right)\bar{\theta}'}\hspace{.25em}\Delta_F\!\left(y-y'^\dagger\right).
\end{eqnarray*}
In the last line, we use the equation
$
 \left[\theta\left(i\Slash{\partial}+m\right)\bar{\theta}'\right]^2 = \frac{1}{2}\theta^2\bar{\theta}'^2\left(-\partial^2-m^2\right),
$
which is shown in appendix C. 
\hspace{.25em}Recalling $y^\mu = x^\mu +
\frac{i}{2}\bar{\theta}\gamma^\mu\theta$ and noting that
$
 \theta\left(i\Slash{\partial}+m\right)\bar{\theta}' = -\bar{\theta}'\left(i\Slash{\partial}-m\right)\theta
$,
then we find
\begin{eqnarray*}
   \left<\Phi^j\!(x,\theta,\bar{\theta})\Phi^{\dagger k}\!(x',\theta',\bar{\theta}')\right>_0
  &=& i\delta^{jk} e^{-\bar{\theta}'\left(i\Slash{\partial}-m\right)\theta +
  \frac{i}{2}\bar{\theta}\Slash{\partial}\theta +
  \frac{i}{2}\bar{\theta}'\Slash{\partial}\theta'}\Delta_F\!\left(x-x'\right).
\end{eqnarray*}
In momentum space, this becomes
\begin{eqnarray}
\left<\Phi^{j}\!(p,\theta,\bar{\theta})\hspace{.25em}\Phi^{{\dagger
 k}}\!(-p,\theta',\bar{\theta}')\right>_0 &=&  e^{-\bar{\theta}'\left(\Slash{p}-m\right)\theta +
  \frac{1}{2}\bar{\theta}\Slash{p}\theta +
  \frac{1}{2}\bar{\theta}'\Slash{p}\theta'}\frac{i}{p^2-m^2+i\epsilon}\hspace{.25em}\delta^{jk}.
\end{eqnarray}
All the propagators of component fields are combined in this superpropagator.

Since we redefine $M$ as $M \to M + m$, the action becomes
\begin{eqnarray*}
 S = \int\! d^3x\hspace{.25em}d^4\theta\left(\Phi^{j\dagger}e^{-m\bar{\theta}\theta}e^{-V}\Phi^j
				     + cV\right).
\end{eqnarray*}
We define $\tilde{\Phi}^j, \tilde{\Phi}^{j\dagger}$ by
\begin{eqnarray*}
 \tilde{\Phi^j} := e^{-\frac{1}{2}m\bar{\theta}\theta}\Phi^j \quad ,\quad
  \tilde{\Phi}^{j\dagger} := e^{-\frac{1}{2}m\bar{\theta}\theta}\Phi^{j\dagger},
\end{eqnarray*}
then we find
\begin{eqnarray}
 S = \int\! d^3x \hspace{.25em}d^4\theta \left(\tilde{\Phi}^{j\dagger}e^{-V}\tilde{\Phi}^j
				       + cV\right), \label{eq21}
\end{eqnarray}
and the propagator of $\tilde{\Phi}^j$ becomes
\begin{eqnarray}
  \left<\tilde{\Phi}^{j}\!(p,\theta,\bar{\theta})\hspace{.25em}\tilde{\Phi}^{{\dagger
 k}}\!(-p,\theta',\bar{\theta}')\right>_0 &=&  e^{-\bar{\theta}'\left(\Slash{p}-m\right)\theta +
  \frac{1}{2}\bar{\theta}\left(\Slash{p}-m\right)\theta +
  \frac{1}{2}\bar{\theta}'\left(\Slash{p}-m\right)\theta'}\frac{i}{p^2-m^2+i\epsilon}\hspace{.25em}\delta^{jk}. \label{eq20}
\end{eqnarray}
Hereafter, we will use $\tilde{\Phi}^j, \tilde{\Phi}^{\dagger j}$ as the dynamical
fields instead of
$\Phi^j,\Phi^{\dagger j}$.

\subsection{Twisted covariant derivatives}

Since the expression (\ref{eq20}) is slightly complicated, we will rewrite
it in terms of differential operators on superspace, obtained by twisting $D_\alpha, \bar{D}_\alpha$. We call them the ``twisted covariant derivatives''.

We first review the propagator of the ordinary massless chiral superfield.
Although the Lagrangian
\begin{eqnarray}
 {\mathcal L}_{\rm chiral}^{\rm massless} = \int \! d^4\theta \hspace{.3em}
  \Phi^\dagger\Phi  \label{eq33}
\end{eqnarray}
does not have any time derivatives, the constraint
$
 \bar{D}_{\alpha}\Phi = 0
$
contains a time derivative and leads to the non-trivial propagation
of $\Phi$. From the Lagrangian (\ref{eq33}), we can show that the propagator
of $\Phi$ becomes
\begin{eqnarray}
 \left<\Phi\!\left(p,\theta,\bar{\theta}\right)\Phi^{\dagger}\!\left(-p,\theta',\bar{\theta}'\right)\right>_0
 =
 \frac{i}{p^2+i\epsilon}e^{-\bar{\theta}'\Slash{p}\theta+\frac{1}{2}\bar{\theta}\Slash{p}\theta+\frac{1}{2}\bar{\theta}'\Slash{p}\theta'}. \label{eq32}
\end{eqnarray}
It is known that this is equivalent to the following expression
\cite{book1, paper4, book2}:
\begin{eqnarray}
  \left<\Phi\!\left(p,\theta,\bar{\theta}\right)\Phi^{\dagger}\!\left(-p,\theta',\bar{\theta}'\right)\right>_0
 = \frac{i}{p^2+i\epsilon}\cdot\frac{1}{4}\bar{D}(p)^2
 \!D(p)^2\delta^{(4)}\!\left(\theta-\theta'\right) \label{eq30}
\end{eqnarray}
where $D(p)$ and $\bar{D}(p)$ are the covariant derivatives in momentum space:
\begin{eqnarray*}
 D(p)_\alpha = -\frac{\partial}{\partial \theta^\alpha} +
  \frac{1}{2}\left(\bar{\theta}\Slash{p}\right)_\alpha \quad ,\quad
\bar{D}(p)_\alpha = -\frac{\partial}{\partial\bar{\theta}^\alpha} +
\frac{1}{2}\left(\theta\Slash{p}\right)_\alpha,
\end{eqnarray*}
and we define
$
 \delta^{(4)}\!\left(\theta-\theta'\right) := \frac{1}{4}\left(\theta-\theta'\right)^2\left(\bar{\theta}-\bar{\theta}'\right)^2.
$
We can show this equivalence of the equations (\ref{eq32}) and
(\ref{eq30}) through a straightforward calculation.
For the calculation of a loop diagram, the expression (\ref{eq30}) is
more useful than (\ref{eq32}) because we can perform the integration by
parts in
supersupace.

We now
look for a constraint for $\tilde{\Phi}$ such that
\begin{eqnarray*}
 \bar{E}_\alpha\tilde{\Phi} = 0.
\end{eqnarray*}
Since $\tilde{\Phi}\!\left(x,\theta,\bar{\theta}\right) =
e^{-\frac{1}{2}m\bar{\theta}\theta}\Phi\!\left(x,\theta,\bar{\theta}\right)$,
we impose
$
 \bar{E}_\alpha \hspace{.25em}e^{-\frac{1}{2}m\bar{\theta}\theta} = e^{-\frac{1}{2}m\bar{\theta}\theta}\bar{D}_\alpha
$, namely
\begin{eqnarray}
 \bar{E}_\alpha := e^{-\frac{1}{2}m\bar{\theta}\theta}\bar{D}_\alpha
  e^{+\frac{1}{2}m\bar{\theta}\theta} = \bar{D}_\alpha +
  \frac{1}{2}m\hspace{.5mm}\theta_\alpha. \label{eq110}
\end{eqnarray}
We define $E_\alpha$ by the same similarity transformation of $D_\alpha$:
\begin{eqnarray}
 E_\alpha := e^{-\frac{1}{2}m\bar{\theta}\theta}D_\alpha
  e^{+\frac{1}{2}m\bar{\theta}\theta} = D_\alpha +
  \frac{1}{2}m\hspace{.125em}\bar{\theta}_\alpha \label{eq111}
\end{eqnarray}
Secondly, we define another set of differential operators $\bar{H}_\alpha, H_\alpha$ as follows:
\begin{eqnarray*}
 \bar{H}_\alpha := e^{+\frac{1}{2}m\bar{\theta}\theta}\bar{D}_\alpha
  e^{-\frac{1}{2}m\bar{\theta}\theta} =\bar{D}_\alpha -
  \frac{1}{2}m\hspace{.125em}\theta_\alpha,\\[1mm]
 H_\alpha := e^{+\frac{1}{2}m\bar{\theta}\theta}D_\alpha
  e^{-\frac{1}{2}m\bar{\theta}\theta} = D_\alpha - \frac{1}{2}m\hspace{.125em}\bar{\theta}_\alpha.
\end{eqnarray*}
Notice that the sign in front of $m$ is opposite to (\ref{eq110}) and (\ref{eq111}).
We call $E_\alpha, \bar{E}_\alpha$ and $H_\alpha, \bar{H}_\alpha$ the
``twisted covariant derivatives'.
Then we can rewrite the expression (\ref{eq20}) through a
straightforward calculation:
\begin{eqnarray}
 \left<\tilde{\Phi}^j\!\left(p,\theta,\bar{\theta}\right)\tilde{\Phi}^{k\dagger}\!\left(-p,\theta',\bar{\theta}'\right)\right>_0
  = \delta^{jk} \frac{i}{p^2-m^2+i\epsilon}\cdot\frac{1}{4}\bar{E}(p)^2
  \! H(p)^2
  \delta^{(4)}\!\left(\theta-\theta'\right), \label{eq38}
\end{eqnarray}
where
\begin{eqnarray*}
 H(p)_\alpha = D(p)_\alpha - \frac{1}{2}m\hspace{.125em}\bar{\theta}_\alpha \quad ,\quad
  \bar{E}(p)_\alpha = \bar{D}(p)_\alpha + \frac{1}{2}m\hspace{.125em}\theta_\alpha.
\end{eqnarray*}
In appendix D, the derivation of
\begin{eqnarray}
 \frac{1}{4}\bar{E}(p)^2\!H(p)^2
  \delta^{(4)}\!\left(\theta-\theta'\right) =
e^{-\bar{\theta}'\left(\Slash{p}-m\right)\theta +
  \frac{1}{2}\bar{\theta}\left(\Slash{p}-m\right)\theta +
  \frac{1}{2}\bar{\theta}'\left(\Slash{p}-m\right)\theta'} \label{eq120} 
\end{eqnarray}
 is shown in detail.

\subsection{Property of the twisted covariant derivatives}

In the previous subsection, we defined the ``twisted covariant
derivatives'' $E_\alpha, \bar{E}_\alpha$ and $H_\alpha,
\bar{H}_\alpha$. We now investigate the property of these differential operators in
detail. 

We first study $E_\alpha$ and $\bar{E}_\alpha$.
We can easily show the anticommutation relations of $E_\alpha, \bar{E}_\alpha$ are those of covariant derivatives:
\begin{eqnarray*}
 \left\{E^\alpha, \bar{E}_\alpha\right\} =
  i\Slash{\partial}^\alpha_{\hspace{.3em}\beta} \quad , \quad \left\{E^\alpha, E_\beta\right\} = \left\{\bar{E}^\alpha,
\bar{E}_\beta\right\} = 0.
\end{eqnarray*}
 They are indeed supercovariant derivatives
when they act on $\tilde{\Phi}^j, \tilde{\Phi}^{j\dagger}$. To see this explicitly,
recall the definition of the ``twisted chiral superfield'' $\tilde{\Phi}^j =
e^{-\frac{1}{2}m\bar{\theta}\theta}\Phi^j$. Under the infinitesimal
supersymmetry transformation, the chiral superfield $\Phi$ transforms as
$
 \Phi^j \to \left(1+\xi Q + \bar{\xi}\bar{Q}\right)\!\Phi^j
$ 
where $\xi, \bar{\xi}$ are the transformation parameters.
So the transformation law for the ``twisted chiral superfield''
$\tilde{\Phi}^j$ becomes as follows:
\begin{eqnarray}
 \tilde{\Phi}^j \quad \to \quad
  e^{-\frac{1}{2}m\bar{\theta}\theta}\left(1+\xi Q +
				      \bar{\xi}\bar{Q}\right)e^{+\frac{1}{2}m\bar{\theta}\theta}\hspace{.25em}\tilde{\Phi}^j. \label{eq35}
\end{eqnarray}
If we define
\begin{eqnarray*}
 R_\alpha &:=& e^{-\frac{1}{2}m\bar{\theta}\theta}\hspace{.25em}Q_\alpha\hspace{.25em}
  e^{+\frac{1}{2}m\bar{\theta}\theta} = Q_\alpha + \frac{i}{2}m\bar{\theta}_\alpha\\
 \bar{R}_\alpha &:=&
  e^{-\frac{1}{2}m\bar{\theta}\theta}\hspace{.125em}\bar{Q}_\alpha\hspace{.25em}e^{+\frac{1}{2}m\bar{\theta}\theta}
  = \bar{Q}_\alpha - \frac{i}{2}m\theta_\alpha,
\end{eqnarray*}
the transformation law (\ref{eq35}) can be written as
$
 \tilde{\Phi}^j \to \left(1 + \xi R + \bar{\xi}\bar{R}\right)\!\tilde{\Phi}^j.
$
We call $R_\alpha, \bar{R}_\alpha$ ``twisted supercharges'', which of course
satisfy the anticommutation relations
\begin{eqnarray*}
 \left\{R^\alpha, \bar{R}_\beta\right\} =
  -i\Slash{\partial}^\alpha_{\hspace{.3em}\beta} = \left\{Q^\alpha,
  \bar{Q}_\beta\right\} \quad , \quad \left\{R^\alpha, R_\beta\right\} =
  \left\{\bar{R}^\alpha, \bar{R}_\beta\right\} = 0.
\end{eqnarray*}
Supersymmetry transformations for twisted chiral superfields
$\tilde{\Phi}^j, \tilde{\Phi}^{j\dagger}$ are generated by $R_{\alpha}, \bar{R}_\alpha$.

We can show explicitly that twisted
covariant derivatives $E_\alpha, \bar{E}_\alpha$ anticommute with
$R_\alpha, \bar{R}_\alpha$:
\begin{eqnarray*}
 \left\{E^\alpha, \bar{R}_\beta\right\} = \left\{\bar{E}^\alpha,
  R_\beta\right\} = \left\{E^\alpha, R_\beta\right\} =
  \left\{\bar{E}^\alpha, \bar{R}_\beta\right\} = 0.
\end{eqnarray*}
Therefore $E_\alpha, \bar{E}_\alpha$ are
indeed supercovariant derivatives when they act on $\tilde{\Phi}^j,
\tilde{\Phi}^{j\dagger}$.
We now find that the ``twisted chiral
condition'' $\bar{E}_\alpha\tilde{\Phi}^j = 0$ is supercovariant. The
``twisted anti-chiral condition'' for $\tilde{\Phi}^{j\dagger}$ becomes
\begin{eqnarray*}
 E_\alpha \tilde{\Phi}^{j\dagger} = 0
\end{eqnarray*}
and 
also supercovariant.

On the other hand, another set of twisted covariant derivatives
$H_\alpha$ and $\bar{H}_\alpha$ do not anticommute with $R_\alpha$ and $\bar{R}_\alpha$. The anticommutation relations are
\begin{eqnarray*}
 \left\{H^\alpha, \bar{R}_\beta\right\} =
  -im\hspace{.125em}\delta^{\alpha}_{\hspace{.3em}\beta} \quad&,& \quad
  \left\{\bar{H}^\alpha, R_\beta\right\} =
  im\hspace{.125em}\delta^{\alpha}_{\hspace{.3em}\beta}\\[1mm]
\left\{H^\alpha, R_\beta\right\}&=&\left\{\bar{H}^\alpha,
 \bar{R}_\beta\right\} = 0
\end{eqnarray*}
and therefore $H_\alpha, \bar{H}_\alpha$ are not supercovariant
derivatives when they act on $\tilde{\Phi}^j,
\tilde{\Phi}^{j\dagger}$. However, since the only difference between
$H_\alpha,\bar{H}_\alpha$ and $E_\alpha, \bar{E}_\alpha$ is the sign
in front of $m$, $H_\alpha, \bar{H}_\alpha$ are supercovariant derivatives
when they act on
\begin{eqnarray*}
 e^{+\frac{1}{2}m\bar{\theta}\theta}\Phi^j \quad , \quad e^{+\frac{1}{2}m\bar{\theta}\theta}\Phi^{j\dagger},
\end{eqnarray*}
while $\tilde{\Phi}^j = e^{-\frac{1}{2}m\bar{\theta}\theta}\Phi^j$ and
$\tilde{\Phi}^{j\dagger} =
e^{-\frac{1}{2}m\bar{\theta}\theta}\Phi^{j\dagger}$. So there
are two ways of ``twisting'' and we can
define two sets of twisted superfields
(twisted supercharges) and twisted covariant derivatives which are
distinguished by the sign in front of $m$.

Anticommutation relations among supercovariant derivatives in different
sets are as follows:
\begin{eqnarray*}
 \left\{E^\alpha, \bar{H}_\beta\right\} =
  \left(i\Slash{\partial}-m\right)^\alpha_{\hspace{.3em}\beta} \quad &,&
  \quad \left\{H^\alpha, \bar{E}_\beta\right\} = \left(i\Slash{\partial}
						 +
						 m\right)^\alpha_{\hspace{.3em}\beta}\\[1mm]
 \left\{E^\alpha, H_\beta\right\}&=&\left\{\bar{E}^\alpha,
  \bar{H}_\beta\right\} = 0,
\end{eqnarray*}
which will be used frequently in appendix E to show useful formulae for
loop calculations.

In the following, we will show a useful formula for the propagator of
the dynamical field.

Note that the only difference between $E_\alpha,
\bar{E}_\alpha$ and $H_\alpha, \bar{H}_\alpha$ is the sign in front of
$m$. 
Therefore, by replacing $\bar{E}_\alpha$ and $H_\alpha$ with
$\bar{H}_\alpha$ and $E_\alpha$ in the equation (\ref{eq120}), we find
\begin{eqnarray}
 \frac{1}{4}\bar{H}(p)^2 E(p)^2
  \delta^{(4)}\!\left(\theta-\theta'\right) =
  e^{-\bar{\theta}'\left(\Slash{p}+m\right)\theta +
  \frac{1}{2}\bar{\theta}\left(\Slash{p}+m\right)\theta +
  \frac{1}{2}\bar{\theta}'\left(\Slash{p}+m\right)\theta'} \label{eq101}
\end{eqnarray}
If we replace $\theta,\bar{\theta}$ with $\theta',\bar{\theta}'$, we find
\begin{eqnarray}
  \frac{1}{4}\bar{H}'(p)^2 E'(p)^2
  \delta^{(4)}\!\left(\theta'-\theta\right) =
  e^{-\bar{\theta}\left(\Slash{p}+m\right)\theta' +
  \frac{1}{2}\bar{\theta}'\left(\Slash{p}+m\right)\theta' +
  \frac{1}{2}\bar{\theta}\left(\Slash{p}+m\right)\theta}, \label{eq121}
\end{eqnarray}
where $E'(p), \bar{H}'(p)$ stand for twisted covariant derivatives with
 $\theta', \bar{\theta}'$.
Recalling the definition of
twisted covariant derivatives
\begin{eqnarray*}
 \bar{H}'(p)_\alpha = -\frac{\partial}{\partial\bar{\theta}'^\alpha} + \frac{1}{2}\left(\theta'\Slash{p}\right)_\alpha - \frac{1}{2}m\theta'_\alpha \quad&,&
  \quad E'(p)_\alpha = -\frac{\partial}{\partial\theta'^\alpha} + \frac{1}{2}\left(\bar{\theta}'\Slash{p}\right)_\alpha + \frac{1}{2}m\bar{\theta}'_\alpha,
\end{eqnarray*}
we can exchange $\bar{H}'(p)$ and $E'(p)$ by the following
 replacement: $\theta' \leftrightarrow \bar{\theta}'
 ,\hspace{1mm}m \to -m$. Therefore by replacing $\theta
 \leftrightarrow \bar{\theta},\hspace{.25em}\theta'
 \leftrightarrow \bar{\theta}' ,\hspace{.25em} m \to -m$ in
 the equation (\ref{eq121}), we find
\begin{eqnarray*}
   \frac{1}{4}E'(p)^2 \bar{H}'(p)^2
  \delta^{(4)}\!\left(\theta'-\theta\right) =
  e^{-\theta\left(\Slash{p}-m\right)\bar{\theta}' +
  \frac{1}{2}\theta'\left(\Slash{p}-m\right)\bar{\theta}' +
  \frac{1}{2}\theta\left(\Slash{p}-m\right)\bar{\theta}}.
\end{eqnarray*}
Then, at last, using the fact
that $\theta\Slash{p}\bar{\theta} = -\bar{\theta}\Slash{p}\theta$ and
$m\theta\bar{\theta} = m\bar{\theta}\theta$, we obtain the following result:
\begin{eqnarray*}
\frac{1}{4}E'(-p)^2 \bar{H}'(-p)^2
  \delta^{(4)}\!\left(\theta'-\theta\right) =
  e^{-\bar{\theta}'\left(\Slash{p}-m\right)\theta +
  \frac{1}{2}\bar{\theta}'\left(\Slash{p}-m\right)\theta' +
  \frac{1}{2}\bar{\theta}\left(\Slash{p}-m\right)\theta}.
\end{eqnarray*}
Since the right-hand side is the same as that of (\ref{eq120}), the
equation (\ref{eq38}) can be written as
\begin{eqnarray}
 \left<\tilde{\Phi}^j\!\left(p,\theta,\bar{\theta}\right)\tilde{\Phi}^{k\dagger}\!\left(-p,\theta',\bar{\theta}'\right)\right>_0
  &=& \delta^{jk} \frac{i}{p^2-m^2+i\epsilon}\cdot\frac{1}{4}\bar{E}(+p)^2 \! H(+p)^2
  \delta^{(4)}\!\left(\theta-\theta'\right)\nonumber \\[1mm]
 &=& \delta^{jk} \frac{i}{p^2-m^2+i\epsilon}\cdot\frac{1}{4}E'\!(-p)^2 \! \bar{H}'\!(-p)^2
  \delta^{(4)}\!\left(\theta-\theta'\right). \label{eq202}
\end{eqnarray}
We will use this formula in the calculation of the loop diagrams of
the twisted chiral superfield.

\vspace{5ex}

\section{Propagator of the auxiliary field}

In the previous section, we studied the propagator of the dynamical
field. In this section, we will investigate that of the auxiliary field. Although the auxiliary field has no kinetic term in the classical
level, the effective action contains the quadratic term of the auxiliary
field induced by quantum 
effects of the dynamical field. We use the quadratic term induced by
1-loop diagrams of the dynamical field, which is the leading order of $1/N$, as a
kinetic term in order to calculate the
propagator of the auxiliary field. Therefore we have to calculate the one-loop diagram of the
dynamical field in order to obtain the propagator of the auxiliary field
in the Large $N$ expansion.

\subsection{One-loop diagram of the dynamical field}

Since the action of the theory is
\begin{eqnarray}
 S = \int\! d^3x \hspace{.25em}d^4\theta
  \left(\tilde{\Phi}^{j\dagger}e^{-V}\tilde{\Phi}^j + cV\right) \qquad
  \left(j=1\sim N\right) \label{eq200}
\end{eqnarray}
the auxiliary field $V$ does not have the kinetic term at the tree
level.
If we perform the path integration over $\tilde{\Phi},
\tilde{\Phi}^\dagger$, we obtain the effective action $S_{\rm eff}$ when the auxiliary
field $V$ are treated as the external background field while the
dynamical fields $\tilde{\Phi},\tilde{\Phi}^\dagger$ fluctuate:
\begin{eqnarray*}
 S_{\rm eff}= S +
  \frac{1}{2}\int\!d^3x\hspace{.125em}d^4\theta
  \hspace{.25em}V\left(iG^{-1}\right)V + \cdots.
\end{eqnarray*}
The quadratic term of the auxiliary field define the inverse propagator
of it, which comes from one-loop diagrams of the dynamical field:
\newsavebox{\boxa}
\newsavebox{\boxb}
\sbox{\boxa}{\includegraphics[width=6cm]{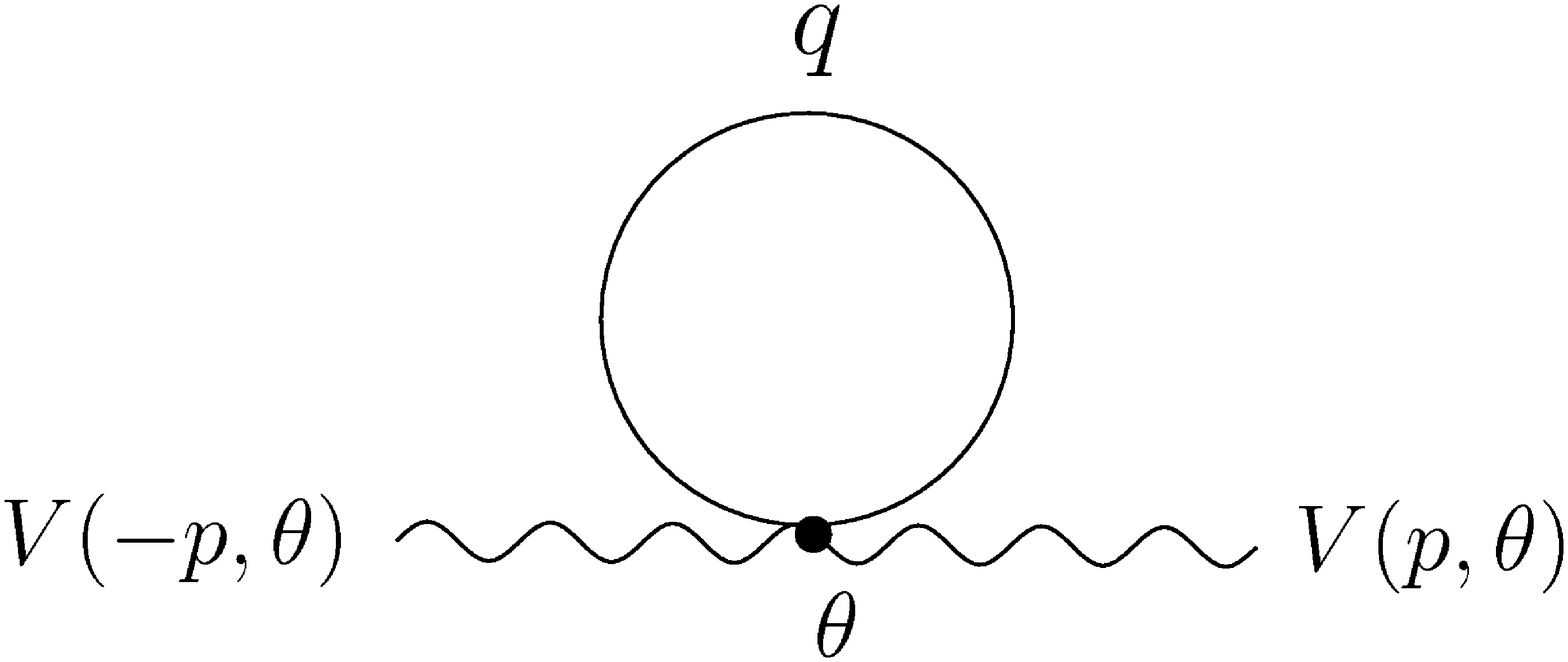}}
\sbox{\boxb}{\includegraphics[width=6cm]{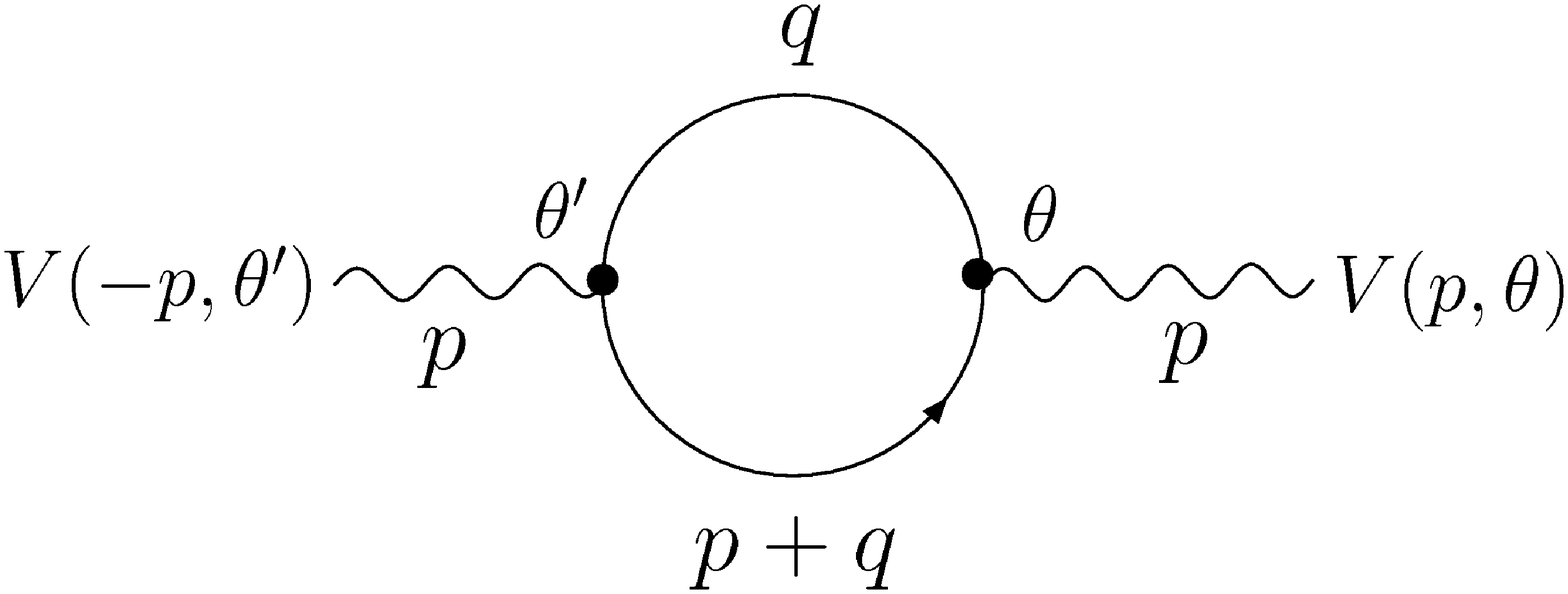}}
\newlength{\la}
\newlength{\lb}
\settowidth{\la}{\usebox{\boxa}}
\settowidth{\lb}{\usebox{\boxb}}
\begin{eqnarray}
 \parbox{\la}{\usebox{\boxa}}
\hspace{.5em} +\hspace{1em}
\parbox{\lb}{\usebox{\boxb}}. \label{eq201}
\end{eqnarray}
Feynman rules are as follows
\sbox{\boxa}{\includegraphics[width=6cm]{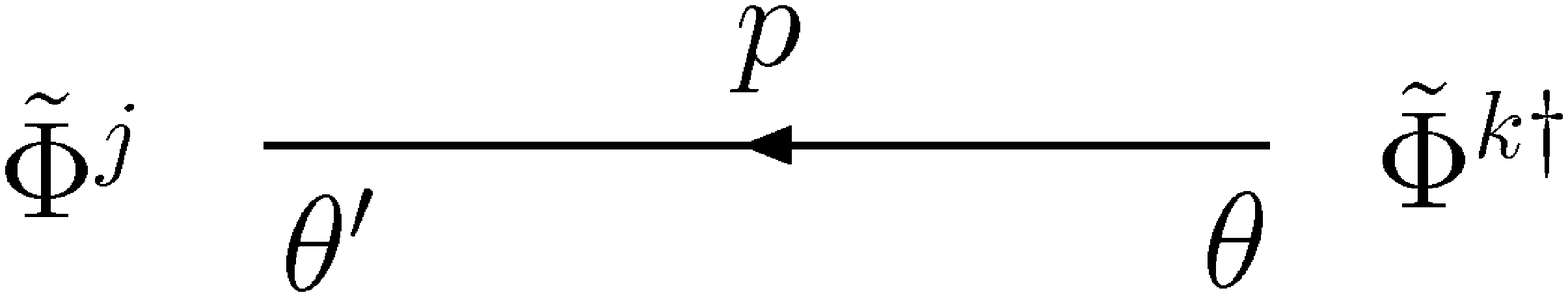}}
\sbox{\boxb}{\includegraphics[width=5cm]{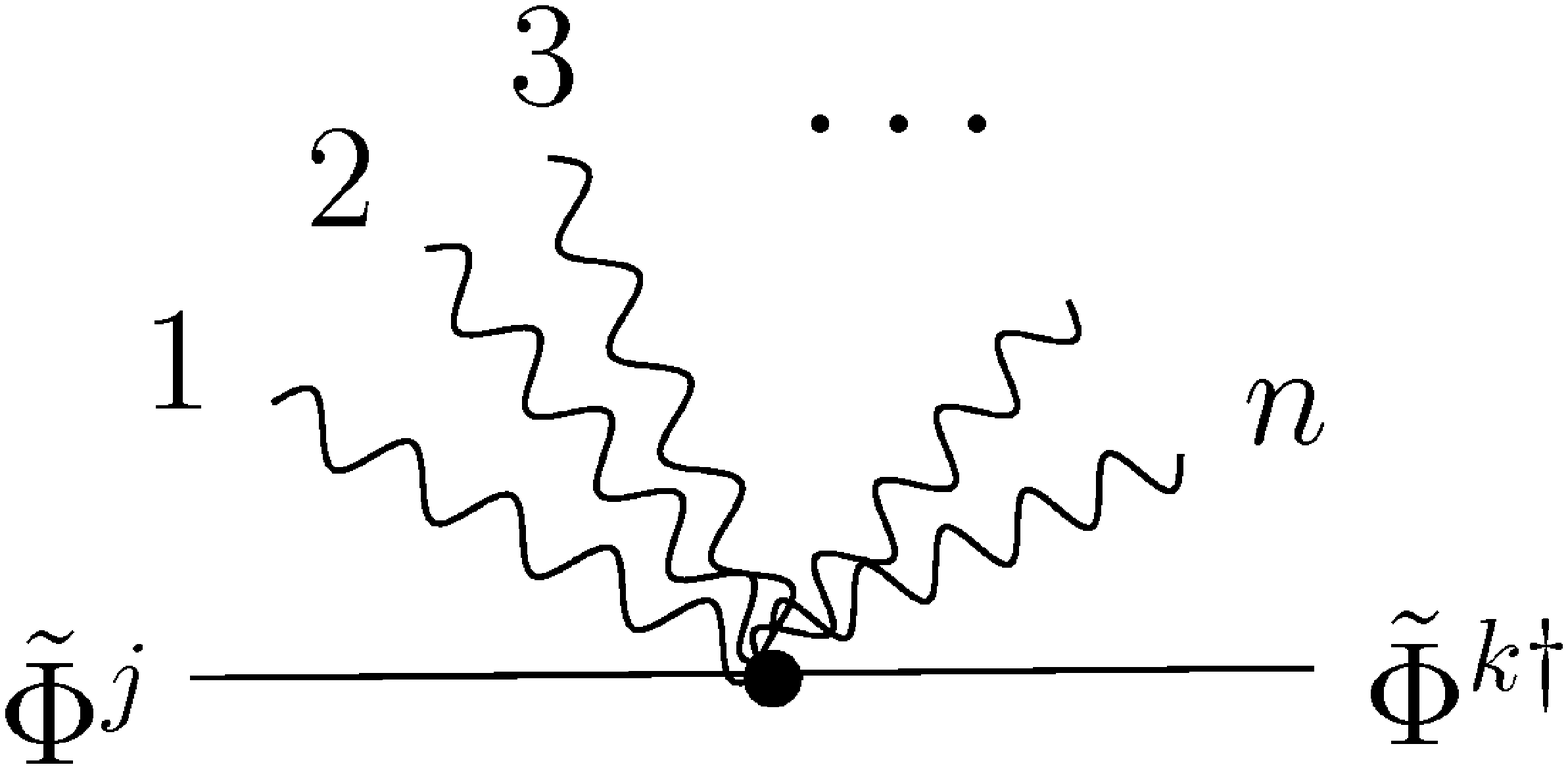}}
\settowidth{\la}{\usebox{\boxa}}
\settowidth{\lb}{\usebox{\boxb}}
\begin{eqnarray*}
\parbox{\la}{\usebox{\boxa}} &=& \delta^{jk}\frac{i}{p^2-m^2+i\epsilon}\cdot\frac{1}{4}\bar{E}(p)^2 H(p)^2\delta^{(4)}\!\left(\theta-\theta'\right),\\[2mm]
\parbox{\lb}{\usebox{\boxb}} &=& i\left(-1\right)^n\delta^{jk}.
\end{eqnarray*}
Then the contribution of the first diagram in (\ref{eq201}) can be evaluated as
\begin{eqnarray*}
\int\!\!\frac{d^3p}{\left(2\pi\right)^3}\int \! d^4\theta
  \hspace{.5em}V(-p,\theta)\cdot i\left(-1\right)^2 \!\int\!\!\frac{d^3q}{\left(2\pi\right)^3}\hspace{.25em}
  \frac{iN}{q^2-m^2+i\epsilon}\left.\left[\frac{1}{4}\bar{E}(q)^2
					  H(q)^2
					  \delta^{(4)}\!\left(\theta-\theta'\right)\right]\right|_{\theta'=\theta, \bar{\theta}'=\bar{\theta}}\cdot V(p,\theta).
\end{eqnarray*}
Since $\delta^{(4)}\!\left(\theta-\theta'\right) =
\frac{1}{4}\left(\theta-\theta'\right)^2\left(\bar{\theta}-\bar{\theta}'\right)^2$,
we can show
\begin{eqnarray}
\hspace{-.5em} \left.\left[\frac{1}{4}\bar{E}(q)^2
					  H(q)^2
					  \delta^{(4)}\!\left(\theta-\theta'\right)\right]\right|_{\theta'=\theta,
 \bar{\theta}'=\bar{\theta}}
 &=&
 \frac{1}{16}\left.\left[\frac{\partial}{\partial\bar{\theta}^\alpha}\frac{\partial}{\partial\bar{\theta}_\alpha}\frac{\partial}{\partial\theta^\beta}\frac{\partial}{\partial\theta_\beta}\left(\theta-\theta'\right)^2\left(\bar{\theta}-\bar{\theta}'\right)^2\right]\right|_{\theta'=\theta,\bar{\theta}'=\bar{\theta}} = \hspace{.5em}1.\hspace{2.5em} \label{eq260}
\end{eqnarray}
Therefore the contribution of the first diagram in (\ref{eq201}) becomes
\sbox{\boxa}{\includegraphics[width=5.5cm]{1.eps}}
\settowidth{\la}{\usebox{\boxa}}
\begin{eqnarray}
\parbox{\la}{\usebox{\boxa}} =  -\int\!\!\frac{d^3p}{\left(2\pi\right)^3}\int\!d^4\theta
  \hspace{.5em}V(-p,\theta)\cdot\int\!\!\frac{d^3q}{\left(2\pi^3\right)}\frac{N}{q^2-m^2+i\epsilon}
  \cdot V(p,\theta). \label{eq255}
\end{eqnarray}
In the right-hand side, the momentum integral over $q$ has a linear
divergence. So we here regularize the integral, for instance by
introducing a momentum cut-off.

\subsection{Integration by parts at vertices}

We now evaluate the contribution to the effective action from the second diagram in (\ref{eq201}):
\sbox{\boxa}{\includegraphics[width=6cm]{2.eps}}
\settowidth{\la}{\usebox{\boxa}}
\begin{eqnarray}
\hspace{-2.5em}&\parbox{\la}{\usebox{\boxa}}& = N\int\!\!\frac{d^3p}{\left(2\pi\right)^3}\frac{d^3q}{\left(2\pi\right)^3}\int\!d^4\theta\hspace{.25em}d^4\theta'\hspace{.25em}V(-p,\theta')\cdot
  \left(-i\right)^2\frac{i}{\left(p+q\right)^2-m^2+i\epsilon}\nonumber
  \\[1mm]
&&\hspace{-12.5em}\times\frac{i}{q^2-m^2+i\epsilon}
\left[\frac{1}{4}\bar{E}'(p+q)^2 H'(p+q)^2\delta^{(4)}\!\left(\theta'-\theta\right)\right]\left[\frac{1}{4}\bar{E}(q)^2
   H(q)^2\delta^{(4)}\!\left(\theta-\theta'\right)\right]
 \cdot V(p,\theta),\hspace{2.5em}  \label{eq203}
\end{eqnarray}
where $\bar{E}'(p+q)$ and $H'(p+q)$ are twisted covariant derivatives with
$\theta', \bar{\theta}'$ and momentum $p+q$. Using the equation
(\ref{eq202}), we can rewrite the term in the first bracket:
\begin{eqnarray}
\frac{1}{4}\bar{E}'(p+q)^2
  H'(p+q)^2\delta^{(4)}\!\left(\theta'-\theta\right) =
  \frac{1}{4}E(-p-q)^2
  \bar{H}(-p-q)^2\delta^{(4)}\!\left(\theta-\theta'\right).\label{eq204}
\end{eqnarray}
In the following, we perform the $\theta,\bar{\theta}$ integration by
parts, then we can apply $E(-p-q)^2\bar{H}(-p-q)^2$ to the second bracket and
$V(p,\theta)$ in the equation (\ref{eq203}). We first note the partial
integration rule for covariant derivatives \cite{book2}:
\begin{eqnarray*}
 \int\! d^4\theta\hspace{.25em} \left\{D(p)_\alpha \hspace{.125em}A\right\} B = -\int\!
  d^4\theta \hspace{.25em}\left(-1\right)^{\left|A\right|}A\left\{D(-p)_\alpha \hspace{.125em}B\right\},
\end{eqnarray*}
where $\left|A\right|=1$ for grassmann-odd $A$ and $\left|A\right|=0$
for grassmann-even $A$. 
Note that the sign in front of
$-\frac{\partial}{\partial\theta^\alpha}$ and
$\frac{1}{2}\left(\bar{\theta}\Slash{p}\right)_\alpha$ become opposite
after the integration by parts in the definition of the covariant
derivative $D(p)_\alpha = - \frac{\partial}{\partial\theta^\alpha} + \frac{1}{2}\left(\bar{\theta}\Slash{p}\right)_\alpha$.
Namely, through the integration by parts, $D(p)_\alpha$ becomes
$D(-p)_\alpha$ and $\bar{D}(p)$ becomes $\bar{D}(-p)$ as well.

We now recall the definition of twisted covariant derivatives:
\begin{eqnarray*}
 E(p)_\alpha = D(p)_\alpha +
  \frac{1}{2}m\hspace{.125em}\bar{\theta}_\alpha \quad &,& \quad
  \bar{E}(p)_\alpha = \bar{D}(p)_\alpha + \frac{1}{2}m
  \hspace{.125em}\theta_\alpha \\
H(p)_\alpha = D(p)_\alpha - \frac{1}{2}m\hspace{.125em}\bar{\theta}_\alpha
 \quad &,& \quad \bar{H}(p)_\alpha = \bar{D}(p)_\alpha - \frac{1}{2}m\hspace{.125em}\theta_\alpha.
\end{eqnarray*}
Then we find in the similar way that $E(p)_\alpha$ becomes $H(-p)_\alpha$ and
$\bar{H}(p)_\alpha$ becomes $\bar{E}(-p)_\alpha$ through the integration
by parts. 

If we take the following operator 
\begin{eqnarray}
 \int\!\!\frac{d^3p}{\left(2\pi\right)^3}\frac{d^3q}{\left(2\pi\right)^3}\int \!\! d^4\theta \hspace{.25em} \left\{E(-p-q)_\alpha
				   \hspace{.125em}\tilde{\Phi}^{\dagger}(-p-q)\right\}
 V\left(p\right)\tilde{\Phi}(q), \label{eq210}
\end{eqnarray}
we can integrate by parts and move $E(-p-q)_\alpha$ to
$V(p)\tilde{\Phi}(q)$ by substitfuting $H(p+q)_\alpha$ for it:
\begin{eqnarray}
 -\int\!\!\frac{d^3p}{\left(2\pi\right)^3}\frac{d^3q}{\left(2\pi\right)^3}
  \int \!\! d^4\theta \hspace{.5em} \tilde{\Phi}^{\dagger}(-p-q)\cdot
  H(p+q)_\alpha\left\{V\left(p\right)\tilde{\Phi}(q)\right\}. \label{eq211}
\end{eqnarray}
Then we can distribute $H(p+q)_\alpha$ to $V(p)$ and $\tilde{\Phi}(q)$
as
\begin{eqnarray}
 H(p+q)\left\{V(p)\tilde{\Phi}(q)\right\} =
  \left\{D(p)V(p)\right\}\tilde{\Phi}(q) +
  V(p)\left\{H(q)\tilde{\Phi}(q)\right\}, \label{eq212}
\end{eqnarray}
where we should recall $H(p+q)_\alpha = -\frac{\partial}{\partial\theta^\alpha} +
\frac{1}{2}\left(\bar{\theta}\Slash{p}\right)_\alpha +
\frac{1}{2}\left(\bar{\theta}\Slash{q}\right)_\alpha -
\frac{1}{2}m\hspace{.125em}\bar{\theta}_\alpha$ and $D(p)_\alpha =
-\frac{\partial}{\partial\theta^\alpha} + \frac{1}{2}\left(\bar{\theta}\Slash{p}\right)_\alpha$. We apply the
Leibnitz rule for $-\frac{\partial}{\partial\theta^\alpha}$ and
distribute $\frac{1}{2}\left(\bar{\theta}\Slash{p}\right)_\alpha$ to
$V(p)$ and $\frac{1}{2}\left(\bar{\theta}\Slash{q}\right) - \frac{1}{2}m\hspace{.5mm}\bar{\theta}_\alpha$ to
$\tilde{\Phi}(q)$. The momenta in the covariant derivatives should be
chosen as the momenta of the fields on which they act. We frequently use this kind of integration by parts at interaction
 vertices. Since each interaction vertex contains one pair of
$\tilde{\Phi}$ and $\tilde{\Phi}^\dagger$, through the integration by
parts, we move 
$\frac{1}{2}m\hspace{.5mm}\bar{\theta}_\alpha$ or
$\frac{1}{2}m\hspace{.5mm}\theta_\alpha$ in twisted covariant
 derivatives from $\tilde{\Phi}$ to
$\tilde{\Phi}^\dagger$, or vice versa. We never distribute
 $\frac{1}{2}m\hspace{.5mm}\bar{\theta}_\alpha,
 \frac{1}{2}m\hspace{.5mm}\theta_\alpha$ to the auxiliary field $V$.

Therefore, the rules for partial integration are as follows: twisted
covariant derivatives $E_\alpha, \bar{E}_\alpha \left(H_\alpha,
\bar{H}_\alpha\right)$ are replaced by another set of twisted covariant
derivatives $H_\alpha, \bar{H}_\alpha \left(E_\alpha,
\bar{E}_\alpha\right)$ when they act on the dynamical fields
$\tilde{\Phi}, \tilde{\Phi}^\dagger$, while they act as ordinary
covariant derivatives $D_\alpha, \bar{D}_\alpha$ on the auxiliary
field. The momenta in covariant derivatives should be chosen as the
momenta of the fields on which they act. We can draw the result of the integration by parts (\ref{eq210})-(\ref{eq212}) as follows:
\sbox{\boxa}{\includegraphics[width=6cm]{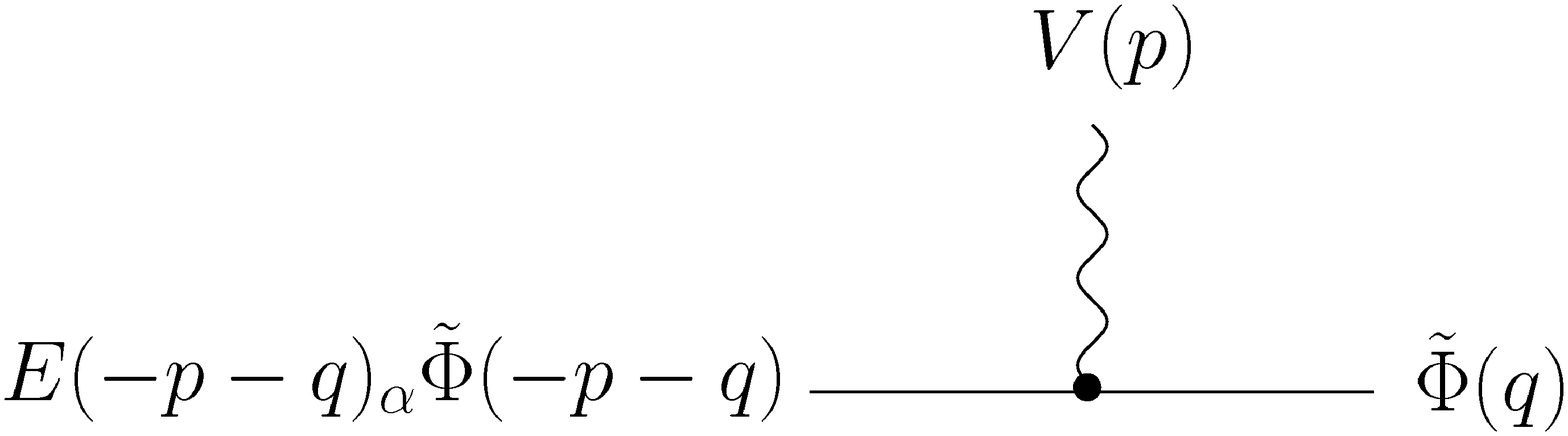}}
\settowidth{\la}{\usebox{\boxa}}
\sbox{\boxb}{\includegraphics[width=5cm]{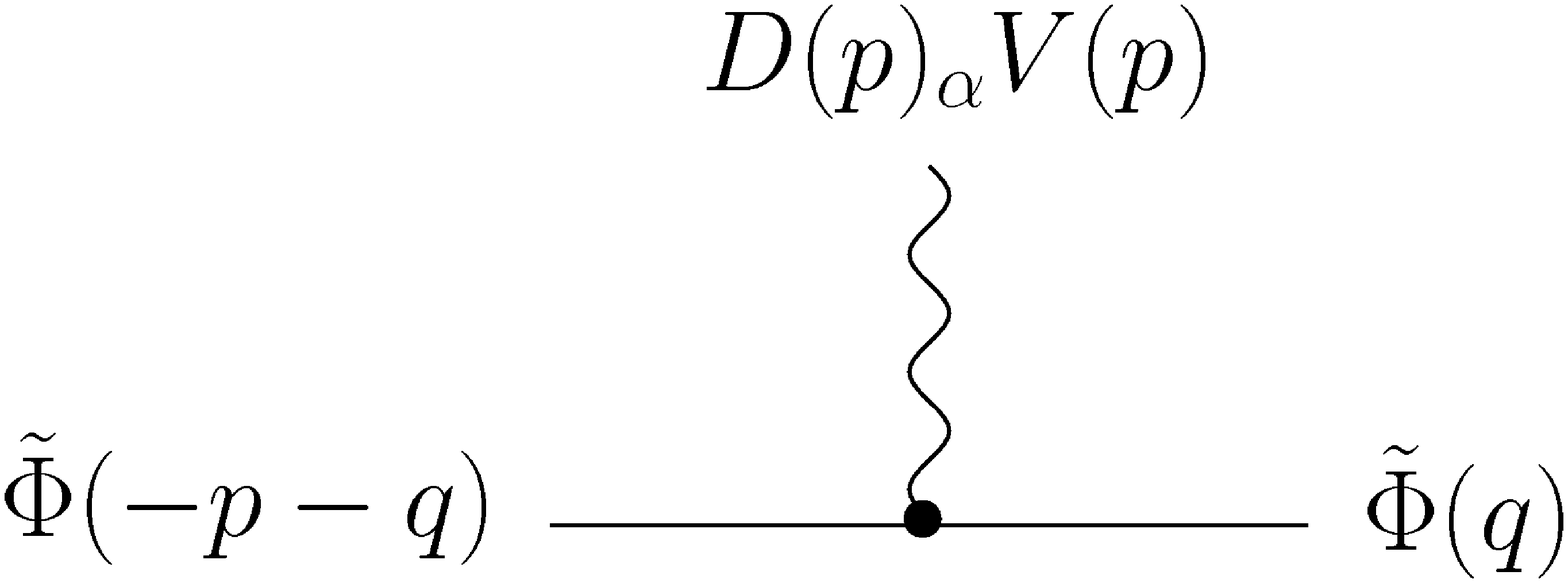}}
\settowidth{\lb}{\usebox{\boxb}}
\newsavebox{\boxc}
\newlength{\lc}
\sbox{\boxc}{\includegraphics[width=5.5cm]{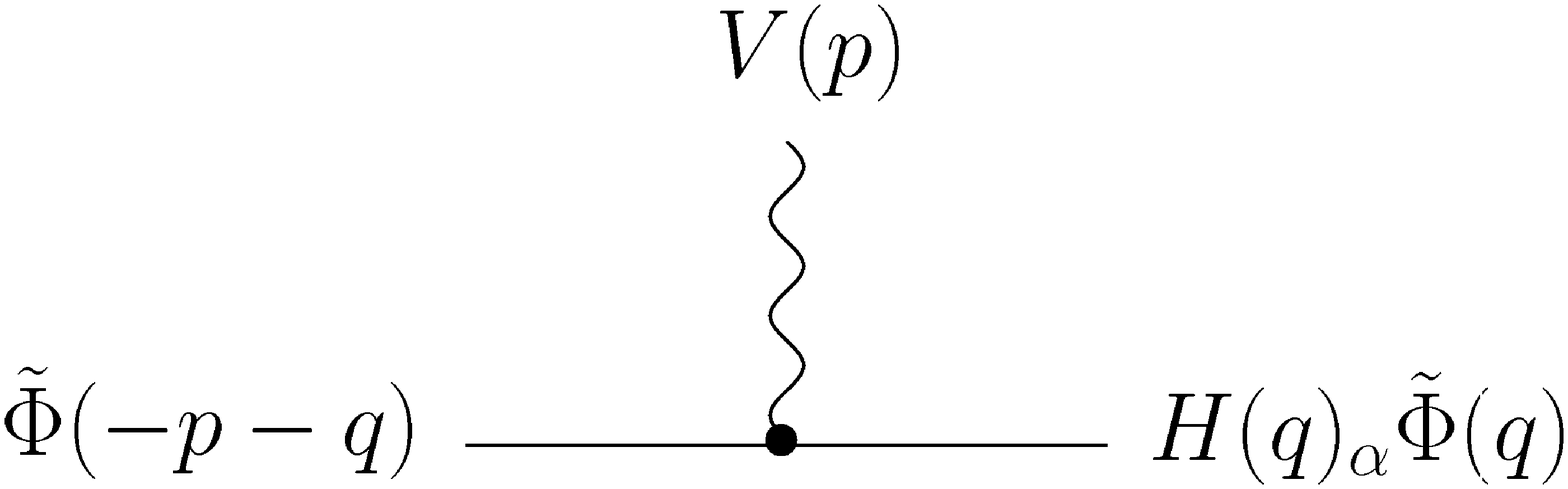}}
\settowidth{\lc}{\usebox{\boxc}}
\begin{eqnarray*}
&&\hspace{-2.5em}\parbox{\la}{\usebox{\boxa}}\\[3mm]
&&\hspace{2.5em}\Longrightarrow\quad\hspace{.5em}-\hspace{.5em}\left[\hspace{.5em}\parbox{\lb}{\usebox{\boxb}}\hspace{.5em}\right]\hspace{.25em}-
\hspace{.5em}\left[\hspace{.25em}\parbox{\lc}{\usebox{\boxc}}\hspace{.5em}\right].
\end{eqnarray*}

Then we perform $\theta,\bar{\theta}$-integration by parts in the
equation (\ref{eq203}). Recall the term in the first bracket has been rewritten
as (\ref{eq204}). We apply, through the integration by parts, $E(p-q)^2
\bar{H}(p-q)^2$ in (\ref{eq204}) to
the second bracket and $V(p,\theta)$ in (\ref{eq203}). We will obtain 16
terms if we perform the integration by parts straightforwardly. To avoid
complicated expressions, 
we rewrite the products of twisted covariant derivatives in (\ref{eq204}) as follows:
\begin{eqnarray}
&& \frac{1}{4}E(-p-q)^2\bar{H}(-p-q)^2\nonumber \\[2mm]
&& \hspace{2.6em}= \frac{1}{4}\bar{H}(-p-q)^2 E(-p-q)^2 -
  \bar{H}(-p-q)\left\{\left(-\Slash{p}-\Slash{q}\right) +
		m\right\}E(-p-q) +
  \left\{\left(-p-q\right)^2-m^2\right\}, \hspace{2.5em}\label{eq250}
\end{eqnarray}
which is shown in appendix E (see proposition E-2). Noting that the
order of $E$s and $\bar{H}$s is reversed, we see that many terms vanish
through the integration by parts.
For instance, if we integrate by parts and move
$\bar{H}(-p-q)_\alpha$ in the first term, it acts as $\bar{E}(q)_\alpha$ on
the second bracket and as $\bar{D}(p)_\alpha$ on $V(p)$.
However, since the second bracket in
(\ref{eq203}) already has $\bar{E}^2$, it vanishes when $\bar{E}(q)_\alpha$
acts on it. Similarly, we will easily find many
terms vanish through the integration by parts if we use the above equation. 

We first evaluate the contribution of the third term in the right-hand side of
(\ref{eq250}). Since this term has no covariant derivatives, we can
easily evaluate it by substituting \hspace{.25em}$\left\{(-p-q)^2-m^2\right\}\delta^{(4)}\!\left(\theta-\theta'\right)$ for the first bracket in (\ref{eq203}):
\begin{eqnarray*}
&& N\int
  \!\!\frac{d^3p}{\left(2\pi^3\right)}\frac{d^3q}{\left(2\pi\right)^3}\int
  \! d^4\theta\hspace{.25em}d^4\theta' \hspace{.5em}V(-p,\theta')
  \frac{1}{q^2-m^2+i\epsilon}\cdot \left[\delta^{(4)}\!\left(\theta-\theta'\right)\frac{1}{4}\bar{E}(q)^2
  H(q)^2\delta^{(4)}\!\left(\theta-\theta'\right)\right] \cdot
  V(p,\theta) \\[2mm]
 &&\hspace{2.5em}= +\int
  \!\!\frac{d^3p}{\left(2\pi^3\right)}\int
  \! d^4\theta \hspace{.5em}V(-p,\theta')
   \cdot \int \!\!\frac{d^3q}{\left(2\pi\right)^3} \frac{N}{q^2-m^2+i\epsilon}
  \cdot V(p,\theta),
\end{eqnarray*}
where we should recall the equation (\ref{eq260}). This exactly cancels
the linearly divergent contribution of (\ref{eq255}).

Then we evaluate the contribution of the second term in
(\ref{eq250}). We first apply $\bar{H}(-p-q)$ to the second
bracket and $V(p,\theta)$ in (\ref{eq203}). However, for the reason
mentioned before, the second bracket vanish if we apply $\bar{H}(-p-q)$
on it as $\bar{E}(q)$. So we apply $\bar{H}(-p-q)$ to $V(p,\theta)$
as $\bar{D}(p)$ through the integration by parts. On the other hand, the twisted
covariant derivative $E(-p-q)$ can be distributed to both the second bracket and
$V(p,\theta)$ in (\ref{eq203}). But if we operate it to the second bracket, we obtain the following factor
\begin{eqnarray}
 \delta^{(4)}\!\left(\theta-\theta'\right) H(q)_\alpha
  \hspace{.125em}\bar{E}(q)^2H(q)^2\hspace{.125em}\delta^{(4)}\!\left(\theta-\theta'\right), \label{eq301}
\end{eqnarray}
and using the commutation relation of $\left[H(q)_\alpha\hspace{.5mm},\hspace{.5mm}
\bar{E}(q)^2\right] =
-2\left[\bar{E}(q)\!\left(\Slash{q}-m\right)\right]_\alpha$ (again see
appendix E) we can rewrite this as
\begin{eqnarray*}
 -2\hspace{.25em}\delta^{(4)}\!\left(\theta-\theta'\right)
  \left[\bar{E}(q)\!\left(\Slash{q}-m\right)\right]_\alpha H(q)^2\hspace{.25em}\delta^{(4)}\!\left(\theta-\theta'\right).
\end{eqnarray*}
We find this is zero due to the property of
grassmann variables. Since $\delta^{(4)}\!\left(\theta-\theta'\right) =
\frac{1}{4}\left(\theta-\theta'\right)^2\left(\bar{\theta}-\bar{\theta}'\right)^2$,
the term containing two delta functions vanishes unless we have
four
derivatives
$\left(\frac{\partial}{\partial\theta^\alpha}\right)^2\!\!\left(\frac{\partial}{\partial\overline{\theta}^\alpha}\right)^2$
between them. Therefore, a non-zero contribution of the second term in
(\ref{eq250}) comes only from the term where
$\bar{H}\!\left(-\Slash{p}-\Slash{q}+m\right)\!E$ is applied to the
auxiliary field. Noting that with grassmann-even
fields $A, B$
\begin{eqnarray}
 \int \! d^4\theta
  \hspace{.25em}\left\{\bar{H}(-p)\!\left(-\Slash{p}+m\right)\!E(-p)
	       \hspace{.5mm}A\right\} \cdot B &=& + \int \! d^4\theta
  \hspace{.25em}\left\{\left(-\Slash{p}+m\right)^\alpha_{\hspace{.3em}\beta}E(-p)^\beta
 A\right\} \cdot \left\{\bar{E}(p)_\alpha
  B\right\}\nonumber \\
 &=& -\left(-\Slash{p}+m\right)^\alpha_{\hspace{.3em}\beta}\int
  \!d^4\theta\hspace{.25em} A\cdot \left\{H(p)^\beta\bar{E}(p)_\alpha
			    \hspace{.125em}B\right\} \nonumber\\
 &=& -\left(-\Slash{p}-m\right)^{\hspace{.3em}\alpha}_{\beta}\int
  \!d^4\theta\hspace{.25em} A\cdot \left\{H(p)^\beta\bar{E}(p)_\alpha
			    \hspace{.125em}B\right\}\nonumber\\
 &=& \int \! d^4\theta\hspace{.25em}A\cdot
  \left\{H(p)\!\left(\Slash{p}+m\right)\!\bar{E}(p)\hspace{.125em}B\right\}, \label{eq775}
\end{eqnarray}
the contribution of the second term of (\ref{eq250}) becomes
\begin{eqnarray}
&& -N\int\!\!\frac{d^3p}{\left(2\pi\right)^3}\frac{d^3q}{\left(2\pi\right)^3}\int\!
  d^4\theta\hspace{.25em}d^4\theta'\hspace{.25em}V(-p,\theta')\hspace{.25em}\delta^{(4)}\!\left(\theta-\theta'\right)\left[\frac{1}{4}\bar{E}(q)^2
							    H(q)^2\delta^{(4)}\!\left(\theta-\theta'\right)\right]\nonumber \\[2mm]
&&\hspace{5em}\times\hspace{.25em}\frac{1}{\left(p+q\right)^2-m^2+i\epsilon}\hspace{.5em}\frac{1}{q^2-m^2+i\epsilon}\hspace{.25em}D(p)\!\left\{\Slash{p}+\Slash{q}+m\right\}\!\bar{D}(p) \label{eq300}
\hspace{.125em}V(p,\theta).
\end{eqnarray}
We can graphically express this integration by parts as follows:
\sbox{\boxa}{\includegraphics[width=6cm]{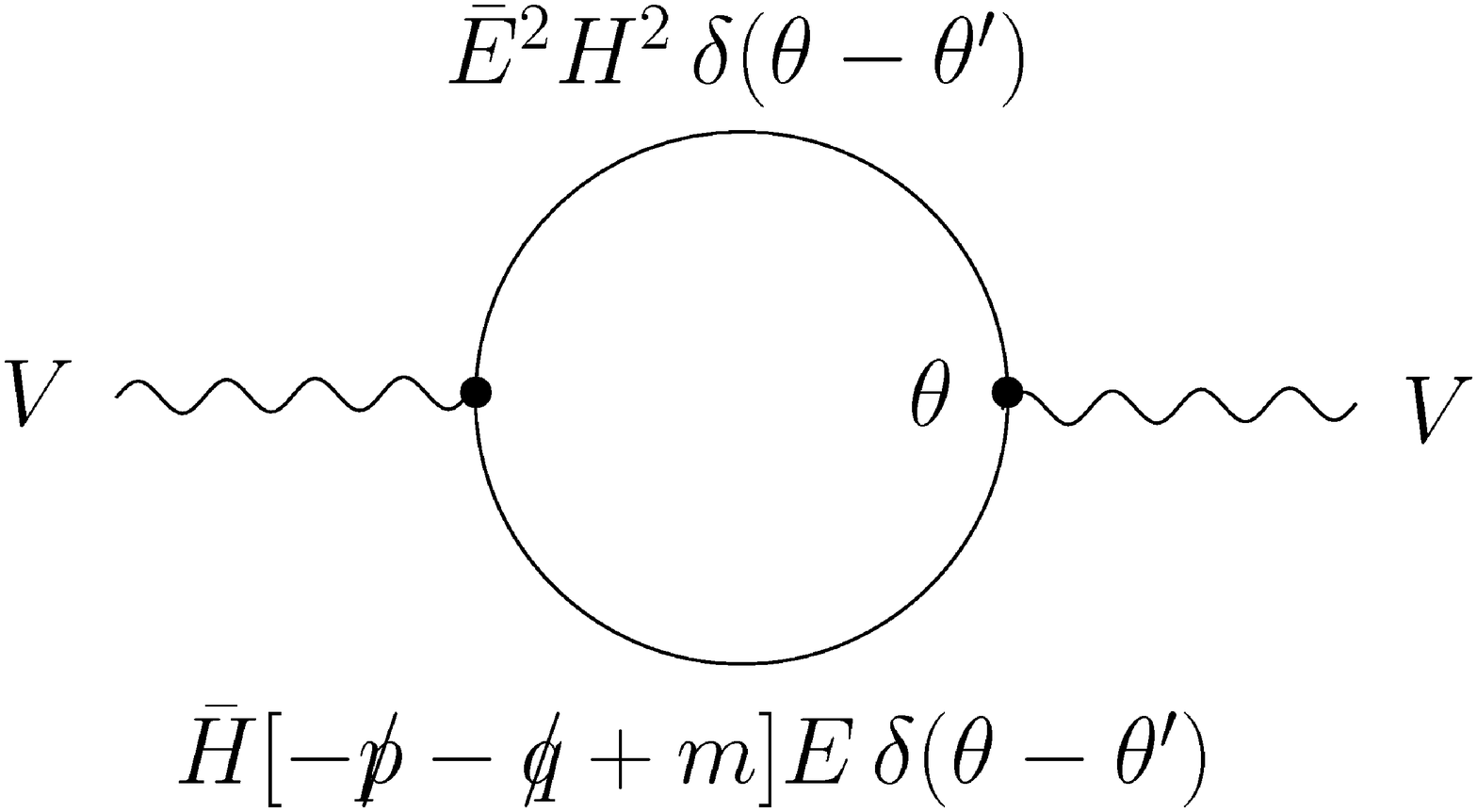}}
\settowidth{\la}{\usebox{\boxa}}
\sbox{\boxb}{\includegraphics[width=8cm]{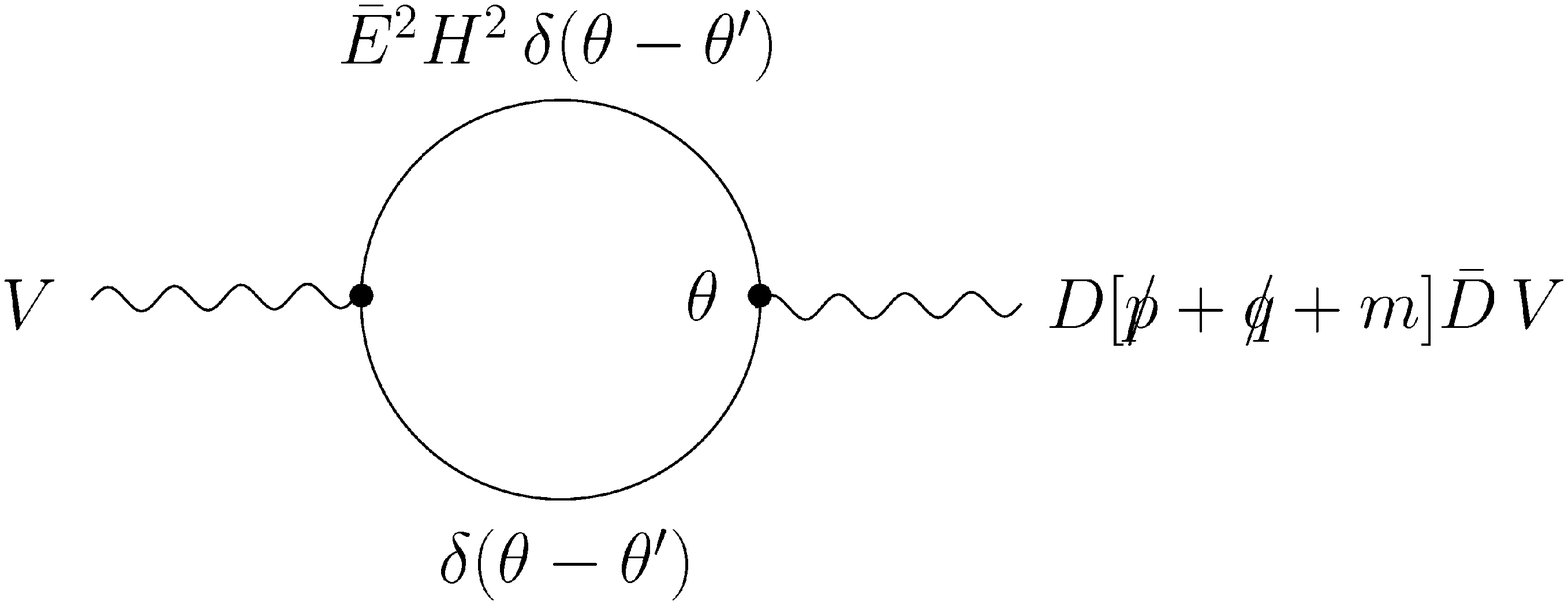}}
\settowidth{\lb}{\usebox{\boxb}}
\begin{eqnarray*}
 \hspace{-.2em}\parbox{\la}{\usebox{\boxa}} \qquad \Longrightarrow \qquad \parbox{\lb}{\usebox{\boxb}}.
\end{eqnarray*}
Recalling the equation (\ref{eq260}) again, we evaluate (\ref{eq300}) as follows:
\begin{eqnarray*}
 && -N\int\!\!\frac{d^3p}{\left(2\pi\right)^3}\frac{d^3q}{\left(2\pi\right)^3}\int\!
  d^4\theta\hspace{.25em}V(-p,\theta)\hspace{.25em}\frac{1}{\left(p+q\right)^2-m^2+i\epsilon}\hspace{2mm}\frac{1}{q^2-m^2+i\epsilon}\hspace{.25em}D(p)\!\left\{\Slash{p}+\Slash{q}+m\right\}\!\bar{D}(p)
\hspace{.125em}V(p,\theta).
\end{eqnarray*}
When we use the identity
\begin{eqnarray*}
 \int\!\!\frac{d^3q}{\left(2\pi\right)^3}\frac{\Slash{q}}{\left[\left(p+q\right)^2-m^2+i\epsilon\right]\left(q^2-m^2+i\epsilon\right)}
  = \int\!\!\frac{d^3q}{\left(2\pi\right)^3}\frac{-\frac{1}{2}\Slash{p}}{\left[\left(p+q\right)^2-m^2+i\epsilon\right]\left(q^2-m^2+i\epsilon\right)},
\end{eqnarray*}
which is easily shown by shifting the integration variables $q \to -q-p$
in the left-hand side, we obtain the following result:
\begin{eqnarray}
 -iN\int\!\!\frac{d^3p}{\left(2\pi\right)^3}\int\!d^4\theta\hspace{.25em}V(-p,\theta)
  D(p)\left(\frac{\Slash{p}}{2}+m\right)\bar{D}(p) V(p,\theta) \cdot
  \frac{1}{4\pi}I(p^2)^{-1}, \label{eq310}
\end{eqnarray}
where we define $I(p^2)^{-1}$ as
\begin{eqnarray*}
 I(p^2)^{-1} &:=&
  \frac{4\pi}{i}\int\!\!\frac{d^3q}{\left(2\pi\right)^3}\hspace{.25em}\frac{1}{\left(p+q\right)^2-m^2+i\epsilon}\hspace{.25em}\frac{1}{q^2-m^2+i\epsilon}
  = \frac{\arctan\sqrt{\frac{-p^2}{4m^2}}}{\sqrt{-p^2}}.
\end{eqnarray*}
The second equality is shown in appendix F. This definition of
$I(p^2)^{-1}$ is the same as that in \cite{paper1}.

We now evaluate the contribution of the frist term in (\ref{eq250})
\begin{eqnarray*}
 \frac{1}{4}\bar{H}(-p-q)^2 E(-p-q)^2\delta^{(4)}\!\left(\theta-\theta'\right).
\end{eqnarray*} 
We integrate by parts and apply $\frac{1}{4}\bar{H}^2 E^2$ to the second bracket
and $V(p,\theta)$ in (\ref{eq203}). For the same reason as before, we
first apply $\bar{H}(-p-q)^2$ only to $V(p,\theta)$ as
$\bar{D}(p)^2$:
\sbox{\boxa}{\includegraphics[width=6cm]{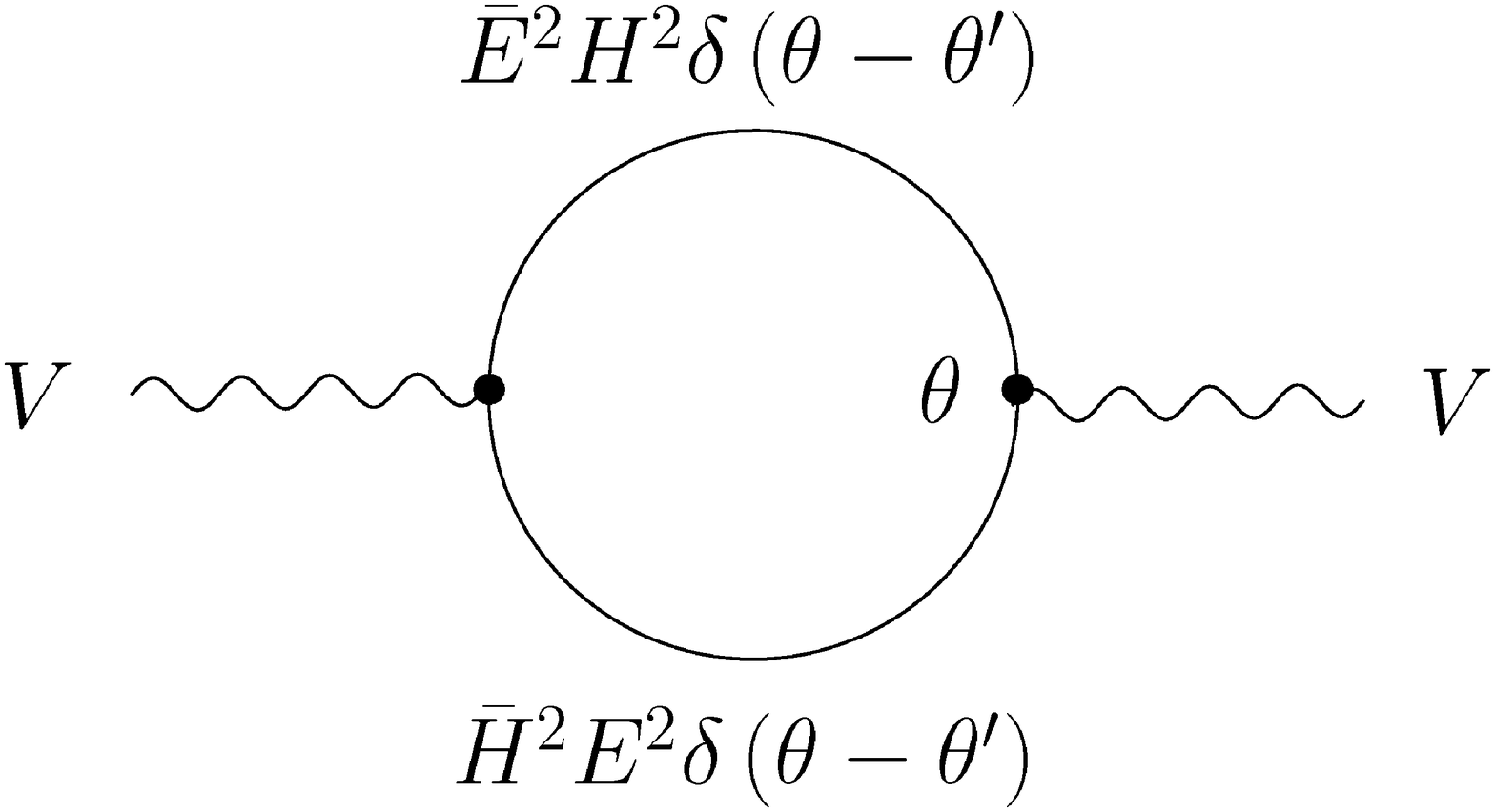}}
\settowidth{\la}{\usebox{\boxa}}
\sbox{\boxb}{\includegraphics[width=6cm]{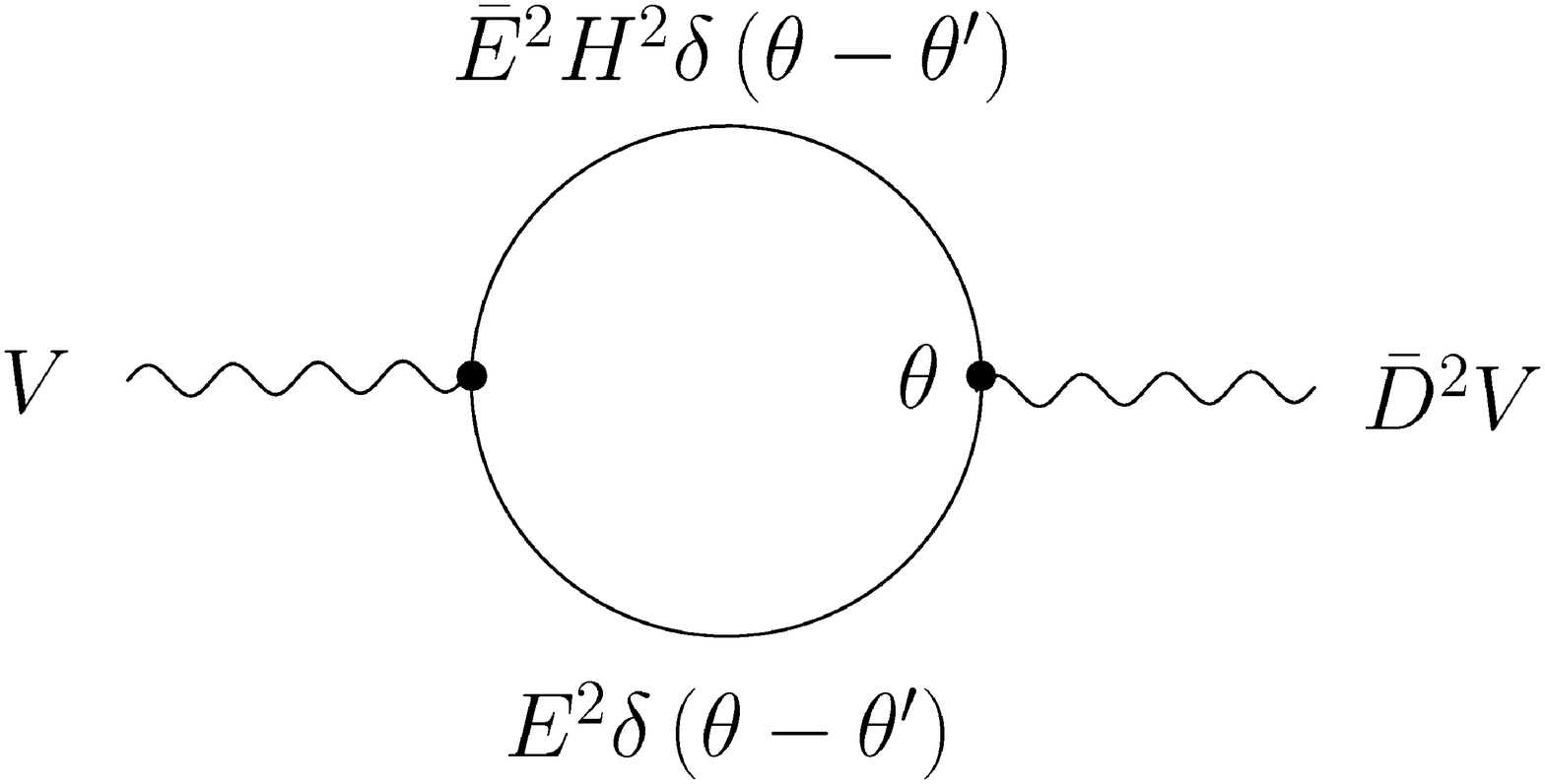}}
\settowidth{\lb}{\usebox{\boxb}}
\begin{eqnarray*}
 \parbox{\la}{\usebox{\boxa}} \qquad\Longrightarrow\qquad \parbox{\lb}{\usebox{\boxb}}.
\end{eqnarray*}
When we moreover apply $E^2(-p-q)$ to the second bracket (the upper
chiral propagator in the above picture) and $V(p,\theta)$ in
(\ref{eq203}), we should note the term (\ref{eq301}) vanish as before
and the term
\begin{eqnarray*}
 \delta^{(4)}\!\left(\theta-\theta'\right) \hspace{.25em}H(q)^2\bar{E}(q)^2
  H(q)^2\hspace{.25em}\delta^{(4)}\!\left(\theta-\theta'\right)
\end{eqnarray*}
also vanish. This is because we have at most two derivatives between two
delta functions since we can show $H(q)^2\bar{E}(q)^2 H(q)^2 =
\left(q^2-m^2\right)H(q)^2$.
We do not have non-zero
contributions unless there are four derivatives between two
delta functions. Therefore again, the non-zero contribution of the first
term in (\ref{eq250}) comes only from the term where
$\frac{1}{4}\bar{H}^2E^2$ is applied to the auxiliary field:
\begin{eqnarray}
\hspace{-1.25em} && N\int\!\!\frac{d^3p}{\left(2\pi\right)^3}\frac{d^3q}{\left(2\pi\right)^3}\int\!
  d^4\theta\hspace{.25em}d^4\theta'\hspace{.5em}V(-p,\theta')\cdot\frac{1}{\left(p+q\right)^2-m^2+i\epsilon}\hspace{.5mm}\frac{1}{q^2-m^2+i\epsilon}\nonumber
  \\[2mm]
&&\hspace{5em}\times\hspace{.5em}\delta^{(4)}\!\left(\theta-\theta'\right)\left[\frac{1}{4}\bar{E}(q)^2
									 H(q)^2\delta^{(4)}\!\left(\theta-\theta'\right)\right]\frac{1}{4}D(p)^2\bar{D}(p)^2 V(p,\theta)\nonumber \\[2mm]
&=& N\int\!\!\frac{d^3p}{\left(2\pi\right)^3}\frac{d^3q}{\left(2\pi\right)^3}\int\!
  d^4\theta\hspace{.5em}V(-p,\theta)\cdot\frac{1}{\left(p+q\right)^2-m^2+i\epsilon}\hspace{.125em}\frac{1}{q^2-m^2+i\epsilon}\cdot\frac{1}{4}D(p)^2\bar{D}(p)^2
  V(p,\theta)\nonumber \\[2mm]
&=& iN \int\!\!\frac{d^3p}{\left(2\pi\right)^3}\int
\!d^4\theta\hspace{.5em}V(-p,\theta)\frac{1}{4}D(p)^2\bar{D}(p)^2
V(p,\theta)\cdot \frac{1}{4\pi}I(p^2)^{-1}. \label{eq311}
\end{eqnarray}
We can graphically express this integration by parts as follows:
\sbox{\boxa}{\includegraphics[width=6cm]{8.eps}}
\settowidth{\la}{\usebox{\boxa}}
\sbox{\boxb}{\includegraphics[width=6cm]{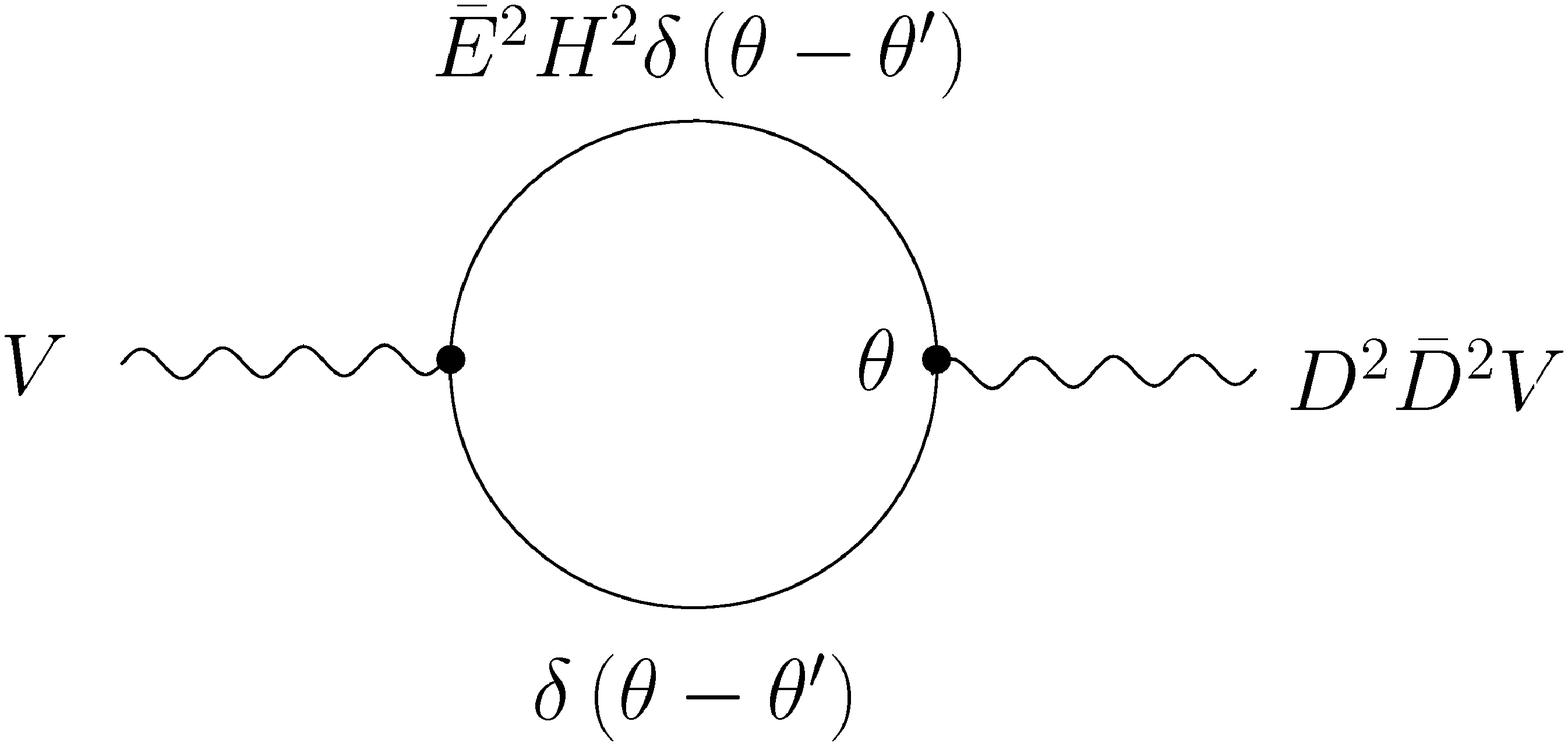}}
\settowidth{\lb}{\usebox{\boxb}}
\begin{eqnarray*}
 \parbox{\la}{\usebox{\boxa}} \quad \Longrightarrow \quad \parbox{\lb}{\usebox{\boxb}}.
\end{eqnarray*}

In summary, we can evaluate the equation (\ref{eq203}) by using the formula
(\ref{eq250}) for one of the chiral propagator. The contribution of the third term of
(\ref{eq250}) cancels the contribution of (\ref{eq255}), the
contributions of the second and third term are given by (\ref{eq310})
and (\ref{eq311}). Considering all the
contributions, the quadratic terms of the auxiliary field in the
effective action becomes
as follows:
\begin{eqnarray}
&&\frac{1}{2}\int\!\!\frac{d^3p}{\left(2\pi\right)^3}\int\!d^4\theta\hspace{.25em}V\!(-p,\theta)\left(iG^{-1}\right)
 V\!(p,\theta)\nonumber \\[1mm]
\hspace{2.5em}&=& \frac{N}{2}\int\!\!\frac{d^3p}{\left(2\pi\right)^3}\int\!
  d^4\theta\hspace{.5em}V(-p,\theta)\left\{\frac{1}{4}D(p)^2\bar{D}(p)^2
				    - \frac{1}{2}D(p)\hspace{.125em}\Slash{p}\hspace{.125em}\bar{D}(p)
				   - m
				   \hspace{.125em}D(p)\bar{D}(p)\right\}
  V(p,\theta)\cdot\frac{1}{4\pi}I(p^2)^{-1}\nonumber \\[2mm]
&=&  \frac{N}{2} \int\!\!\frac{d^3p}{\left(2\pi\right)^3}\int\!
  d^4\theta\hspace{.5em}V(-p,\theta)\left\{\frac{1}{4}D(p)\bar{D}(p)^2 D(p) -m
				   \hspace{.125em}D(p)\bar{D}(p)\right\}
  V(p,\theta)\cdot\frac{1}{4\pi}I(p^2)^{-1}, \label{eq312}
\end{eqnarray}
where the last equality is shown
in appendix G. This inverse propagator of the auxiliary field is the same
as that of the super Yang-Mills field except for the mass term
$m\hspace{.5mm}D\bar{D}$ and non-local factor $I(p^2)^{-1}$.
\vspace{1.25em}
 
\subsection{Propagator of the auxiliary field}

\hspace{1.3em}In order to derive the propagator of the auxiliary field from
(\ref{eq312}), we have to evaluate the inverse of the differentical
operator in it. To do so, we first study the algebra of $D_\alpha$ and
$\bar{D}_\alpha$ in detail. Because of the fact that $D^2 = \bar{D}^2 = 0$,
all the differential operators composed of $D_\alpha$ and $\bar{D}_\alpha$ can be
written as linear combinations of the following six operators (similar
to the case of ${\mathcal N}=1$ in four dimensions \cite{book1}), namely, a
set of projection operators of chiral and anti-chiral superfield
\begin{eqnarray*}
P_2 := \frac{\bar{D}^2 D^2}{4p^2} \quad , \quad P_1 := \frac{D^2\bar{D}^2}{4p^2},
\end{eqnarray*}
and other four operators:
\begin{eqnarray*}
 P_+ := -\frac{iD^2}{2\sqrt{-p^2}} \quad , \quad P_- :=
  -\frac{i\bar{D}^2}{2\sqrt{-p^2}} \quad , \quad P_T :=
  -\frac{D\bar{D}^2D}{2p^2} \quad , \quad P_D := -\frac{iD\bar{D}}{\sqrt{-p^2}},
\end{eqnarray*}
where we omit to write explicitly the momentum dependence of covariant derivatives. Note that the term
$D\Slash{p}\bar{D}$ can be written as a linear combination of $P_T$ and
$P_1$. Through a straightforward
calculation, we can show that
\begin{eqnarray}
 P_1 + P_2 + P_T = 1. \label{eq335}
\end{eqnarray}
The multiplication rules of these operators
are indicated in table \ref{table1}, where the blanks mean zero.
\begin{table}[h]
\begin{center}
\begin{tabular}{|c||c|c|c|c|c|c|}\hline
left $\backslash$ right & $P_1$ & $P_2$ & $P_+$ & $P_-$ & $P_T$ & $P_D$
 \\ \hline\hline
$P_1$ & $P_1$ &  & $P_+$ &  &  &  \\ \hline
$P_2$ &  & $P_2$ &  & $P_-$ &  &  \\ \hline
$P_+$ &  & $P_+$ &  & $P_1$ &  &  \\ \hline
$P_-$ & $P_-$ &  & $P_2$ &  &  &  \\ \hline
$P_T$ &  &  &  &  & $P_T$ & $P_D$ \\ \hline
$P_D$ &  &  &  &  & $P_D$ & $P_T$ \\ \hline
\end{tabular}
\caption{The multiplicative property of oparators} \label{table1}
\end{center}
\end{table}
The derivations of this table and the equation (\ref{eq335}) are shown in appendix G.

We now want to derive the inverse of
\begin{eqnarray}
 \frac{1}{4}D\bar{D}^2D - m\hspace{.1em}D\bar{D} \quad =\quad
  -\frac{p^2}{2}P_T - im \sqrt{-p^2}\hspace{.125em}P_D \label{eq340}
\end{eqnarray}
by using table \ref{table1}.
But this operator is non-invertible because this annihilates arbitrary
anti-chiral superfields (indeed also annihilates arbitrary chiral superfields). Note here that
\begin{eqnarray*}
 D\bar{D} = \bar{D}D + \left\{D_\alpha,\bar{D}^\alpha\right\} = \bar{D}D
  + i{\rm tr}\left(\Slash{\partial}\right) = \bar{D}D.
\end{eqnarray*}
This singularity is of course due to the gauge symmetry. So we need to introduce a
gauge-fixing term to define the inverse of (\ref{eq340}). We here introduce
the following supersymmetric gauge-fixing term in the action
\begin{eqnarray*}
 {S}_{\rm GF} = \frac{N}{2\alpha}\int\!\!\frac{d^3p}{\left(2\pi\right)^3}\int\!
  d^4\theta \hspace{.5em}V(-p,\theta)\cdot\frac{1}{8}\left[D^2\bar{D}^2 + \bar{D}^2
				     D^2\right]V(p,\theta)\cdot \frac{1}{4\pi}I(p^2)^{-1},
\end{eqnarray*}
where we omit to write momenta $p$ of covariant derivatives
explicitly.  With this gauge fixing, the inverse propagator of the auxiliary
field is given by
\begin{eqnarray*}
 \frac{1}{2}\int\!\!\frac{d^3p}{\left(2\pi\right)^3}\hspace{.125em}d^4\theta\hspace{.25em}
  V(-p)\left(iG^{-1}\right)V(p) \hspace{.5em} +\hspace{.5em} S_{\rm GF} \quad = \quad \frac{N}{2}\int\!\!\frac{d^3p}{\left(2\pi\right)^3}\int\!d^4\theta\hspace{.5em}V(-p,\theta)
 \nabla_V V(p,\theta)\cdot\frac{1}{4\pi}I(p^2)^{-1},
\end{eqnarray*}
where
\begin{eqnarray*}
\nabla_V &=& \frac{1}{4}D\bar{D}^2D - mD\bar{D} +
  \frac{1}{8\alpha}\!\left(D^2\bar{D}^2 + \bar{D}^2 D^2\right)
  \quad =\quad -\frac{P_T}{2}-im\sqrt{-p^2}\hspace{.125em}P_D + \frac{p^2}{2\alpha}\left(P_1+P_2\right).
\end{eqnarray*}
Then we can evaluate the inverse of $\nabla_V$. Indeed, by supposing $\nabla_V^{-1} = aP_1 + bP_2 +cP_+ +
dP_- + eP_T + fP_D$, we can easily show
\begin{eqnarray*}
 \nabla_V \nabla_V^{-1} = \frac{p^2}{2\alpha}\left(aP_1+bP_2\right) +
  \frac{p^2}{2\alpha}\left(cP_++dP_-\right) -\frac{1}{2} \left(p^2e +
		      2im\sqrt{-p^2}\hspace{.25em}f\right)P_T -\frac{1}{2}\left(p^2f +
							    2im\sqrt{-p^2}\hspace{.25em}e\right)P_D.
\end{eqnarray*}
If we impose $a=b=\frac{\alpha}{p^2}$, $c=d=0$, $e=
-\frac{2}{p^2-4m^2}$ and $f=-\frac{4im}{\sqrt{-p^2}\left(p^2-4m^2\right)}$,
\begin{eqnarray*}
 \nabla_V\nabla_V^{-1} = P_1 + P_2 + P_T =1.
\end{eqnarray*}
Therefore  the inverse operator of $\nabla_V$ is
\begin{eqnarray*}
 \nabla_V^{-1} &=& -\frac{2}{p^2-4m^2}\left(P_T +
				     \frac{2im}{\sqrt{-p^2}}P_D\right) +
 \frac{2\alpha}{p^2}\left(P_1+P_2\right) \\[2mm]
 &=& \frac{1}{p^2-4m^2}\cdot\frac{D\bar{D}^2D-4mD\bar{D}}{p^2} +
 \frac{\alpha}{2p^4}\left(D^2\bar{D}^2 + \bar{D}^2D^2\right).
\end{eqnarray*}
Using this inverse operator, the superpropagator of the auxiliary field
$V$ can be written as
\begin{eqnarray}
\left<V(-p,\theta',\bar{\theta}')V(p,\theta,\bar{\theta})\right>_0 = \frac{4\pi
 i}{N}I(p^2)\cdot\nabla_V^{-1}\delta^{(4)}\!\left(\theta-\theta'\right). \label{eq501}
\end{eqnarray}
Note that this propagator has a pole at $p^2=4m^2$, which implies that
a one-particle state of the auxiliary field is a bound state of the
dynamical field.

If we expand this propagator in components, we obtain propagators of
component fields. However, it leads to a complicated expression to expand
(\ref{eq501}) straightforwardly since the auxiliary superfield
$V(p,\theta,\bar{\theta})$ has many unphysical component fields which
can be eliminated if we choose the non-supersymmetric gauge such as Wess-Zumino
gauge. We can, nevertheless, easily obtain the propagators of $v_{\mu}$ and $M$ by taking
the coefficient of $\bar{\theta}'\theta'\bar{\theta}\theta$ in the
expansion of (\ref{eq501}), namely
\begin{eqnarray}
 \left<v_\mu\!\left(-p\right)v_{\nu}\!\left(p\right)\right>_0 &=&
  \frac{4\pi
  i}{N}I(p^2)\left\{\frac{1}{p^2-4m^2}\left[-\eta_{\mu\nu}+\left(1+\alpha\cdot\frac{p^2-4m^2}{p^2}\right)\frac{p_\mu
	   p_\nu}{p^2}-\frac{2mi}{p^2}\epsilon_{\mu\nu\rho}p^\rho\right]\right\}\label{v}\\[1mm]
\left<M(-p)M(p)\right> &=& \frac{4\pi
i}{N}I(p^2)\frac{1}{p^2-4m^2}.\label{M}
\end{eqnarray}
These propagators of component fields coincide with
the result in \cite{paper1}. The derivation of these expressions are
shown in appendix H.

\vspace{5ex}

\section{Divergent diagrams and renormalization}

In this section, we investigate divergent diagrams and the
renormalizability. We first study the superficial degree of divergence
and show that there are two types of divergent diagrams. We can prove all
divergences can be eliminated by renormalizations of the coupling
constant $g$ and the wavefunction of the
dynamical field $\tilde{\Phi}$.

\vspace{5ex}
\subsection{Superficial degree of divergence}

We first evaluate the superficial degree of divergence. Recall the 
superpropagators of the dynamical field and the auxiliary field
\begin{eqnarray*}
 \left<\tilde{\Phi}^{\dagger k}\!\left(-p,\theta',\bar{\theta}'\right)\tilde{\Phi}^j\!\left(p,\theta,\bar{\theta}\right)\right>_0 &=& \delta^{jk}\frac{i}{p^2-m^2+i\epsilon}\cdot\frac{1}{4}\bar{E}^2
  H^2\delta^{(4)}\!\left(\theta-\theta'\right)\\[2mm]
\left<V\!\left(-p,\theta',\bar{\theta}'\right)V\!\left(p,\theta,\bar{\theta}\right)\right>_0
&=& \frac{4\pi
i}{N}I(p^2)\left[\frac{1}{p^2-4m^2}\cdot\frac{D \bar{D}^2\!D -
	    4mD\bar{D}}{p^2}\right.\\
&&\hspace{7.5em} \left.+ \frac{\alpha}{2p^2}\left(D^2\!\bar{D}^2+\bar{D}^2\!D^2\right)\right]\delta^{(4)}\!\left(\theta-\theta'\right),
\end{eqnarray*}
where momenta of covariant derivatives are all equal to $p$. Postponing
 the discussion on momentum dependence of covariant derivatives,
 we can evaluate high-energy
behaviors of above superpropagators as follows:
\begin{eqnarray*}
  \left<\tilde{\Phi}^{\dagger
   k}\!\left(-p,\theta',\bar{\theta}'\right)\tilde{\Phi}^j\!\left(p,\theta,\bar{\theta}\right)\right>_0
   &\sim& \frac{1}{p^2}\times\bar{E}^2\!H^2\\[2mm]
\left<V\!\left(-p,\theta',\bar{\theta}'\right)V\!\left(p,\theta,\bar{\theta}\right)\right>_0
&\sim& \frac{\sqrt{-p^2}}{p^4}\times\left(D\bar{D}^2\! D \hspace{.5em}{\rm
				     or}\hspace{.5em} mD\bar{D}
				     \hspace{.5em}{\rm
				     or}\hspace{.5em}D^2\!\bar{D}^2
				     \hspace{.5em}{\rm or}\hspace{.5em}\bar{D}^2\!D^2\right).
\end{eqnarray*}
Note that $I(p^2) =
\frac{\sqrt{-p^2}}{\arctan\sqrt{-\frac{p^2}{4m^2}}}\sim \sqrt{-p^2}$ at
high energy. 

Then we evaluate high-energy behaviors of covariant
derivatives. In any loop diagram, we can integrate by parts and
reduce the number of integrations over grassmann coordinates by
virtue of the delta function $\delta^{(4)}\!(\theta-\theta')$. Then the
final expression of grassmann integrations on each loop
has a factor
\begin{eqnarray*}
 \delta^{(4)}\!\left(\theta-\theta'\right)\hspace{.25em}\left({\rm product\hspace{.5em}of\hspace{.5em}covariant\hspace{.5em}derivatives}\right)\hspace{.5em}\delta^{(4)}\!\left(\theta-\theta'\right)
\end{eqnarray*}
in the integrand. As we have seen in the previous section, however,
this factor will give a vanishing result unless there are four
derivatives between two delta functions. What can be obtained if we have six covariant
derivatives between two delta functions? The answer turns out to be zero when we
note
\begin{eqnarray*}
 D^2\!\bar{D}^2\!D^2 = \left[D^2,\bar{D}^2\right]D^2 = 4p^2 D^2
\end{eqnarray*}
since we have only two derivatives between delta functions.
How about
the case in which we have eight derivatives between delta functions? In
such a case, we obtain a factor
\begin{eqnarray*}
 D^2\!\bar{D}^2\!D^2\!\bar{D}^2 = 4p^2D^2\!\bar{D}^2
\end{eqnarray*}
and this gives a non-zero contribution. Similarly, if we have twelve
covariant derivatives, we obtain $\left(D^2\!\bar{D}^2\right)^3 =
\left(4p^2\right)^2D^2\!\bar{D}^2$. Therefore we find that
$D^2\!\bar{D}^2\sim p^2$ unless they are used to differentiate a delta
function $\delta^{(4)}\!\left(\theta-\theta'\right)$. In every loop, we use one
$D^2\!\bar{D^2}$ to differentiate a delta function in the formula
\begin{eqnarray*}
 \int d^4\theta' \hspace{.5em}\delta^{(4)}\!\left(\theta-\theta'\right)\left[\frac{1}{4}D^2\!\bar{D}^2\right]\delta^{(4)}\!\left(\theta-\theta'\right)
  \quad =\quad 1.
\end{eqnarray*}
This formula is easily shown in the same way as (\ref{eq260}). When the
diagram contains $L$ loops, $L$ sets of $D^2\bar{D}^2$ are used to
differentiate delta functions, reducing the degree of divergence by
$2L$.

\begin{figure}[h]
\begin{center}
\includegraphics[width=7cm]{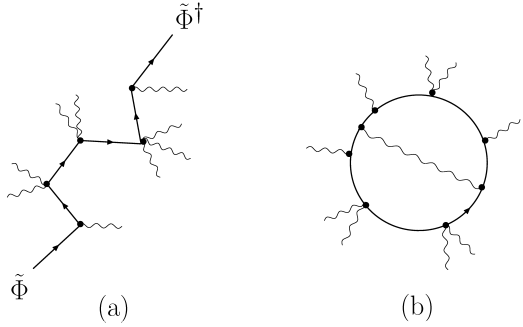}
\caption{Typical examples of two types of $\tilde{\Phi}$-lines} \label{pic2}
\end{center}
\end{figure}

By counting $D\bar{D}\sim \Slash{p}$, 
we then obtain the complete behavior of superpropagators at high
energy as
\begin{eqnarray*}
   \left<\tilde{\Phi}^{\dagger
   k}\!\left(-p,\theta',\bar{\theta}'\right)\tilde{\Phi}^j\!\left(p,\theta,\bar{\theta}\right)\right>_0
   &\sim& 1\\[2mm]
\left<V\!\left(-p,\theta',\bar{\theta}'\right)V\!\left(p,\theta,\bar{\theta}\right)\right>_0
&\sim& \frac{1}{\sqrt{-p^2}},
\end{eqnarray*}
and the degree of divergence has to be reduced by $2L$ if the diagram has
$L$ loops.   
Then we have the superficial degree of divergence $d$
as
\begin{eqnarray*}
 d &=& 3L - P_V -2L,
\end{eqnarray*}
where $L$ denotes the number of loops and $P_V$ denotes the number of propagators of
the auxiliary field. The first term comes from the fact that each loop
has three momentam integration. The last term comes from the fact that
we use four covariant derivatives at each loop to defferentiate a delta
function. Using the relation $L=\left(P_V+P_{\Phi}\right)-V+1$,
where $P_\Phi$ is the number of propagators of the dynamical field and
$V$ denotes the number of vertices, we
find
\begin{eqnarray}
 d =  P_\Phi - V + 1. \label{eq601}
\end{eqnarray}
Then we should notice that all vertices in this theory contains exactly
one $\tilde{\Phi}^\dagger\tilde{\Phi}$. This means that
\begin{eqnarray}
 V = P_\Phi + \frac{E_\Phi}{2}, \label{eq600}
\end{eqnarray}
where $E_\Phi$ denotes the number of external lines of
$\tilde{\Phi}$. The formula (\ref{eq600}) can be shown as
follows. Noting the symmetry of $\tilde{\Phi}\to
e^{i\alpha}\tilde{\Phi}, \hspace{1mm}\tilde{\Phi}^\dagger\to
e^{-i\alpha}\tilde{\Phi}^\dagger$, we find that the internal lines of
$\tilde{\Phi}$ are not branched. This means there are two types of
$\tilde{\Phi}$ lines. The first type goes from one external
line to another external line without branches: (a). The second type is an
internal circle of $\tilde{\Phi}$ which has no external lines: (b).
 Typical examples of These two types are indicated in figure \ref{pic2}.
We now count the number of vertices on the line of $\tilde{\Phi}$. For type (a), we easily find $V = P_\Phi + 1 = P_\Phi +
\frac{E_\Phi}{2}$. On the other hand, for type (b), $V = P_\Phi$ is
satisfied. But this can be also written as $V = P_\Phi + \frac{E_\Phi}{2}$
because $E_\Phi=0$ for type (b). Since both types of (a) and (b) satisfy
(\ref{eq600}), diagrams containing both types also satisfy (\ref{eq600}).

Combining (\ref{eq601}) and (\ref{eq600}), we find the final result:
\begin{eqnarray*}
 d = 1- \frac{E_\Phi}{2}.
\end{eqnarray*}
Therefore there are only two types of divergent diagrams:
\sbox{\boxa}{\includegraphics[width=3.2cm]{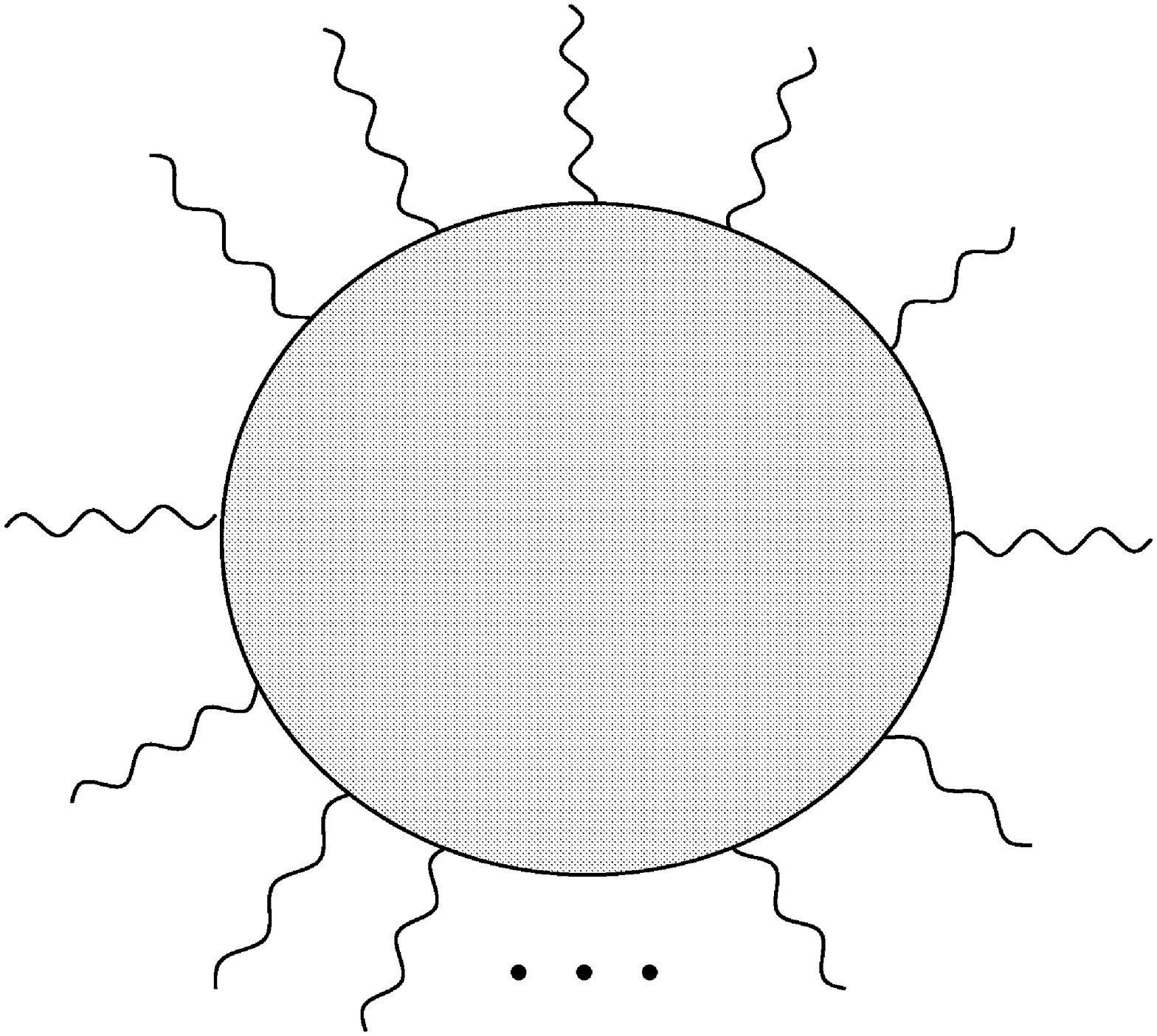}}
\settowidth{\la}{\usebox{\boxa}}
\sbox{\boxb}{\includegraphics[width=5cm]{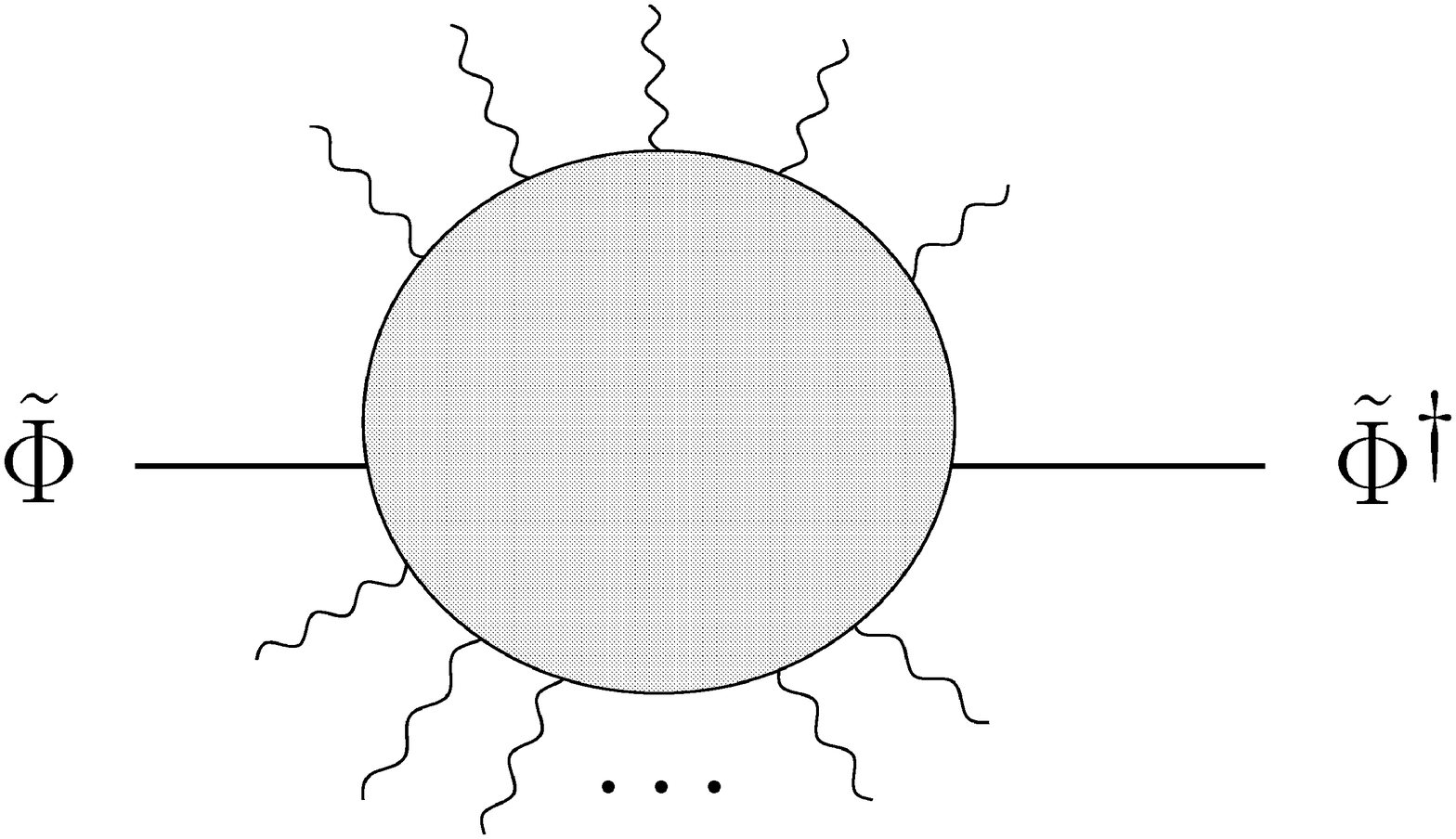}}
\settowidth{\lb}{\usebox{\boxb}}
\begin{eqnarray}
 \parbox{\la}{\usebox{\boxa}}: d=1\qquad, \qquad 
\parbox{\lb}{\usebox{\boxb}}:d=0. \label{eq700}
\end{eqnarray}
In the following, we study these two types of diagrams in detail. We
will show that all the divergences can be absorbed into the bare
coupling constant and the wavefunction of $\tilde{\Phi}$.

\vspace{5ex}

\subsection{Renormalization of the coupling constant}

In this subsection, we study the amplitudes without external
$\tilde{\Phi}$-lines shown in the left diagram
in (\ref{eq700}). Since its superficial degree of divergence is 1, it may
contain linear and logarithmic divergences. To see these divergence explicitly, one
might expand the amplitude in powers of the external momentum $p^\mu$ in the same way as in
the ordinary field theory. However, since we now work in the superfield
perturbation theory, each field has grassmann coordinates in addition to
spacetime coordinates. Therefore we have to expand the amplitude in
powers of $D_\alpha, \bar{D}_\alpha$ as well as $p^\mu$. The reason for
expanding it by $D_\alpha, \bar{D}_\alpha$ rather than
$\frac{\partial}{\partial\theta^\alpha},
\frac{\partial}{\partial\overline{\theta}^\alpha}$ is supersymmetry.

Although
there are no terms linear in $p^\mu$ due to the
Lorentz invariance, there may be terms linear in $D^2, \bar{D}^2$
or $\bar{D}D$, which can be logarithmically divergent:
\begin{eqnarray*}
 \parbox{\la}{\usebox{\boxa}} = a\hspace{.125em}\Lambda +
  \log\Lambda\left(b\hspace{.125em}\bar{D}D + c\hspace{.125em}
  D^2 + d
  \hspace{.125em}\bar{D}^2\right) + \hspace{1mm}{\rm finite}\hspace{.25em}{\rm terms},
\end{eqnarray*}
where $a,b,c$ and $d$ are constants independent of external momenta and
grassmann coordinates. The differential
operators $\bar{D}D, D^2, \bar{D}^2$ acts on external
auxiliary fields and independent of internal momenta. Terms linear in
$p^2, p^4$ or $p^2\bar{D}D$ are included in ``finite terms''.

Therefore, the effective action might need counter terms of the form
\begin{eqnarray}
 \int\!d^3x\int\!d^4\theta \left[\alpha_n V^n + \beta_n
			    V^{n-2}\!\left(\bar{D}V\right)\!\left(DV\right) +
		  \gamma_n
V^{n-2}\!\left(DV\right)^2 + \delta_n V^{n-2}\!\left(\bar{D}V\right)^2\right], \label{eq730}
\end{eqnarray}
where $n$ is a positive integer and $\alpha_n, \beta_n, \gamma_n,
\delta_n$ are constants. When $n=1$, the second, third, and fourth
terms should be considered as $\bar{D}\!DV,
\hspace{.5mm}D^2V,$ and $\bar{D}^2V$, respectively.

If we assume the existence of a gauge invariant regularization,
$\gamma_n$ and $\delta_n$ must be zero because
operators $V^{n-1}D^2V, V^{n^1}\bar{D}^2V$ are not gauge invariant. Similarly, we can show $\beta_n = 0$ unless $n\leq2$ and $\alpha_n=0$
unless $n=1$. However, we can explicily show these results by
analyzing loop integrations without assuming the existence of a gauge
invariant regularization.

\vspace{5ex}

\subsubsection{Operators of the form $V^{n-1}\!\bar{D}DV,\hspace{1mm}V^{n-1}\!D^2V,\hspace{1mm}V^{n-1}\!\bar{D}^2V$}

We first show explicitly that $\beta_n,\gamma_n,\delta_n=0$ for all
$n$. Note that all covariant derivatives acting on external fields come
from partial integrals over grassmann coordinates. In order
to obtain operatos $V^{n-1}\bar{D}DV, V^{n-1}D^2V, V^{n-1}\bar{D}^2V$,
we have to move two covariant derivatives from propagators to external
fields through integrations by parts. Suppose there are $k$ propagators
in the diagram. Since every term in every propagator has four
covariant derivatives except for $mD\bar{D}$ in the superpropagator of
the auxiliary field,
$4k-2$ covariant derivatives remain in loops after moving two
covariant derivatives to external fields, assuming there is no $mD\bar{D}$ in the diagram. We perform integrations over grassmann coordinates and
shrink all grassmann loops using the formula
\begin{eqnarray*}
 \int d^4\theta' \hspace{.5em}\delta^{(4)}\!\left(\theta-\theta'\right)\left[\frac{1}{4}D^2\!\bar{D}^2\right]\delta^{(4)}\!\left(\theta-\theta'\right)
  \quad =\quad 1
\end{eqnarray*}
or the similar one which has $E^2\bar{H}^2$ instead of $D^2\bar{D}^2$.
We use $4L$ covariant derivatives to shrink all grassmann loops when the
diagrams has $L$ loops. Then $4(k-L)-2$
covariant derivatives remain. But, since all grassmann loops have been shrinked, these
$4(k-L)-2$ derivatives must be changed into internal momenta by using
the anticommutation relation
\begin{eqnarray*}
 \left[D^2,\bar{D}^2\right] = 4q^2 - 4\bar{D}\Slash{q}D
\end{eqnarray*}
unless they vanish for the reason that there are less than four
derivatives between two delta functions. Therefore, if we move two
covariant derivatives to external fields, at most $4(k-L)-2$ covariant
derivatives are changed into $q^{2(k-L)-1}$, where $q$ is a typical
internal momentum. Recall here that $D^2\bar{D}^2 \sim p^2$. However, assuming Lorentz invariance, this factor
have to be written as $(q^2)^{k-L}\Slash{q}$, which contains an odd number
of internal momenta. Notice that each propagator is invariant under
$p\to -p$ except for covariant derivatives and vertex factors are
independent of momenta. Then we find that the total integrand is an odd
function of internal momenta.
We know that the degree
of divergence is reduced at least by 1 if the Feynman integrand is an
odd function of internal momenta. Then these integrals are not divergent
because their original superficial degree of divergence is zero.

Let us consider what happens if we have some $mD\bar{D}$s in the diagram.
Noting that we count all of $D^2\!\bar{D}^2, D\bar{D}^2D, mD\bar{D}$ as $\sim p^2$ when we evaluate the
superficial degree of divergence, we find that the degree of divergence
is again reduced at least by 1 if the diagram contains some $mD\bar{D}$s.

It is proved that in the effective acton there is no quantum correction to the operator
of the form of $V^{n-1}D^2V, V^{n-1}\bar{D}^2V$ or $V^{n-1}\bar{D}DV$.

\vspace{5ex}

\subsubsection{Operators of the form $V^n$}

We now show explicitly that $\alpha_n$ in (\ref{eq730}) vanishes unless $n=1$.
Recall that no $V^2$ term arose when we evaluated the inverse propagator
of the auxiliary field in subsection 4.2. The reason
for this is the cancellation between (\ref{eq255}) and the contribution
of the third term
in (\ref{eq250}). The contribution of the third term of (\ref{eq250}) is
a part of (\ref{eq203}). The contribution proportional to
$V^2$ induced by the partial integration over grassmann coordinates in
(\ref{eq203}) exactly cancelled (\ref{eq255}) which is also proportional
to $V^2$. This cancellation is due to the fact that the dynamical field
is a chiral superfield. To see this, we examine the following diagram:
\sbox{\boxa}{\includegraphics[width=11cm]{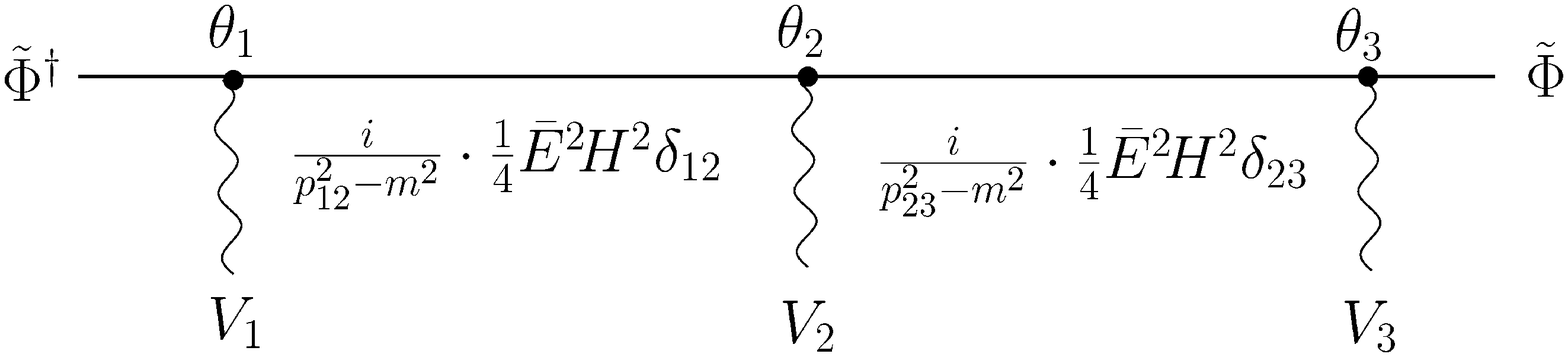}}
\settowidth{\la}{\usebox{\boxa}}
\begin{eqnarray}
 \parbox{\la}{\usebox{\boxa}}, \label{eq770}
\end{eqnarray}
where $\delta_{ij}$ stands for $\delta^{(4)}\!(\theta_i-\theta_j)$ and
$p_{ij}$ denotes the momentum which flows from $\theta_i$ to
$\theta_j$. Therefore $p_{ji}$ is equal to $-p_{ij}$. When we write $\bar{E}^2\!H^2\delta_{ij}$,
the covariant derivatives in front of $\delta_{ij}$ stands for covariant
derivatives with grassmann coordinates $\theta_i, \bar{\theta}_i$ and
momentum $p_{ij}$.  On the other hand, if we write $\bar{E}^2\!H^2\delta_{ji}$,
they are covariant derivatives with $\theta_j, \bar{\theta}_j, p_{ji}$.  We
distinguish $\delta_{ij}$ from $\delta_{ji}$. Then we can write the
formula (\ref{eq204}) as
\begin{eqnarray*}
 \frac{1}{4}\bar{E}^2\!H^2\delta_{12} = \frac{1}{4}E^2\!\bar{H}^2\delta_{21}.
\end{eqnarray*}
By using this formula, we can rewrite (\ref{eq770}) as
\sbox{\boxa}{\includegraphics[width=11cm]{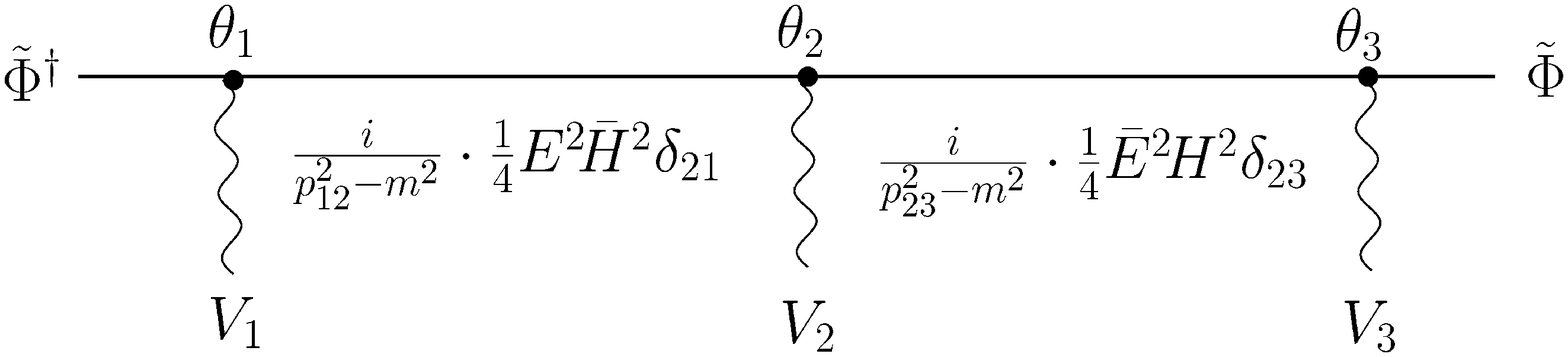}}
\settowidth{\la}{\usebox{\boxa}}
\begin{eqnarray}
 \parbox{\la}{\usebox{\boxa}}. \label{eq771}
\end{eqnarray}
Then we use the formula
\begin{eqnarray*}
 \frac{1}{4}E^2\bar{H}^2\delta_{21} = \frac{1}{4}\bar{H}^2E^2\delta_{21}
  - \bar{H}\!\left(\Slash{p}_{21}+m\right)\!E\delta_{21} + (p_{21}^2-m^2)\delta_{21},
\end{eqnarray*}
which is easily shown by the commutation relations of $E^2$ and $\bar{H}^2$.
We can graphically express this formula as follows:
\sbox{\boxa}{\includegraphics[width=11cm]{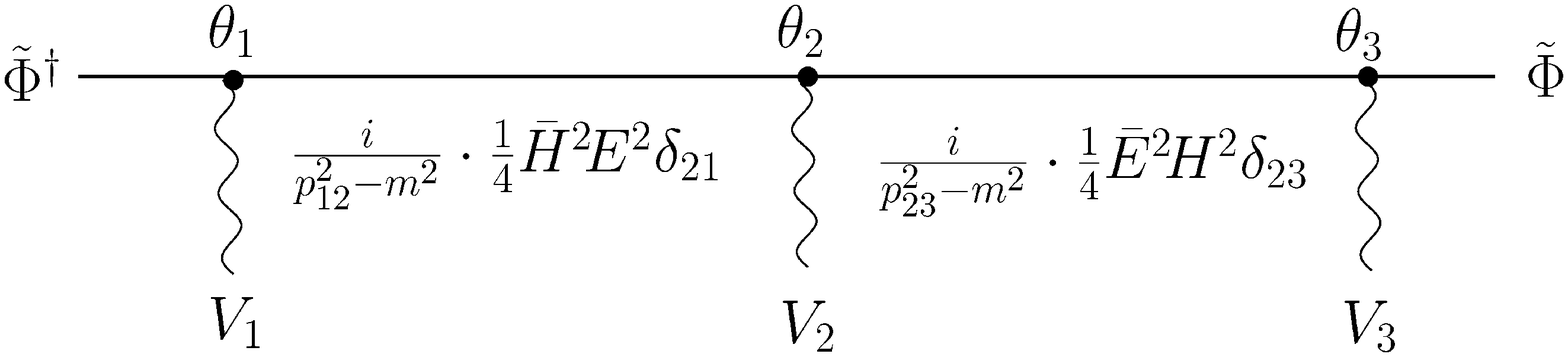}}
\settowidth{\la}{\usebox{\boxa}}
\sbox{\boxb}{\includegraphics[width=12cm]{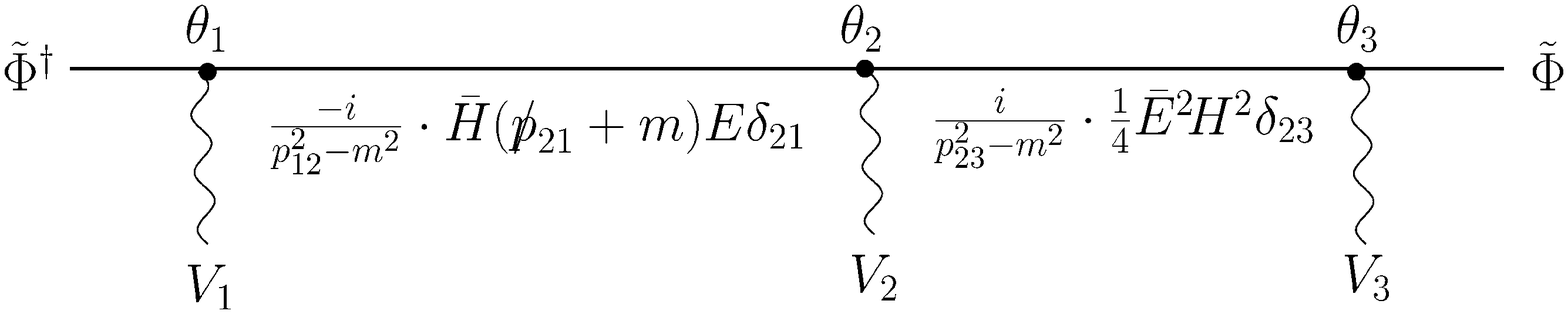}}
\settowidth{\lb}{\usebox{\boxb}}
\sbox{\boxc}{\includegraphics[width=11cm]{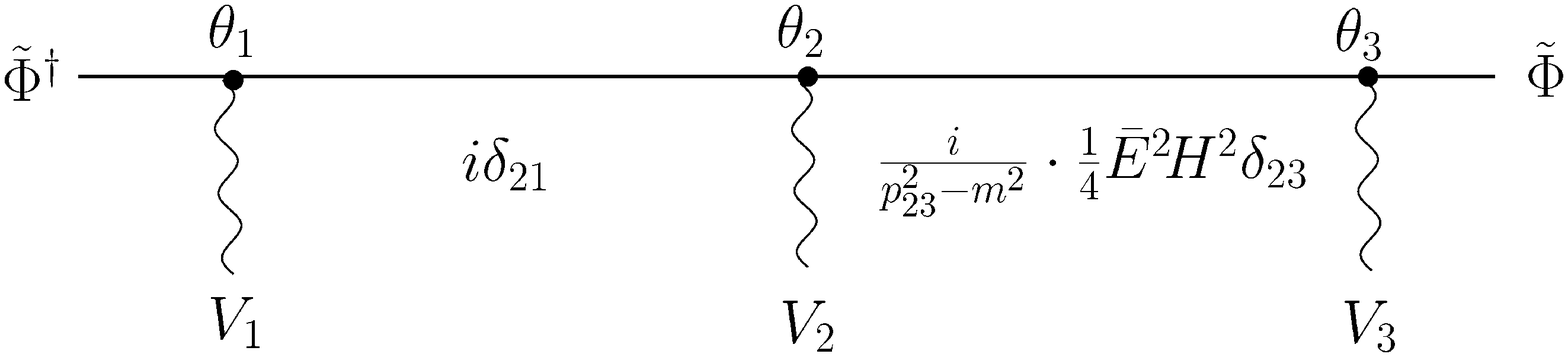}}
\settowidth{\lc}{\usebox{\boxc}}
\begin{eqnarray}
 (\ref{eq771}) &=& \parbox{\la}{\usebox{\boxa}} \nonumber \\[2mm]
&&\hspace{10mm}+\parbox{\lb}{\usebox{\boxb}}\nonumber \\[2mm]
&&\hspace{20mm}+\parbox{\lc}{\usebox{\boxc}}. \label{eq772}
\end{eqnarray}
In the third term in the right-hand side, we can easily perform the
integration over $\theta_1$. Then we obtain the following contribution
\newsavebox{\boxd}
\newlength{\ld}
\sbox{\boxd}{\includegraphics[width=8cm]{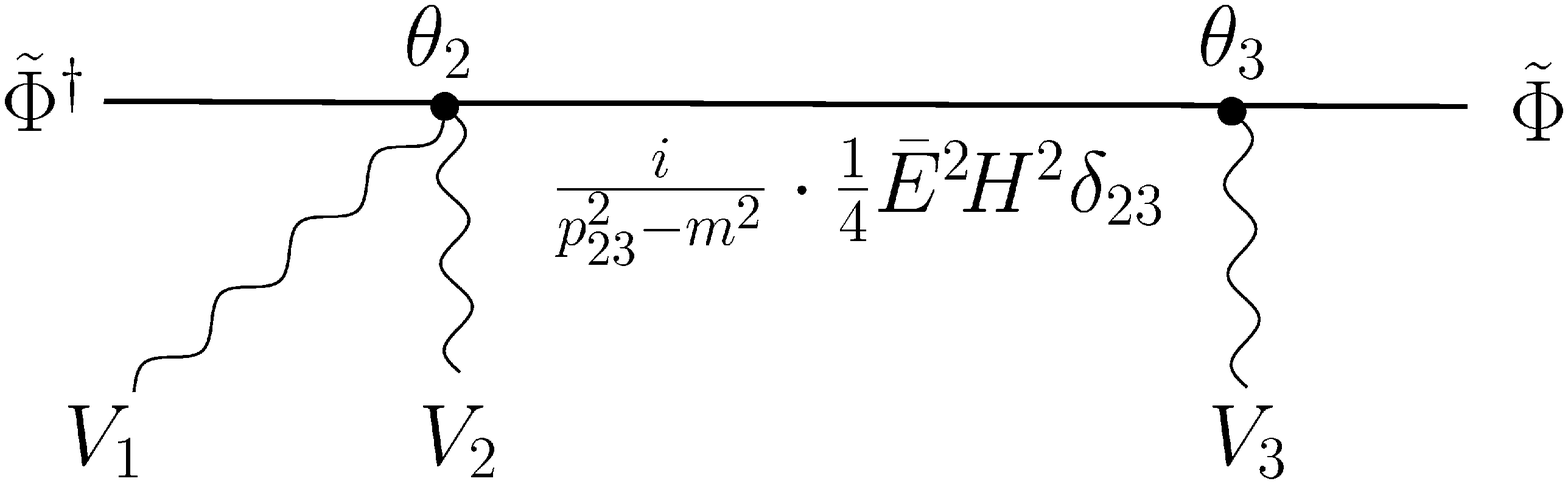}}
\settowidth{\ld}{\usebox{\boxd}}
\begin{eqnarray}
-\quad \left(\parbox{\ld}{\usebox{\boxd}}\right). \label{eq773}
\end{eqnarray}
In order to understand the minus sign, we should recall that the vertex factor is $i(-1)^n$
when the vertex is attached to $n$ lines of auxiliary field. Although performing the
integration over $\theta_1$ does not change the number of lines of
auxiliary field, it reduces the number of vertices by one, leaving a
factor $i$ which has been attached to the annihilated
vertex. Moreover, there is another $i$ in front of $\delta_{21}$. The minus sign in (\ref{eq773}) comes
from these two factors of $i$. Then the contribution from the third
term in (\ref{eq772})
exactly cancels that of the following diagram:
\sbox{\boxd}{\includegraphics[width=7cm]{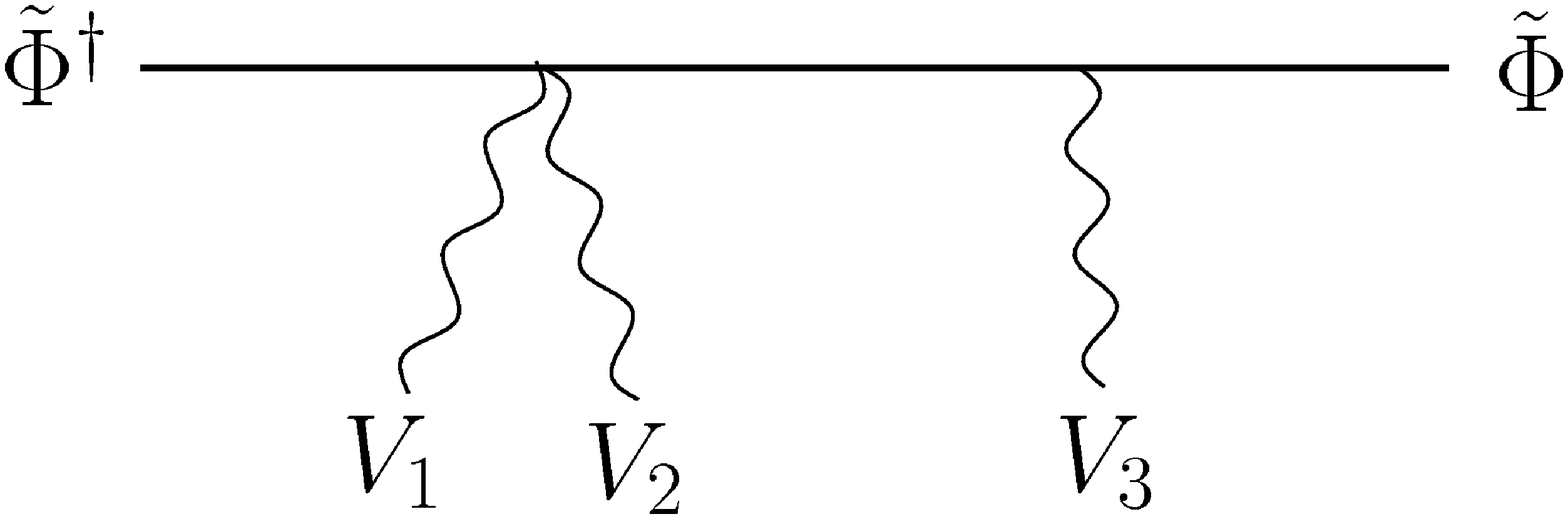}}
\settowidth{\ld}{\usebox{\boxd}}
\begin{eqnarray}
\parbox{\ld}{\usebox{\boxd}}.\label{eq777}
\end{eqnarray}

We now consider the remaining
terms, namely, the first and second terms in (\ref{eq772}). Performing
the partial integration over $\theta_2$, we can show as before that all vanish except for contributions in
which all covariant derivatives between $\theta_1$ and $\theta_2$ are applied to external auxiliary fields. For
instance, see the second term in
(\ref{eq772}). If we move $\bar{H}_\alpha$ from the left chiral propagator to
the right one exchanging it for $\bar{E}_\alpha$, it vanish because
$\bar{E}_\alpha\bar{E}^2=0$. We have to apply $\bar{H}_\alpha$ to $V_2$ to
obtain a non-zero contribution. Then we move $E_\alpha$ in the left
chiral propagator and perform the integration over $\theta_1$ by virtue
of $\delta_{21}$. The result is as follows:
\sbox{\boxc}{\includegraphics[width=9cm]{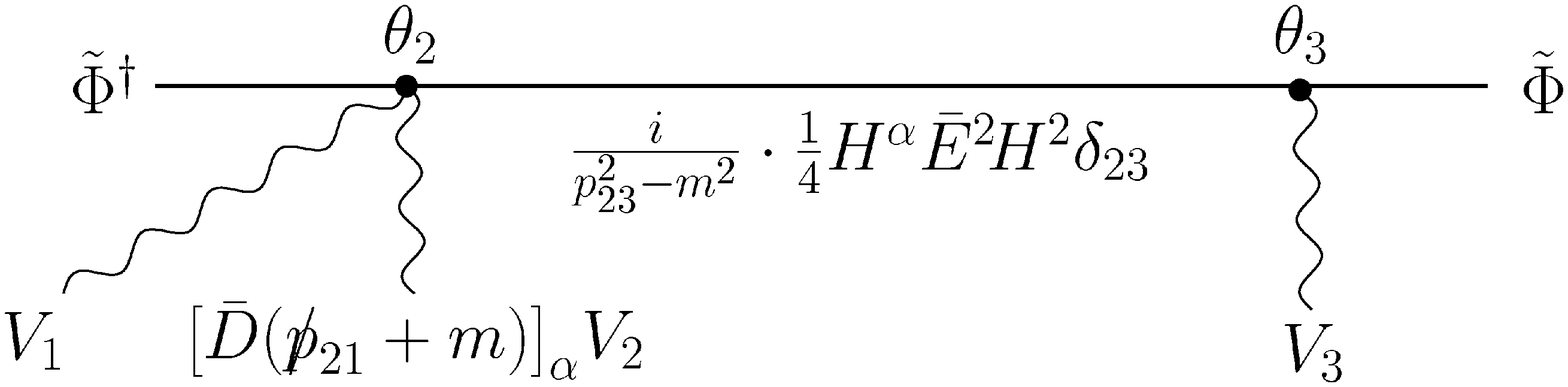}}
\settowidth{\lc}{\usebox{\boxc}}
\sbox{\boxd}{\includegraphics[width=9cm]{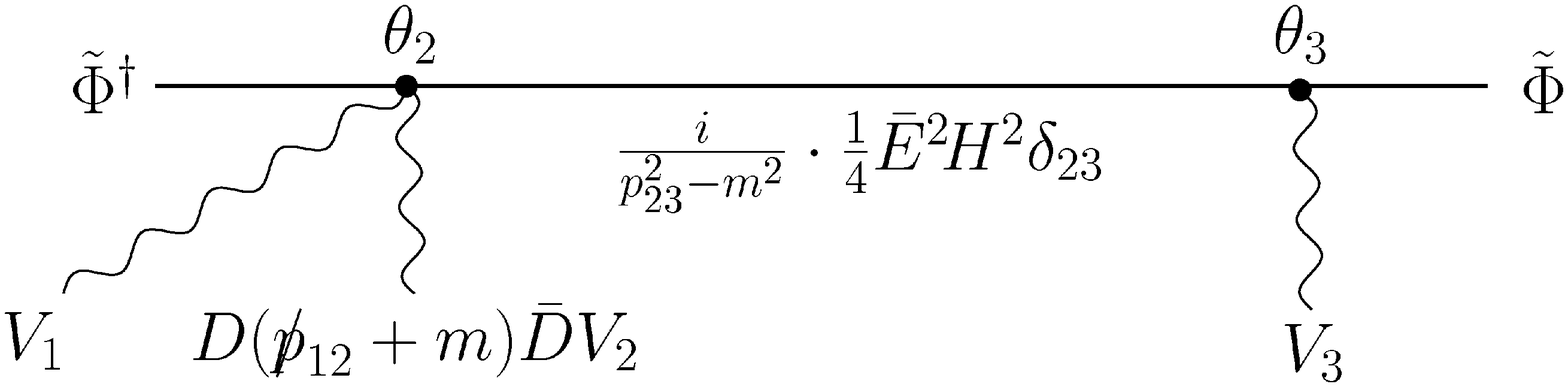}}
\settowidth{\ld}{\usebox{\boxd}}
\begin{eqnarray*}
 \frac{-i}{p_{12}^2-m^2}&\times&\left[\parbox{\lc}{\usebox{\boxc}}\right.\\[3mm]
&& \hspace{5em}\left.+ \parbox{\ld}{\usebox{\boxd}}\right],
\end{eqnarray*}
where we use the similar equation as (\ref{eq775}). 

In the second term
in the bracket, all covariant derivatives moved to the external
auxiliary field $V_2$. On the other hand, in the
first term, $H_\alpha$ is applied to the other chiral
propagator. This $H_\alpha$ can, however, move to $V_1$ or $V_2$ if we again perform the
partial integration over $\theta_2$. Noting that $E_\alpha\tilde{\Phi}^\dagger
= 0$, we find that the contribution of applying $H_\alpha$ to the external
$\tilde{\Phi}^\dagger$ vanish. So we move it only to external auxiliary fields exchanging it
for $D_\alpha$. The result is the same if there is an another chiral
propagator instead of the external $\tilde{\Phi}^\dagger$ because
$E_\alpha\!\!<\tilde{\Phi}^\dagger\!(p,\theta,\bar{\theta})\tilde{\Phi}\!(-p,\theta',\bar{\theta}')>_0
= 0$. This vanishing occurs due to the fact that the dynamical
field is a (twisted) chiral superfield. Therefore, through the integration by parts, the non-zero contributions of the
second term in (\ref{eq772}) come only when
all the covariant derivatives are applied to external auxiliary fields $V_1, V_2$.

In the same way, we can show that non-zero contributions of the first
term in (\ref{eq772}) arise only when all the covariant
derivatives are applied to external auxiliary fields, by performing a partial integration over $\theta_2$
to move $\bar{H}^2\!E^2$ in front of $\delta_{21}$. Therefore, all
non-zero contributions from
the first and second terms in (\ref{eq772}) have at least one covariant
derivative applied to external auxiliary fields. Only the third term
has no covariant derivative applied to external auxiliary fields but it
was cancelled by the contribution of (\ref{eq777}). We can express whole argument given above by the following simple graphical equation:
\sbox{\boxa}{\includegraphics[width=6cm]{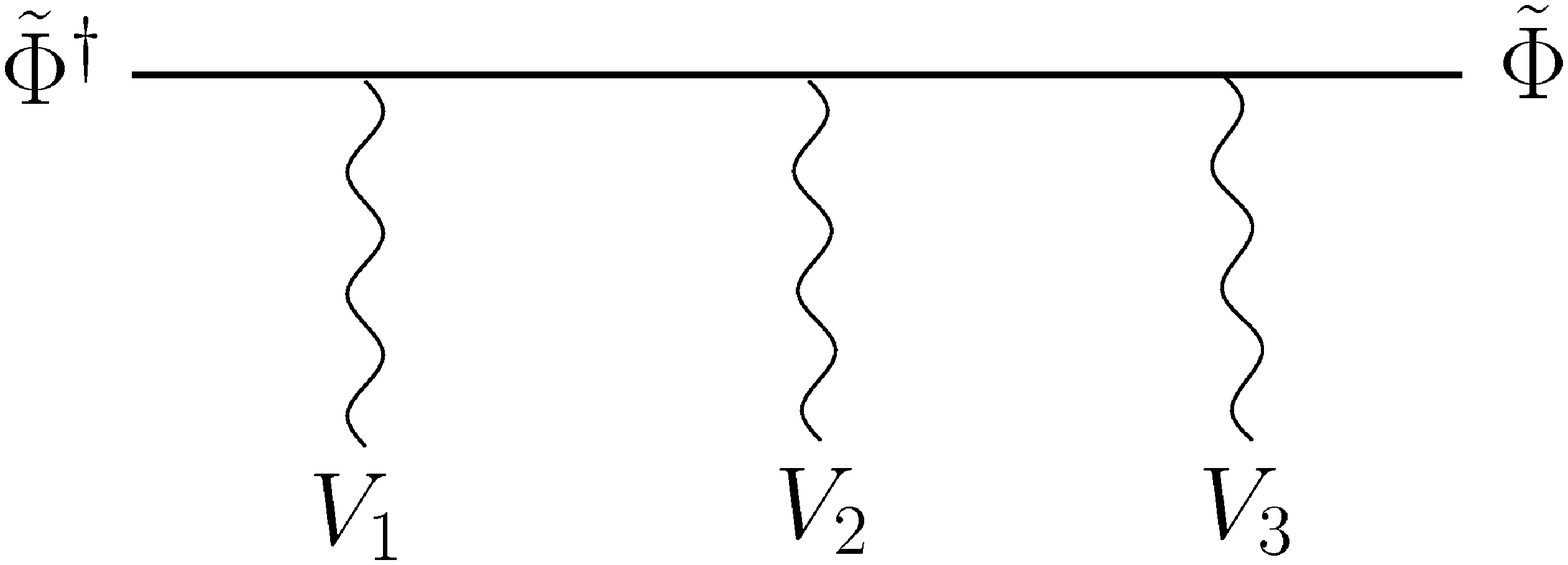}}
\settowidth{\la}{\usebox{\boxa}}
\sbox{\boxb}{\includegraphics[width=6cm]{22.eps}}
\settowidth{\lb}{\usebox{\boxb}}
\begin{eqnarray*}
 \parbox{\la}{\usebox{\boxa}} \quad+\quad \parbox{\lb}{\usebox{\boxb}} \sim\quad 0,
\end{eqnarray*}
where ``$\sim$'' means that both sides
are equal up to terms with at least one covariant derivative applied to
external fields.

Recalling the purpose of this subsection, in the following, we
consider only one-particle irreducible amplitudes with no external chiral
superfields and no covariant derivatives applied to external auxiliary fields:
\sbox{\boxb}{\includegraphics[width=3cm]{14.eps}}
\settowidth{\lb}{\usebox{\boxb}}
\begin{eqnarray*}
 \left[\parbox{\lb}{\usebox{\boxb}}\right]_{\rm 1PI}.
\end{eqnarray*}
For instance, consider the following amplitude with two external
auxiliary fields
\sbox{\boxa}{\includegraphics[width=5cm]{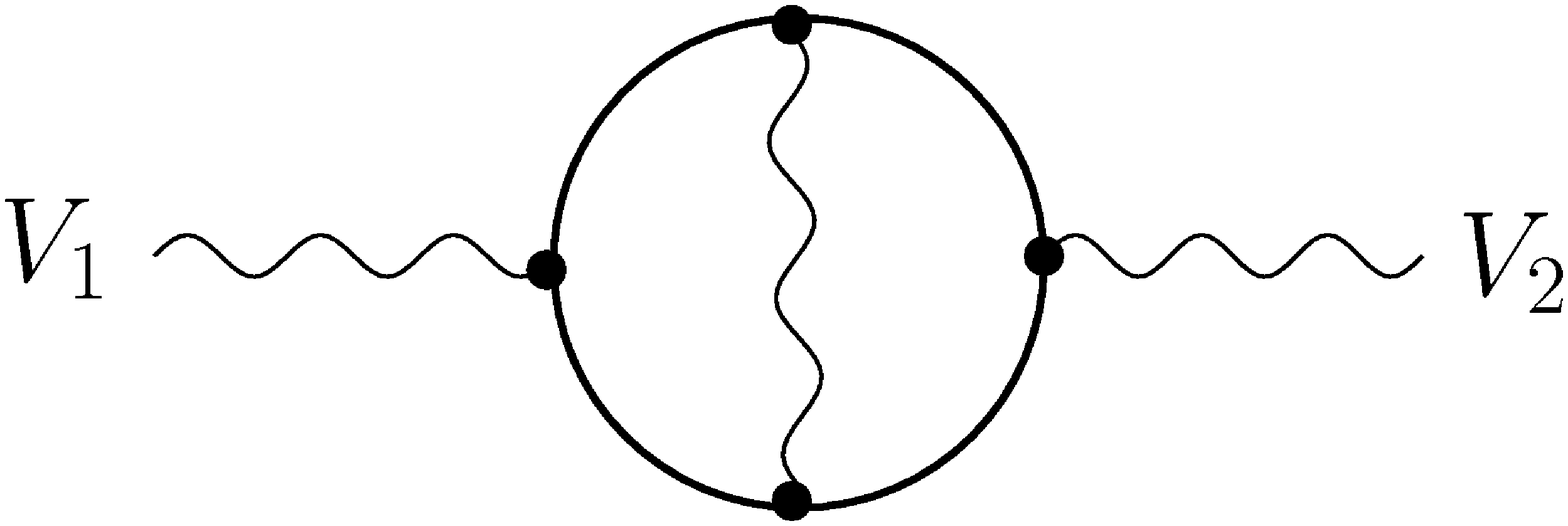}}
\settowidth{\la}{\usebox{\boxa}}
\begin{eqnarray*}
\parbox{\la}{\usebox{\boxa}},
\end{eqnarray*}
This induces an order $1/N^2$ correction to the inverse propagator of the
auxiliary field. By the same argument as above, we can show this
amplitude cancels the
following one
\sbox{\boxb}{\includegraphics[width=3cm]{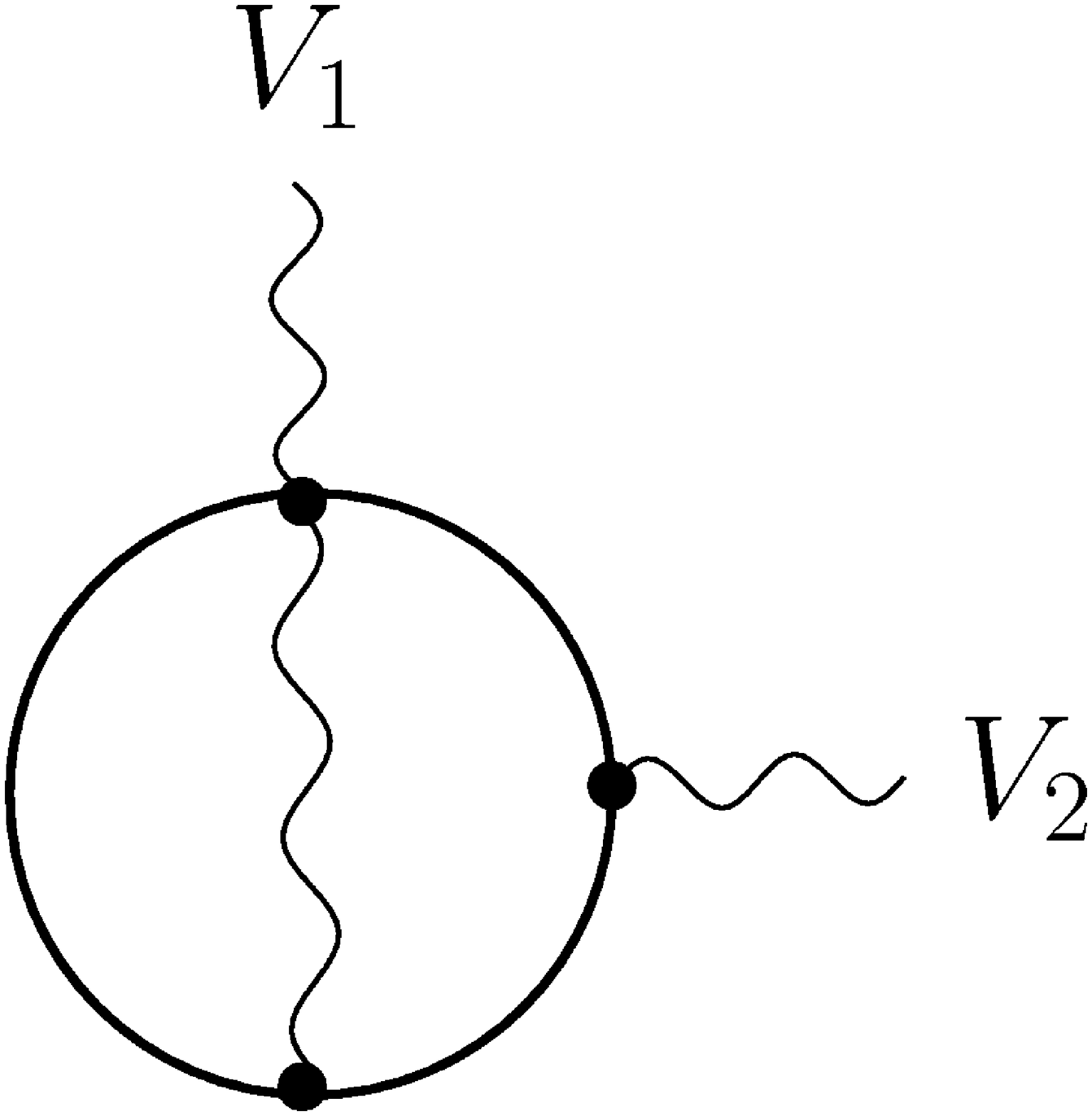}}
\settowidth{\lb}{\usebox{\boxb}}
\begin{eqnarray*}
 \parbox{\lb}{\usebox{\boxb}}.
\end{eqnarray*}
 except for terms at least one covariant derivative
applied to external auxiliary fields. Note that covariant derivatives
do not act on the internal line of the auxiliary field, since $V_1$ is
attached on the propagator of dynamical fields. Shifting the vertex attached to
$V_1$ along the chiral loop in a clockwise direction, We can also show that
\sbox{\boxa}{\includegraphics[width=3cm]{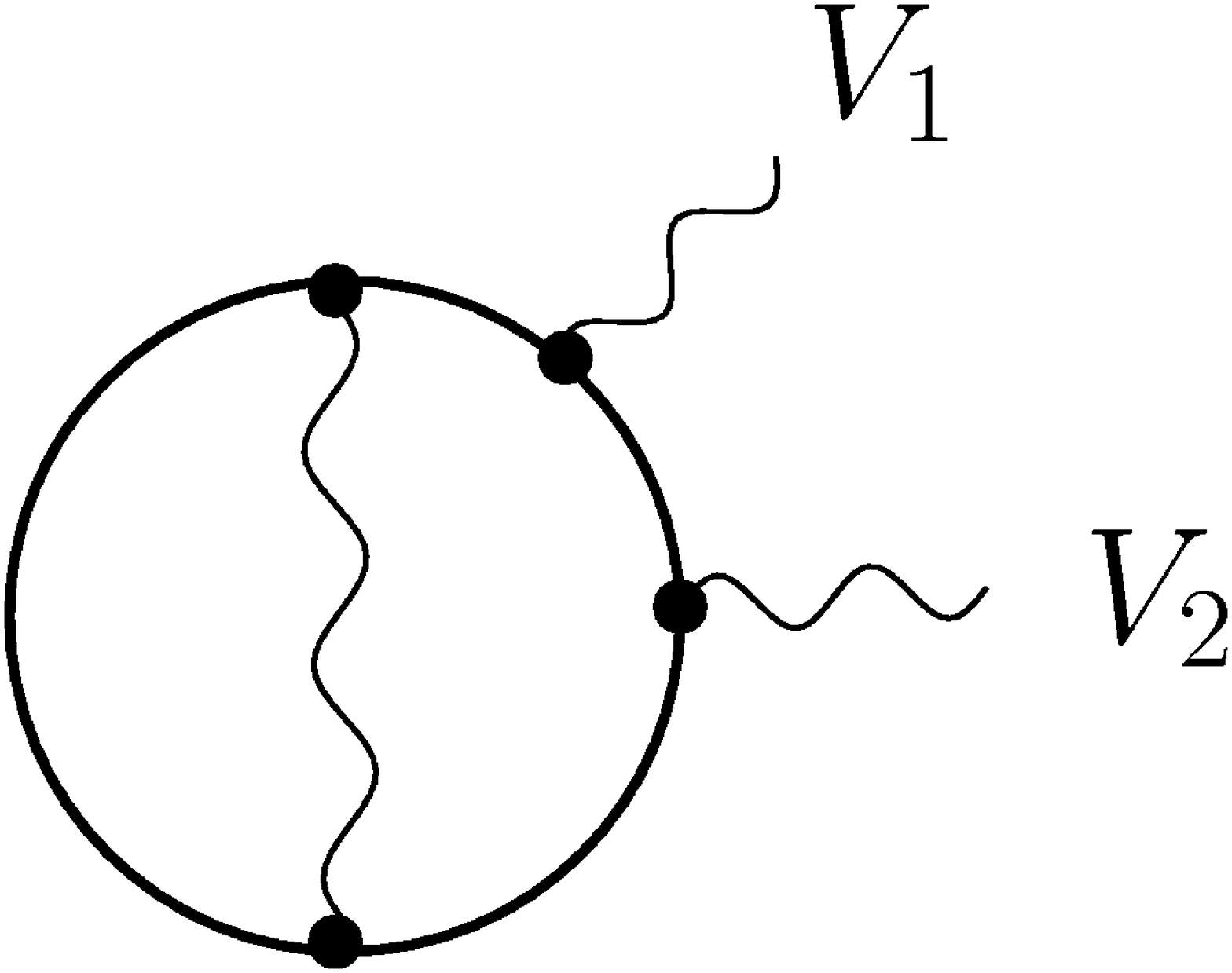}}
\settowidth{\la}{\usebox{\boxa}}
\sbox{\boxb}{\includegraphics[width=3cm]{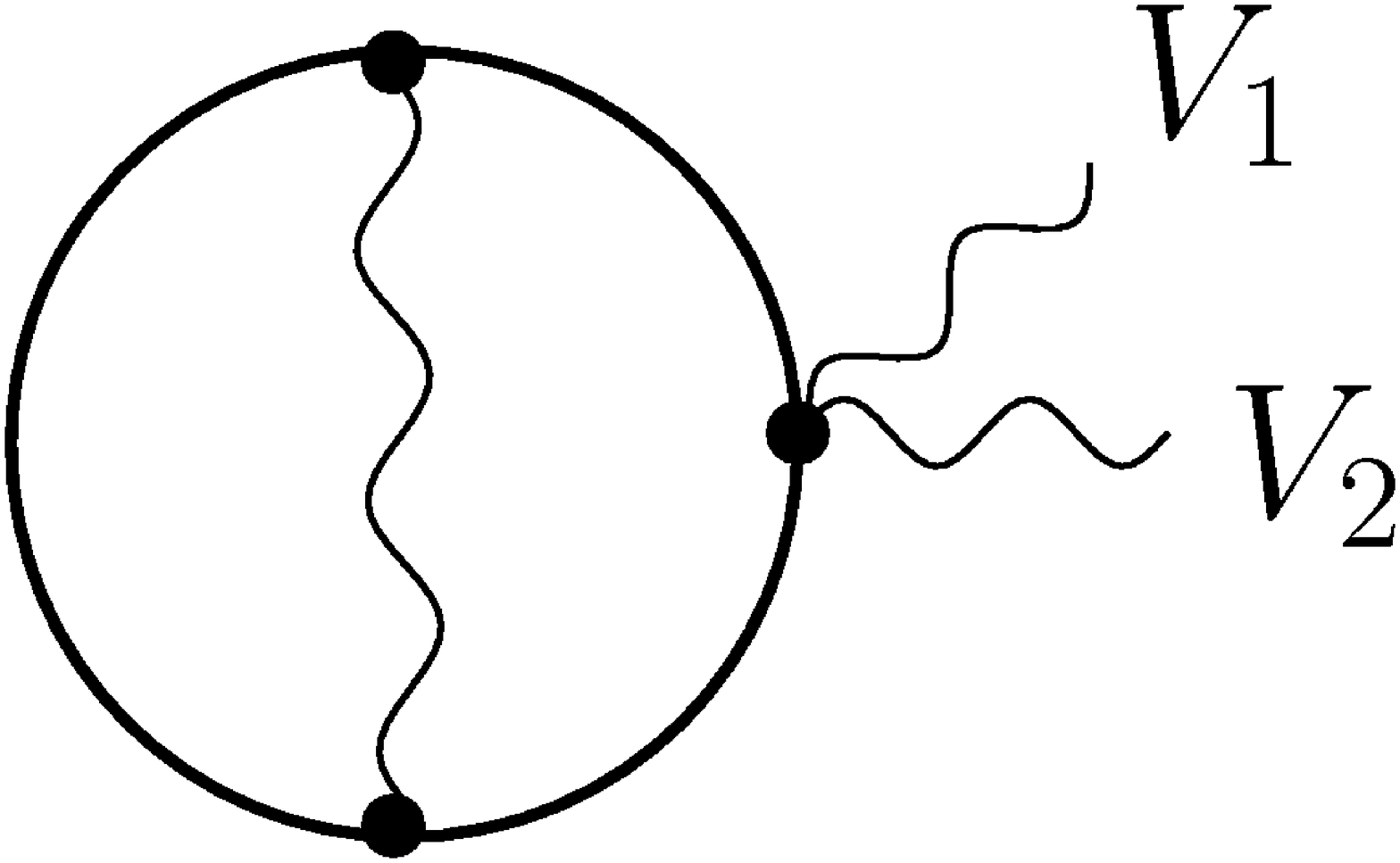}}
\settowidth{\lb}{\usebox{\boxb}}
\sbox{\boxc}{\includegraphics[width=3cm]{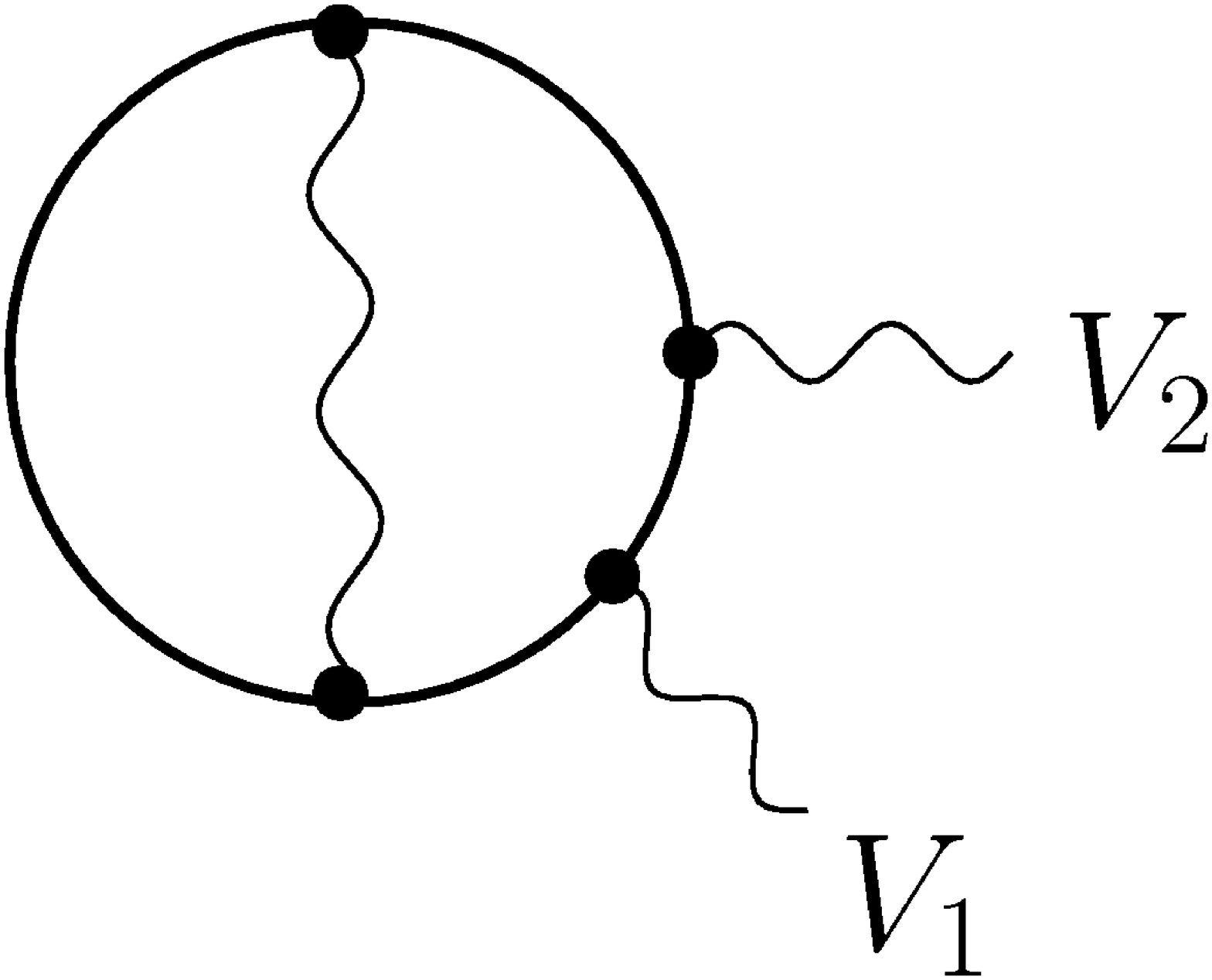}}
\settowidth{\lc}{\usebox{\boxc}}
\sbox{\boxd}{\includegraphics[width=3cm]{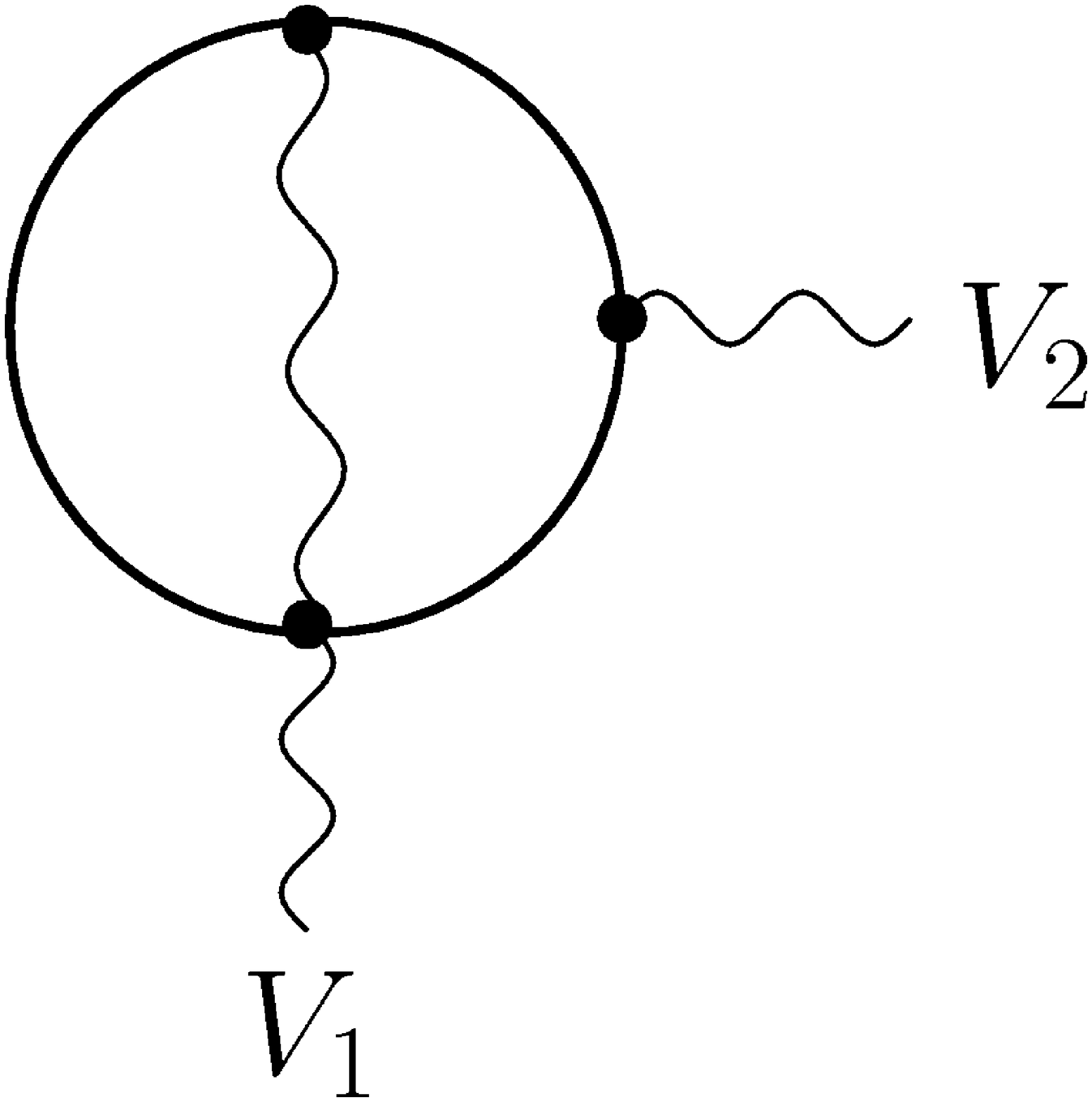}}
\settowidth{\ld}{\usebox{\boxd}}
\begin{eqnarray*}
 \parbox{\la}{\usebox{\boxa}} \quad + \quad \parbox{\lb}{\usebox{\boxb}}
  \quad &\sim& \quad 0\\[3mm]
\parbox{\lc}{\usebox{\boxc}} \quad + \quad \parbox{\ld}{\usebox{\boxd}}
\quad&\sim&\quad 0
\end{eqnarray*}
These six amplitudes are given by inserting a vertex
\sbox{\boxd}{\includegraphics[width=2.5cm]{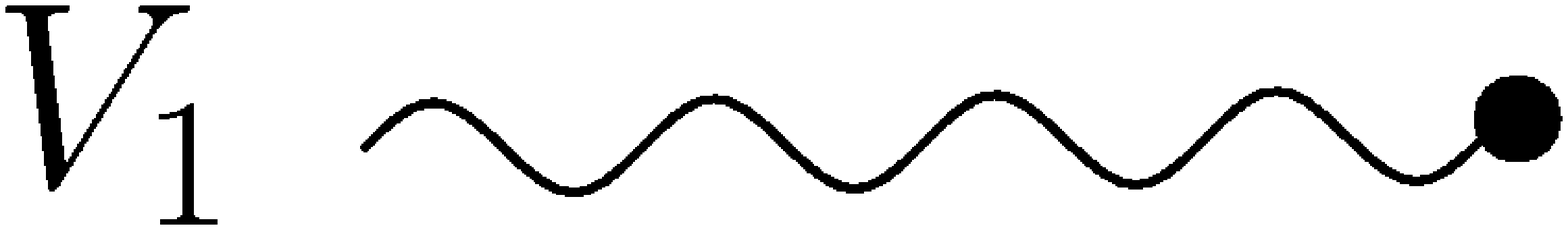}}
\settowidth{\ld}{\usebox{\boxd}}
\begin{eqnarray}
\parbox{\ld}{\usebox{\boxd}} \label{eq780}
\end{eqnarray}
into the following diagram
\sbox{\boxd}{\includegraphics[width=3cm]{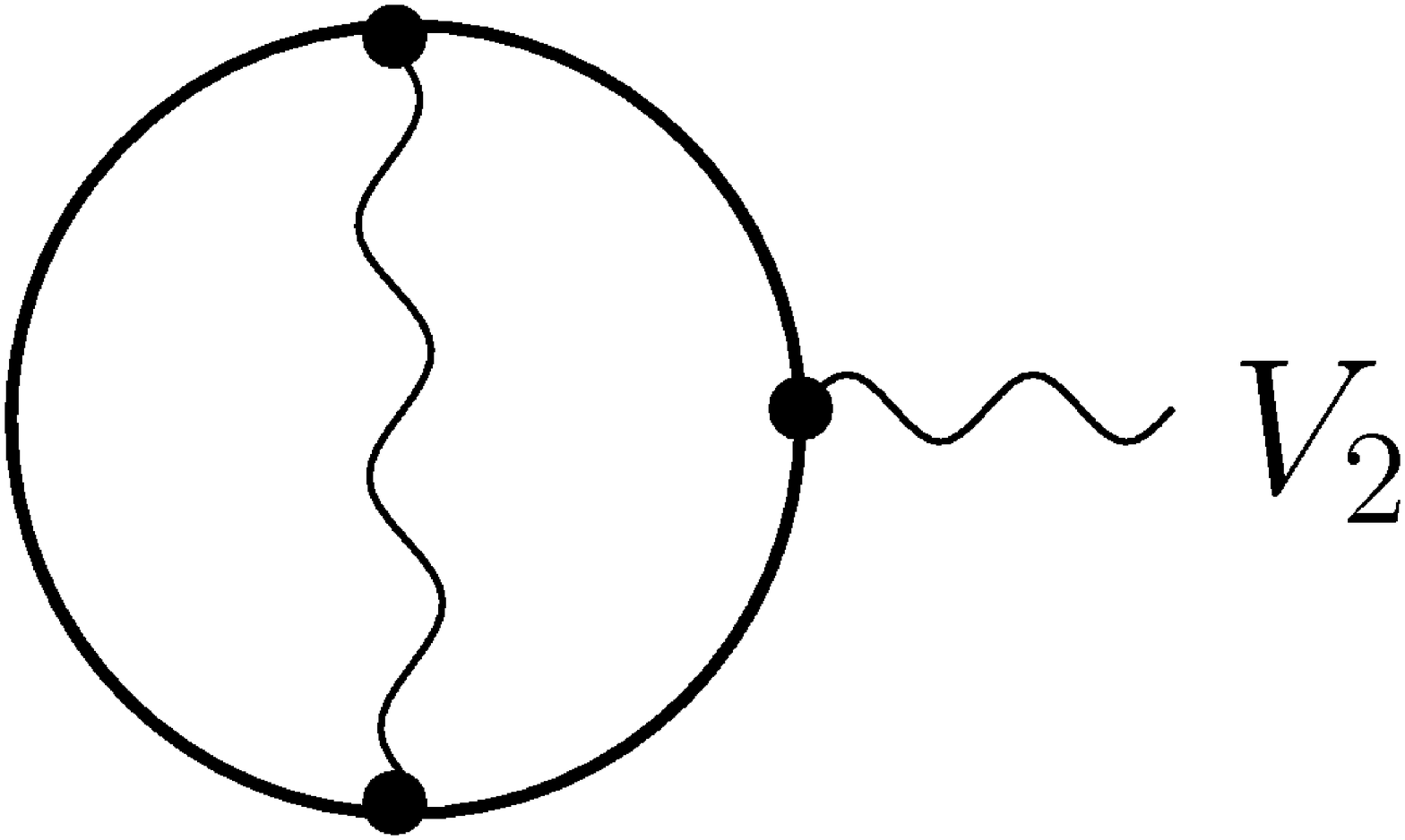}}
\settowidth{\ld}{\usebox{\boxd}}
\begin{eqnarray}
\parbox{\ld}{\usebox{\boxd}}, \label{eq781}
\end{eqnarray}
where there are six possible ways of insertion and summing up all these
amplitudes leads to a vanishing result. Similarly, all possible insertions of (\ref{eq780})
into the following diagrams
\sbox{\boxd}{\includegraphics[width=8cm]{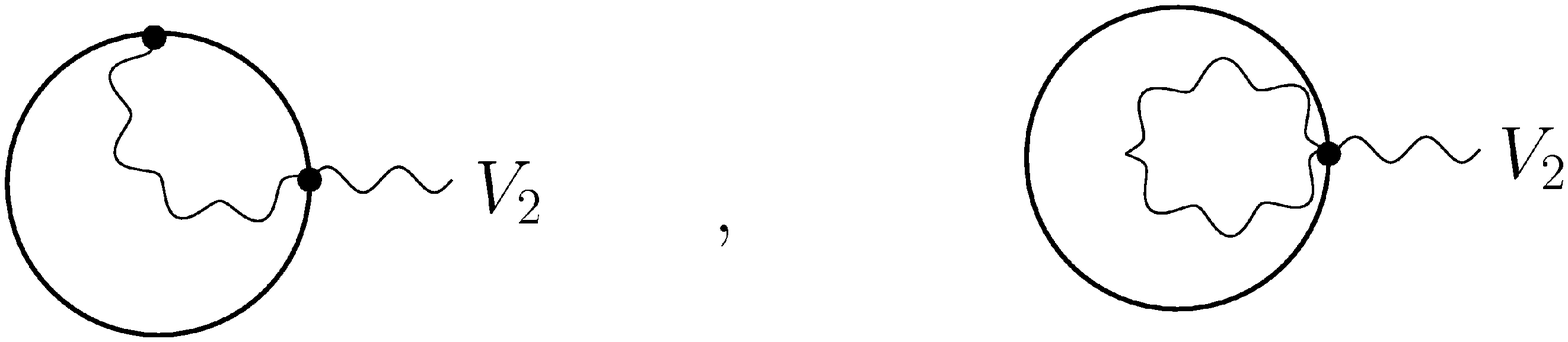}}
\settowidth{\ld}{\usebox{\boxd}}
\begin{eqnarray}
\parbox{\ld}{\usebox{\boxd}} \label{eq782}
\end{eqnarray}
also give vanishing results.

Considering all possible insertions into (\ref{eq781}) and
(\ref{eq782}), we can obtain all amplitudes of order $1/N^2$ with two
external auxiliary fields. This means that in the effective action there is no
quantum correction to the operator $V^2$ in order $1/N^2$. We can also
prove that amplitudes of order $1/N^m$ with two external auxiliary fields vanish when we
sum them up, with an arbitrary positive integer $m$, except for terms
with at least one covariant derivative applied to external fields.

In the same way, if we fixed the number of external auxiliary fields and
the order of $1/N$, except for only one case, we can show that all contributions vanish if
we sum them up, up to terms which have at least one covariant
derivatives applied to external fields. For example, if the diagram has
$n$ external auxiliary fields and consider the contribution of order $1/N^m$, we choose one external auxiliary field
and consider all diagrams of order $1/N^m$ without it. Then we consider all
possible insertions of the chosen external auxiliary field into them. The insertions have to be made at 
propagators or vertices on loops of $\tilde{\Phi}$. Since any chiral loop has the same number of
propagators and vertices, it gives a vanishing result to sum up
all insertions. The only one exception is the following one-loop amplitude:
\sbox{\boxa}{\includegraphics[width=3cm]{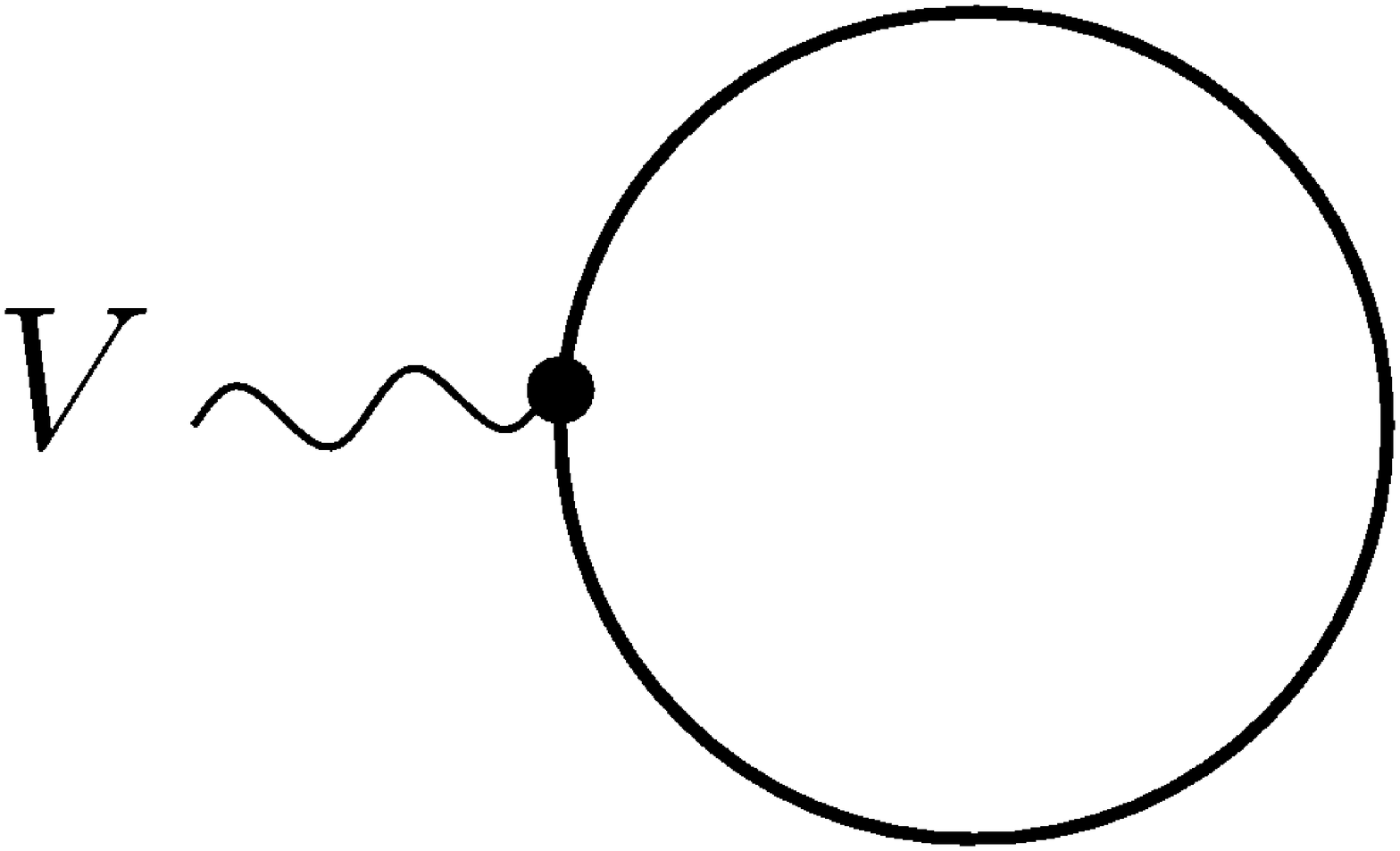}}
\settowidth{\la}{\usebox{\boxa}}
\begin{eqnarray}
 \parbox{\la}{\usebox{\boxa}}. \label{eq1105}
\end{eqnarray}
Since this is only one amplitude of order $N$ with one external
auxiliary field, it has no counterpart to cancel. However, this
contribution was already considered when we evaluated the vacuum structure of the theory
in subsection 2.2. It gave a linearly divergent contribution to be eliminated
by the renormalization of the coupling constant.

In summary, it is proved that no counter terms of the form of
$V^n$ are necessary, except for the case $n=1$. Namely, $\alpha_n$ in
(\ref{eq730}) vanishes unless $n=1$. In the case $n=1$, there is a
linearly divergent term proportional to $N$ but it is cancelled by a
counter term induced by the renormalization of the coupling
constant. Notice here that the counter term is also proportional to
$N$. 

We now find that all divergent amplitudes of
the left type in (\ref{eq700}) become finite, at each order of the
$1/N$-expansion, only by renormalizing the coupling
constant.

\vspace{5ex}

\subsection{Wavefunction renormalization of $\tilde{\Phi}$}

We now show that all divergences from diagrams of the second
type in (\ref{eq700})
\sbox{\boxb}{\includegraphics[width=5cm]{15.eps}}
\settowidth{\lb}{\usebox{\boxb}}
\begin{eqnarray*}
 \left[\parbox{\lb}{\usebox{\boxb}}\right]_{\rm 1PI}
\end{eqnarray*}
can be eliminated by the renormalization of the
wavefunction of $\tilde{\Phi}$. Since its superficial degree of
divergence is zero, it may contain logarithmic
divergences. If we expand the amplitudes in powers of covariant
derivatives and external momenta, only the lowest order, which has no
external momenta and no covariant derivative acting on external
superfields, can be
divergent.  Therefore, in this subsection, we only consider terms with no
covariant derivatives acting on external superfields.

Notice here that we have to expand
amplitudes in
powers of $E_\alpha, \bar{E}_\alpha$ acting on external $\tilde{\Phi},
\tilde{\Phi}^\dagger$ as well as in powers of $D_\alpha, \bar{D}_\alpha$
acting on external $V$. The reason for this is
supersymmetry. Differential operators $E_\alpha, \bar{E}_\alpha$ are
supercovariant when they act on $\tilde{\Phi}, \tilde{\Phi}^\dagger$,
while $D_\alpha, \bar{D}_\alpha$ are supercovariant when they act on $V$.
 In order to study divergent amplitudes, it is
enough to investigate amplitudes which have no $E_\alpha,
\bar{E}_\alpha$ acting on external $\tilde{\Phi}, \tilde{\Phi}^\dagger$
and no $D_\alpha, \bar{D}_\alpha$ on external $V$.

We have already seen that the following type of amplitudes
\sbox{\boxb}{\includegraphics[width=3cm]{14.eps}}
\settowidth{\lb}{\usebox{\boxb}}
\begin{eqnarray}
 \left[\parbox{\lb}{\usebox{\boxb}}\right]_{\rm 1PI} \label{eq1112}
\end{eqnarray}
has no divergence. Especially when we neglect terms with covariant
derivatives acting on external superfields, it led to a vanishing result to sum up
all the diagrams of the above type. The reason for this was as
follows. Suppose the following diagram and take one external auxiliary
field, which is always attached to a loop
of a twisted chiral superfield:
\sbox{\boxa}{\includegraphics[width=2.5cm]{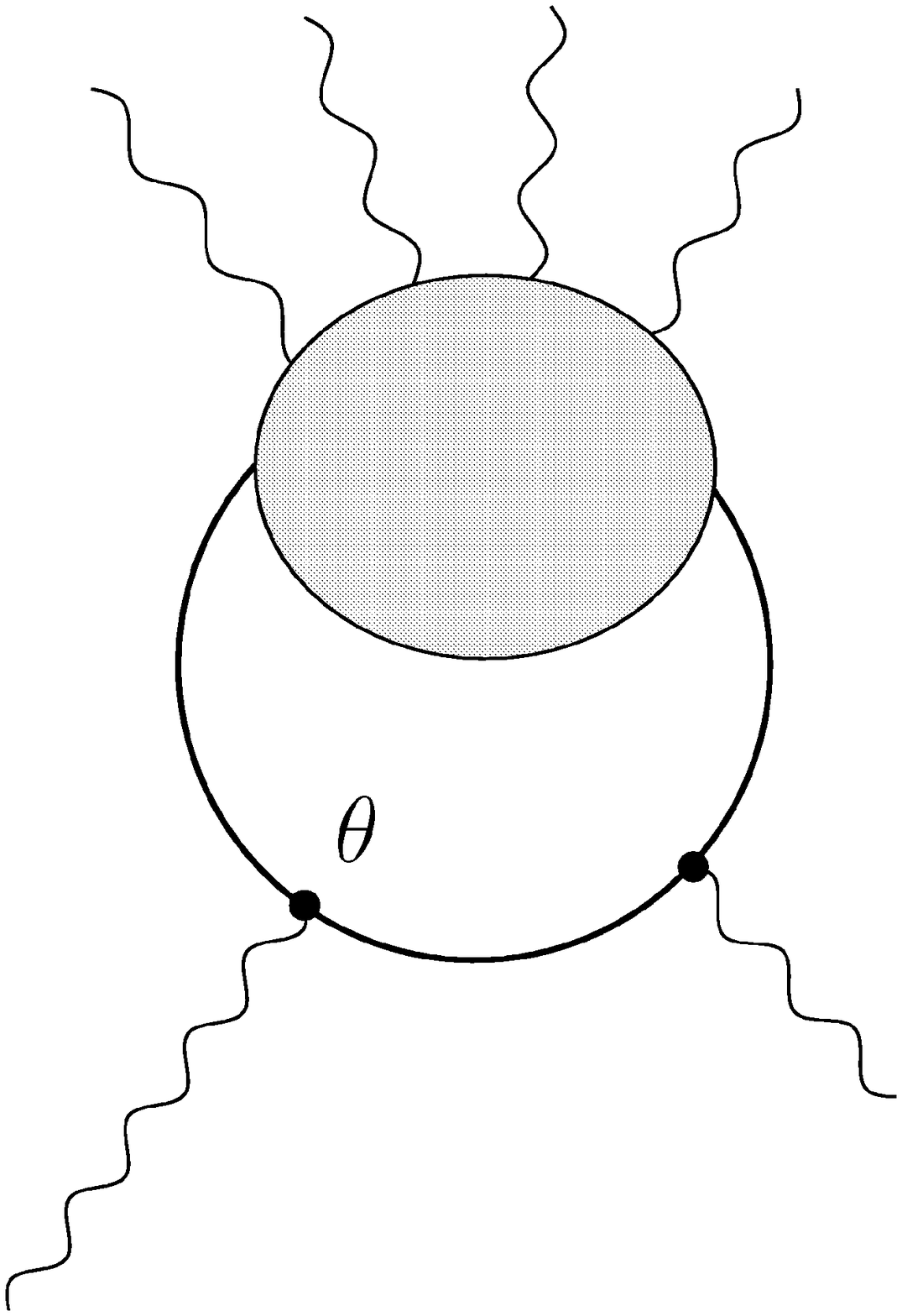}}
\settowidth{\la}{\usebox{\boxa}}
\begin{eqnarray}
 \parbox{\la}{\usebox{\boxa}} \label{eq1110}
\end{eqnarray}
When we perform a partial integration at
a vertex $\theta$ where the chosen external auxiliary field is attached, we use the
formula (\ref{eq771}) {\bf at} a chiral propagator next to the
vertex. Keeping only the contribution with no $D_\alpha,
\bar{D}_\alpha$ acting on external auxiliary fields, we find that it
exactly cancels another diagram
\sbox{\boxa}{\includegraphics[width=2.5cm]{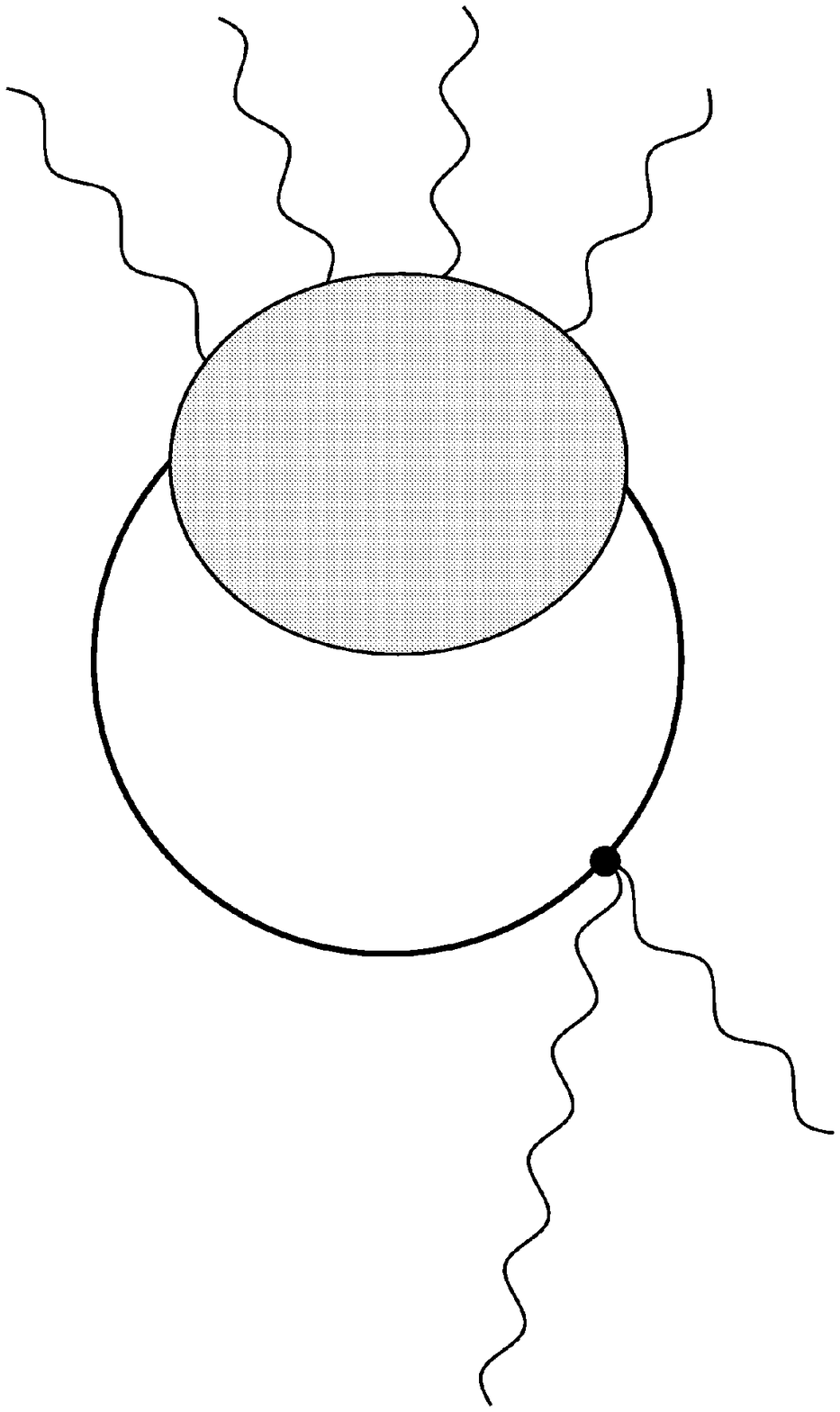}}
\settowidth{\la}{\usebox{\boxa}}
\begin{eqnarray}
 \parbox{\la}{\usebox{\boxa}}. \label{eq1111}
\end{eqnarray}
These two diagrams can be obtained by inserting one external auxiliary
 field in the following diagram:
\sbox{\boxa}{\includegraphics[width=2.5cm]{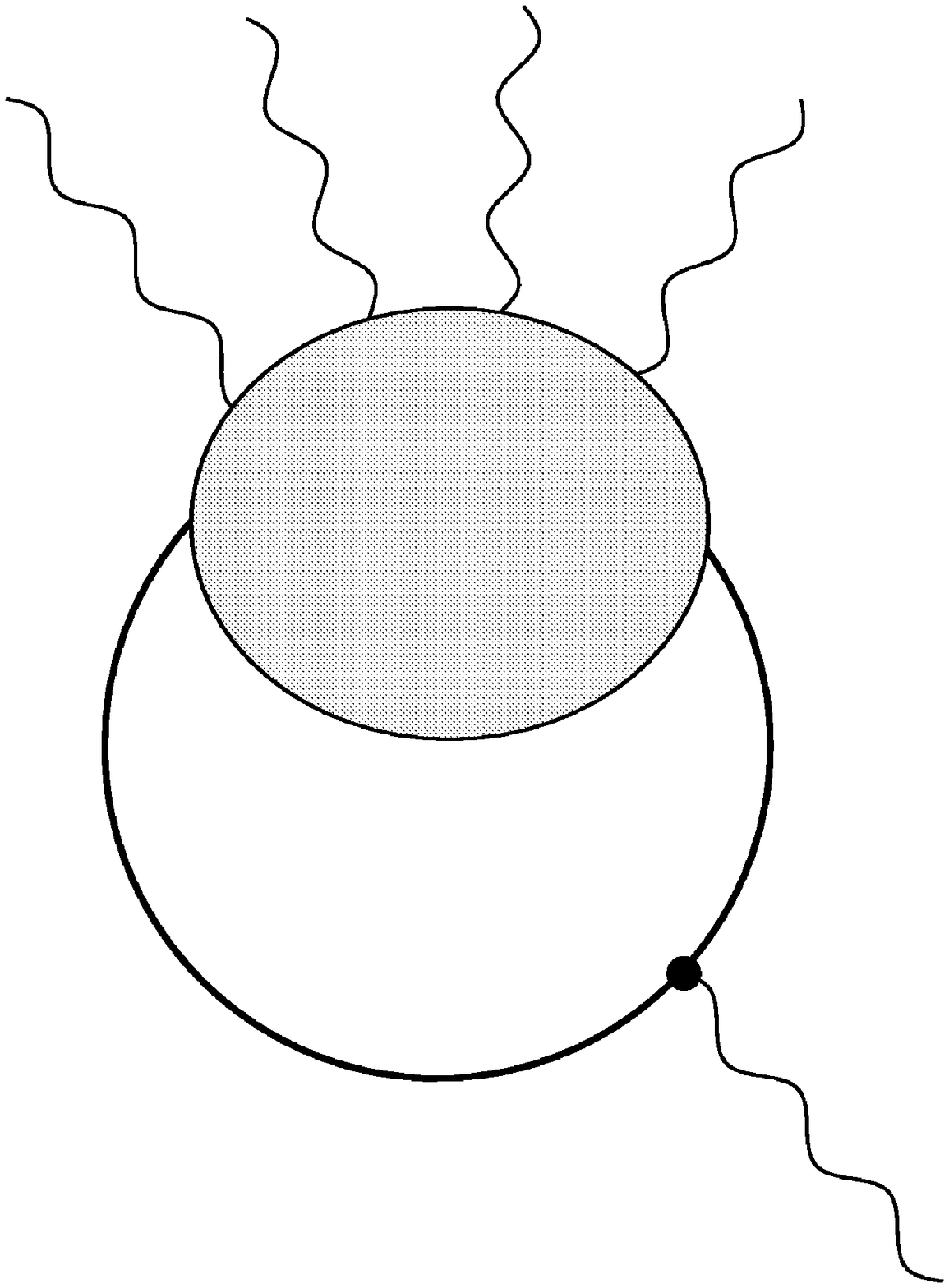}}
\settowidth{\la}{\usebox{\boxa}}
\begin{eqnarray*}
\parbox{\la}{\usebox{\boxa}}.
\end{eqnarray*}
The diagram (\ref{eq1110}) can be obtained by inserting an external $V$ into the
 chiral propagator in the loop while the diagram (\ref{eq1111}) can be obtained by
 inserting it into the vertex in the loop. The diagrams obtained by these
 two insertions cancel each other.
Since any loop
has the same number of propagators and vertices, the sum of
 contributions of diagrams obtained by moving one external auxiliary line along the
chiral loop cancel each other.
 This kind of cancellation occurs when other external lines
 and all internal lines are fixed. Therefore, considering all diagrams of the form
 (\ref{eq1112}) leads to
 a vanishing result.

In the same way, it gives a vanishing result to sum up all the diagrams
of the following form
\sbox{\boxa}{\includegraphics[width=4cm]{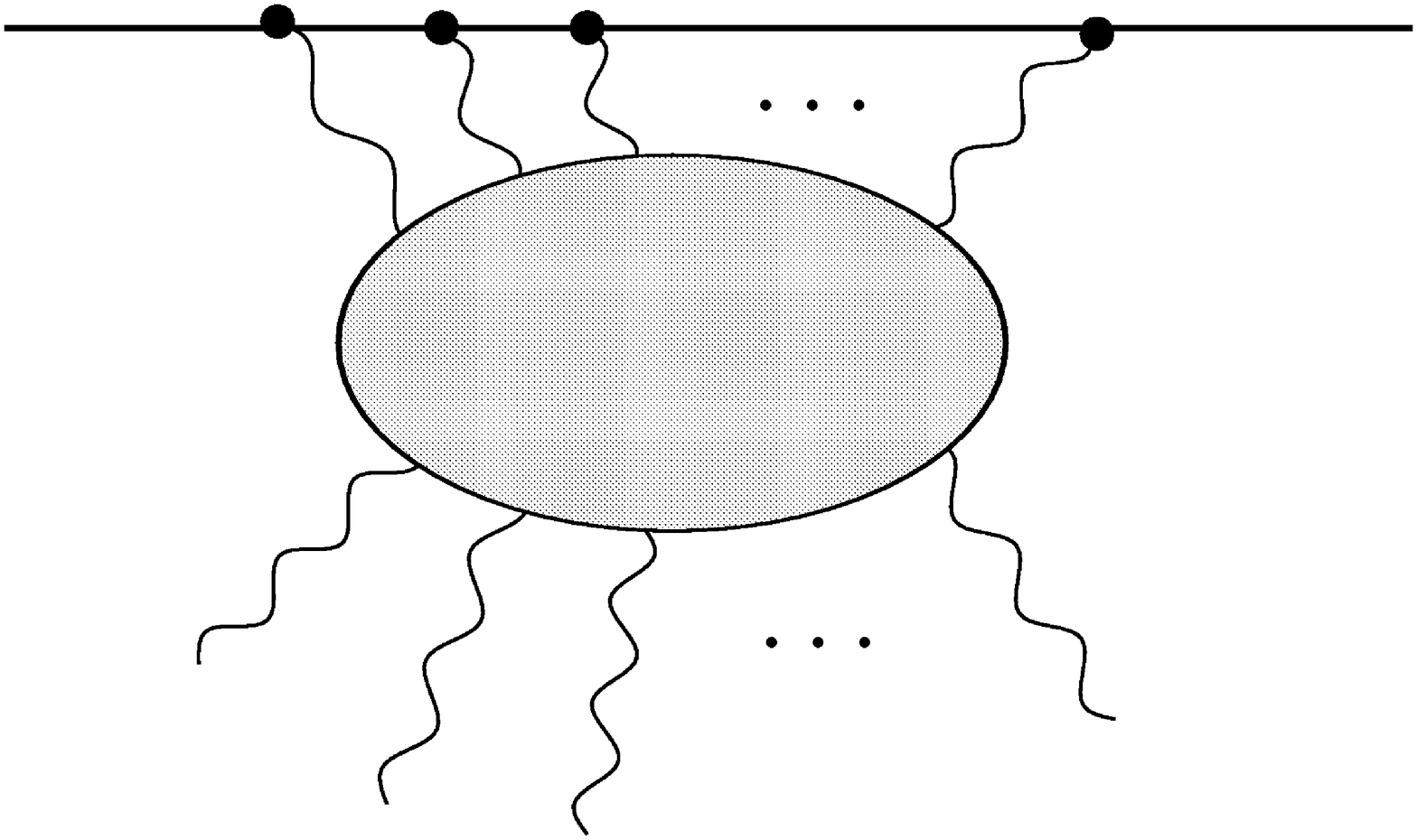}}
\settowidth{\la}{\usebox{\boxa}}
\begin{eqnarray*}
 \parbox{\la}{\usebox{\boxa}}
\end{eqnarray*}
when we neglect terms with covariant derivatives acting on external
superfields. But we cannot use the same argument to evaluate the following
diagrams
\sbox{\boxa}{\includegraphics[width=4cm]{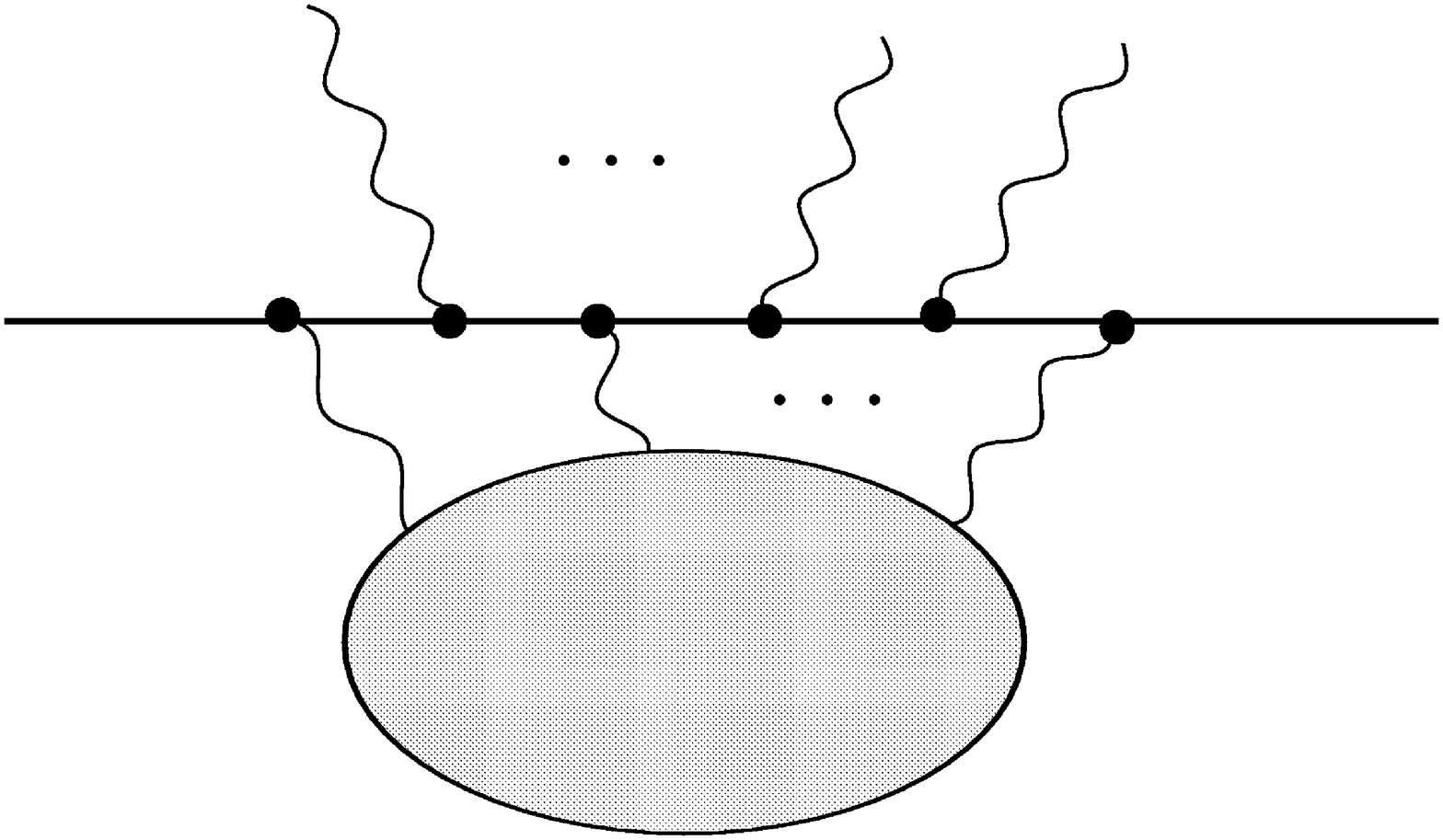}}
\settowidth{\la}{\usebox{\boxa}}
\begin{eqnarray*}
 \parbox{\la}{\usebox{\boxa}}
\end{eqnarray*}
because there is no external auxiliary field attached to an internal chiral
loop. All external auxiliary fields are attached to the chiral line
which connects two external twisted chiral superfields. In this case,
even if we neglect terms with covariant derivatives acting on external
superfields and move one external auxiliary field along the chiral line, with other external
lines and all internal lines being fixed, it gives a non-vanishing result to sum up
all contributions. The reason for this is as
follows. The chiral line has one more
vertices than propagators, and it has also two
external chiral lines. Then, there are two types of insertion of the chosen
external auxiliary field: (A) insertion in a vertex on the chiral
line; (B) insertion in a chiral propagator or an external chiral
line. Each diagram of type (A) has a counterpart of type (B) to
cancel out. However, since there are one more diagram of type (B) than
that of (A), without cancellation when we sum up
all insertion of the chosen external auxiliary field, only one diagram
of type (B) remains.

For instance, suppose the following diagram:
\sbox{\boxa}{\includegraphics[width=4cm]{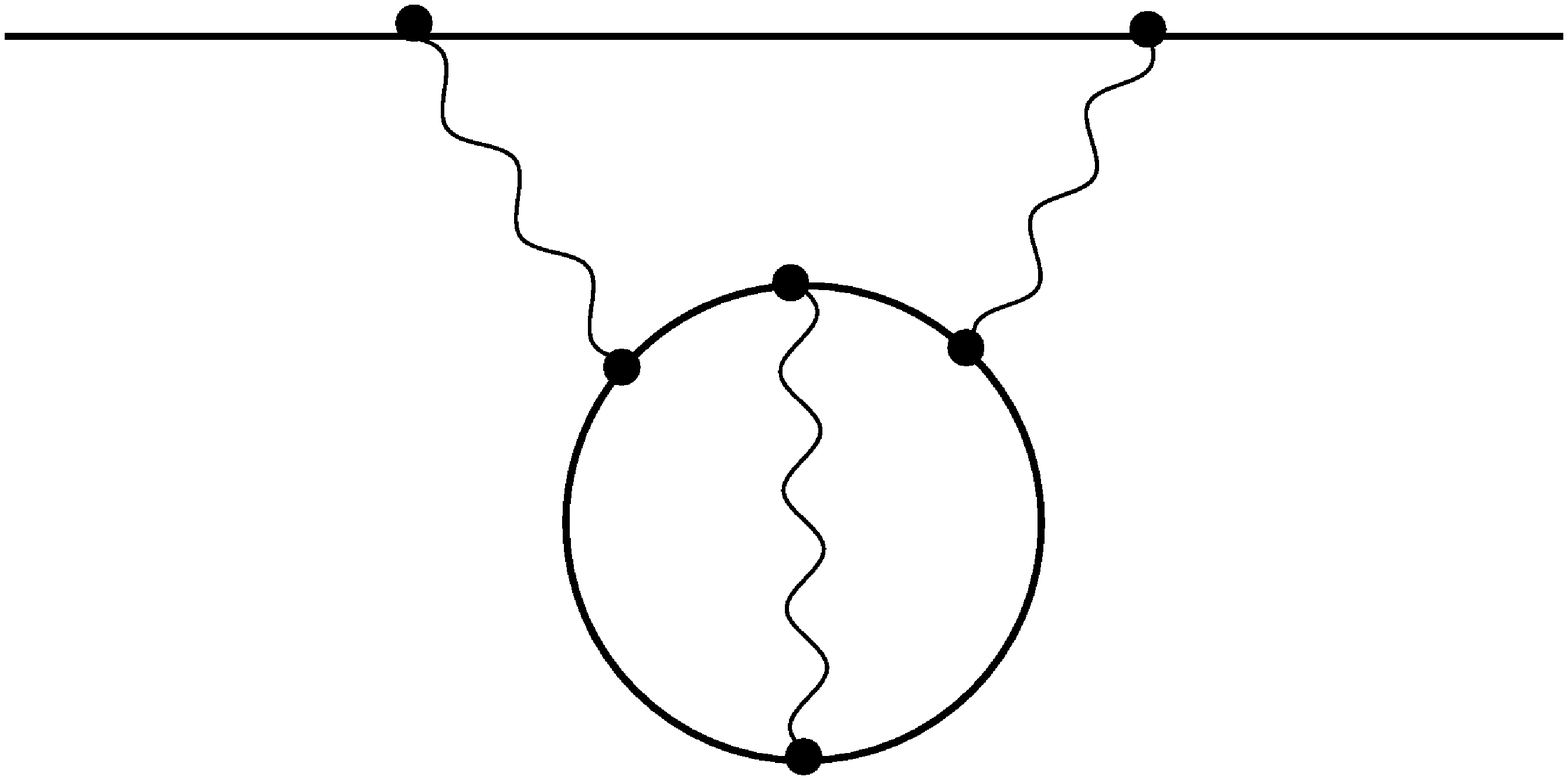}}
\settowidth{\la}{\usebox{\boxa}}
\begin{eqnarray}
 \parbox{\la}{\usebox{\boxa}}. \label{eq1101}
\end{eqnarray}
We insert one external auxiliary field in the above diagram. Since the
chiral loop in the above diagram has four vertices and four propagators,
considering all insertion of the external auxiliary field in the chiral
loop, we obtain a vanishing result up to terms with covariant
derivatives acting on external superfields. On the other hand, the chiral line
in the above diagram has two vertices, one propagator, and two
external lines. We first find that the
following two diagrams vanish when they are summed:
\sbox{\boxa}{\includegraphics[width=4cm]{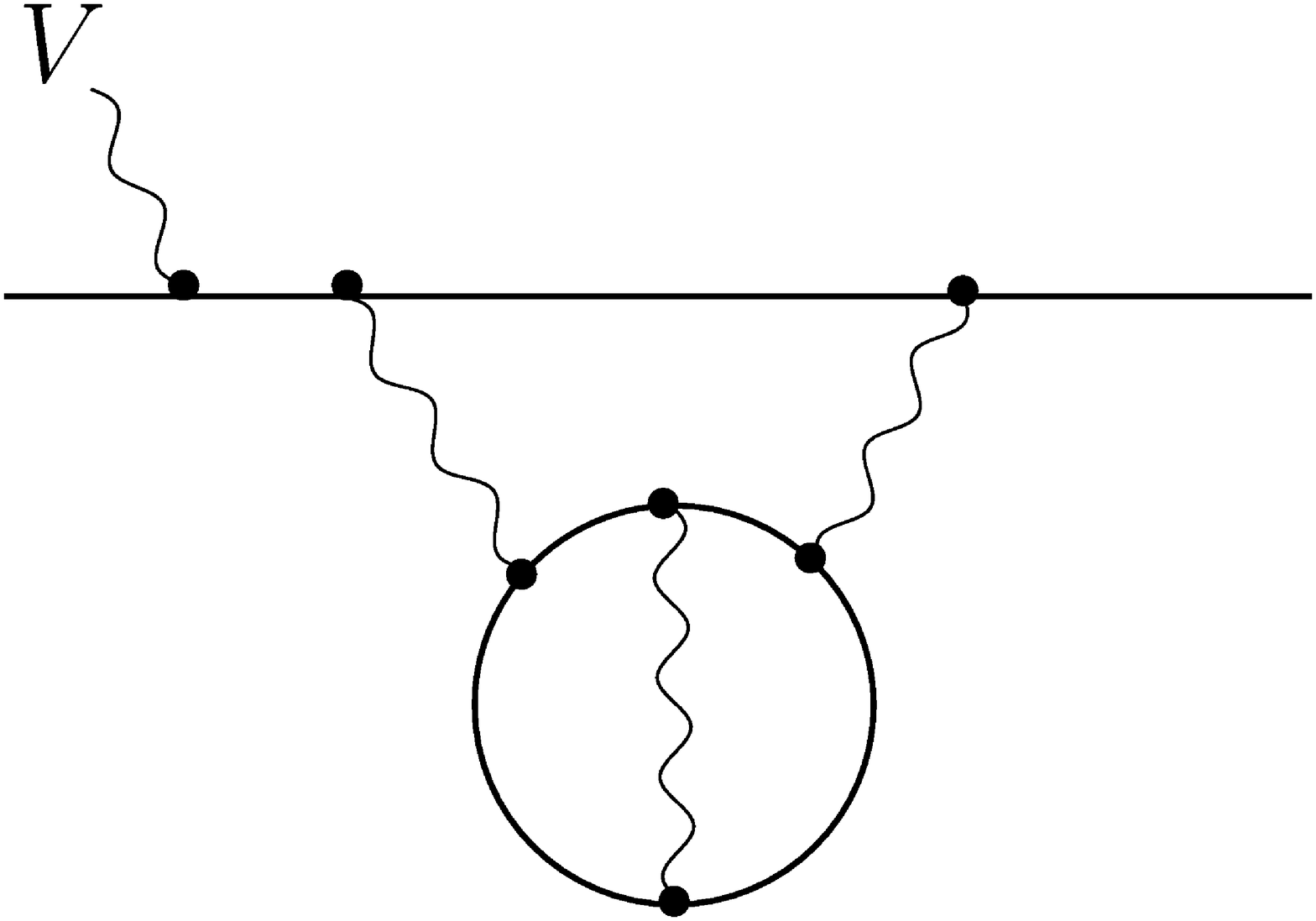}}
\settowidth{\la}{\usebox{\boxa}}
\sbox{\boxb}{\includegraphics[width=4cm]{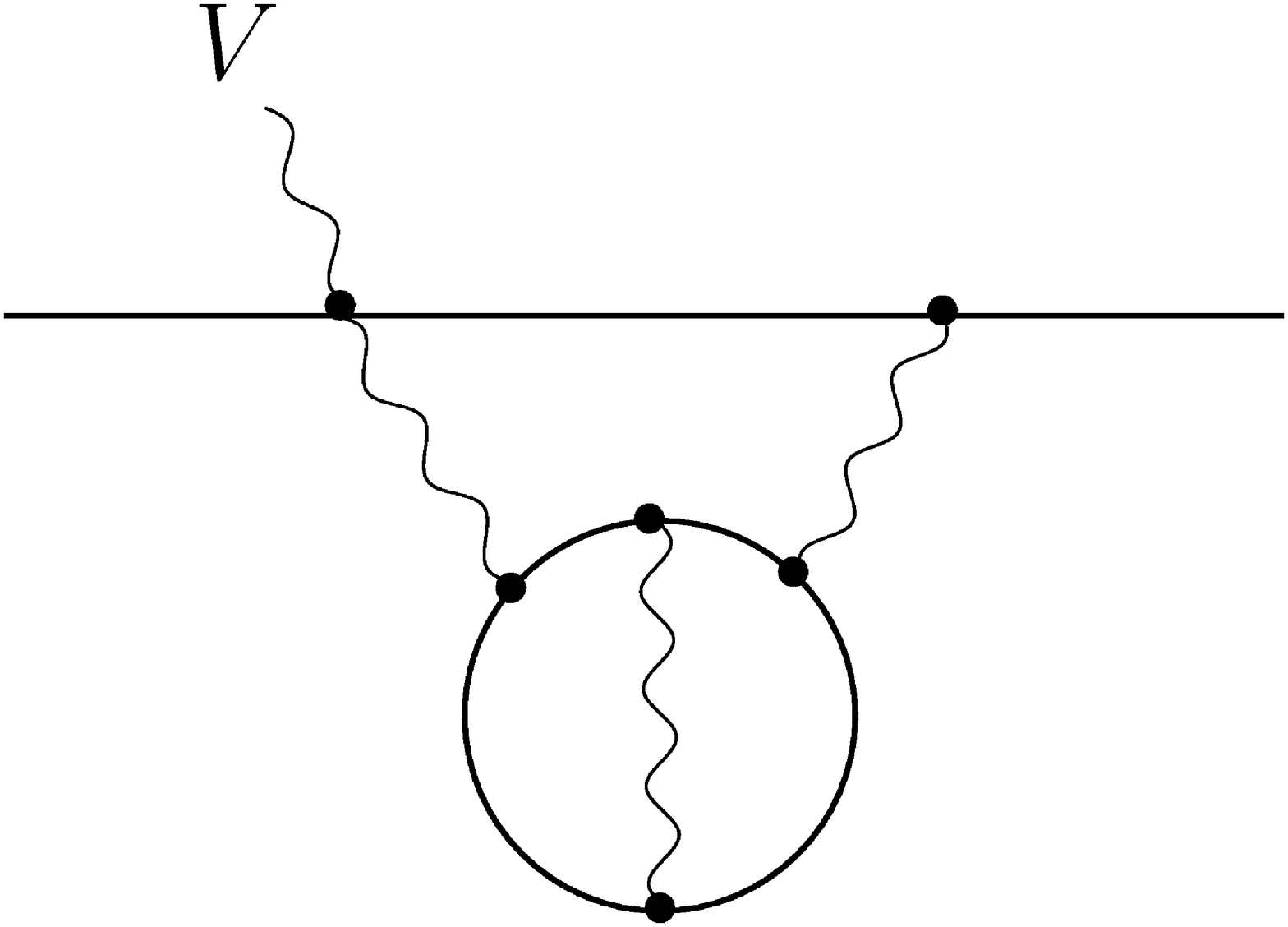}}
\settowidth{\lb}{\usebox{\boxb}}
\begin{eqnarray*}
\parbox{\la}{\usebox{\boxa}} \quad+\quad \parbox{\lb}{\usebox{\boxb}}
 \quad \sim \quad 0,
\end{eqnarray*}
where we neglect terms with covariant derivatives acting
on external superfields. In the same way,
we find that the sum of the following diagrams vanishes:
\sbox{\boxa}{\includegraphics[width=4cm]{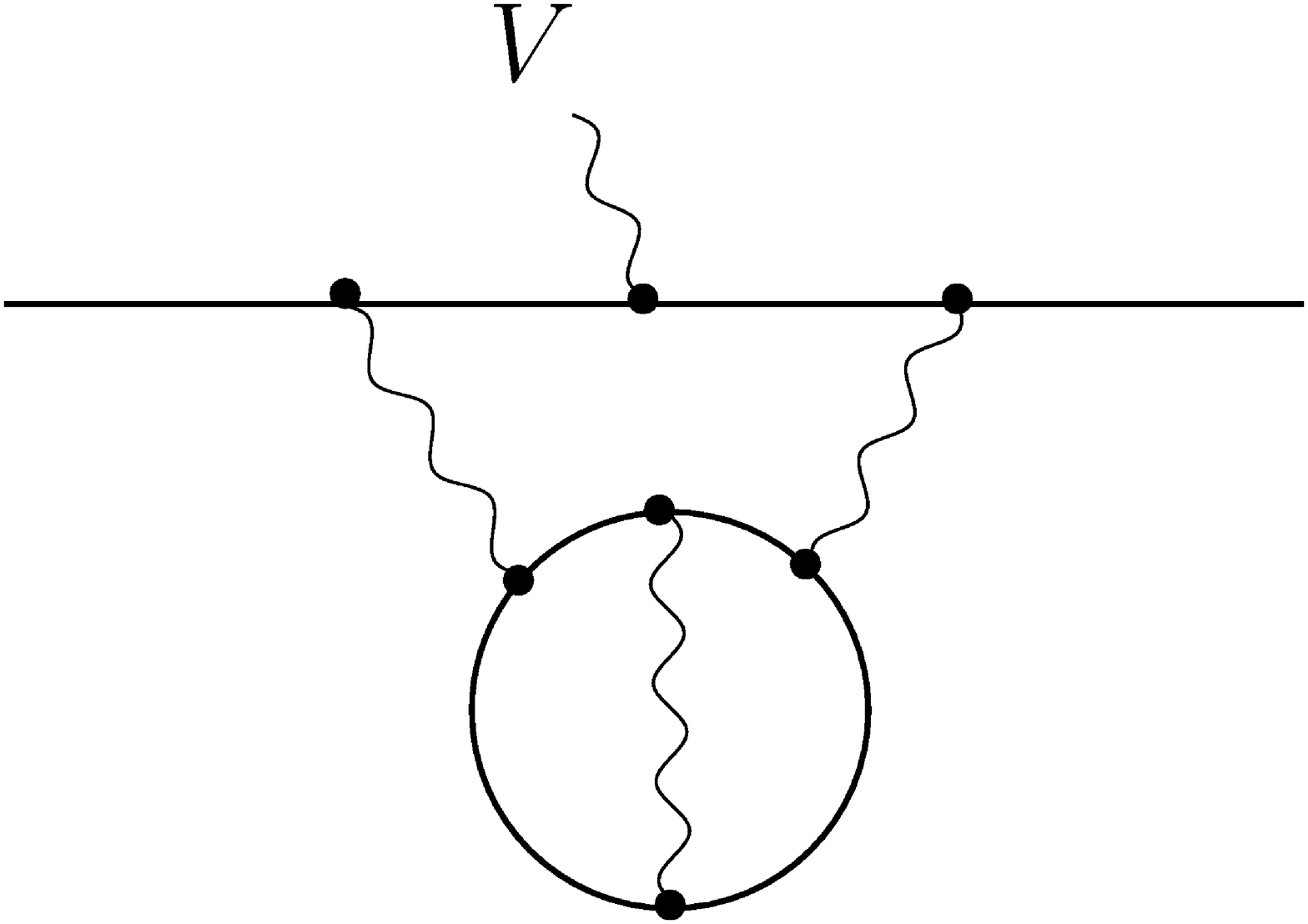}}
\settowidth{\la}{\usebox{\boxa}}
\sbox{\boxb}{\includegraphics[width=4cm]{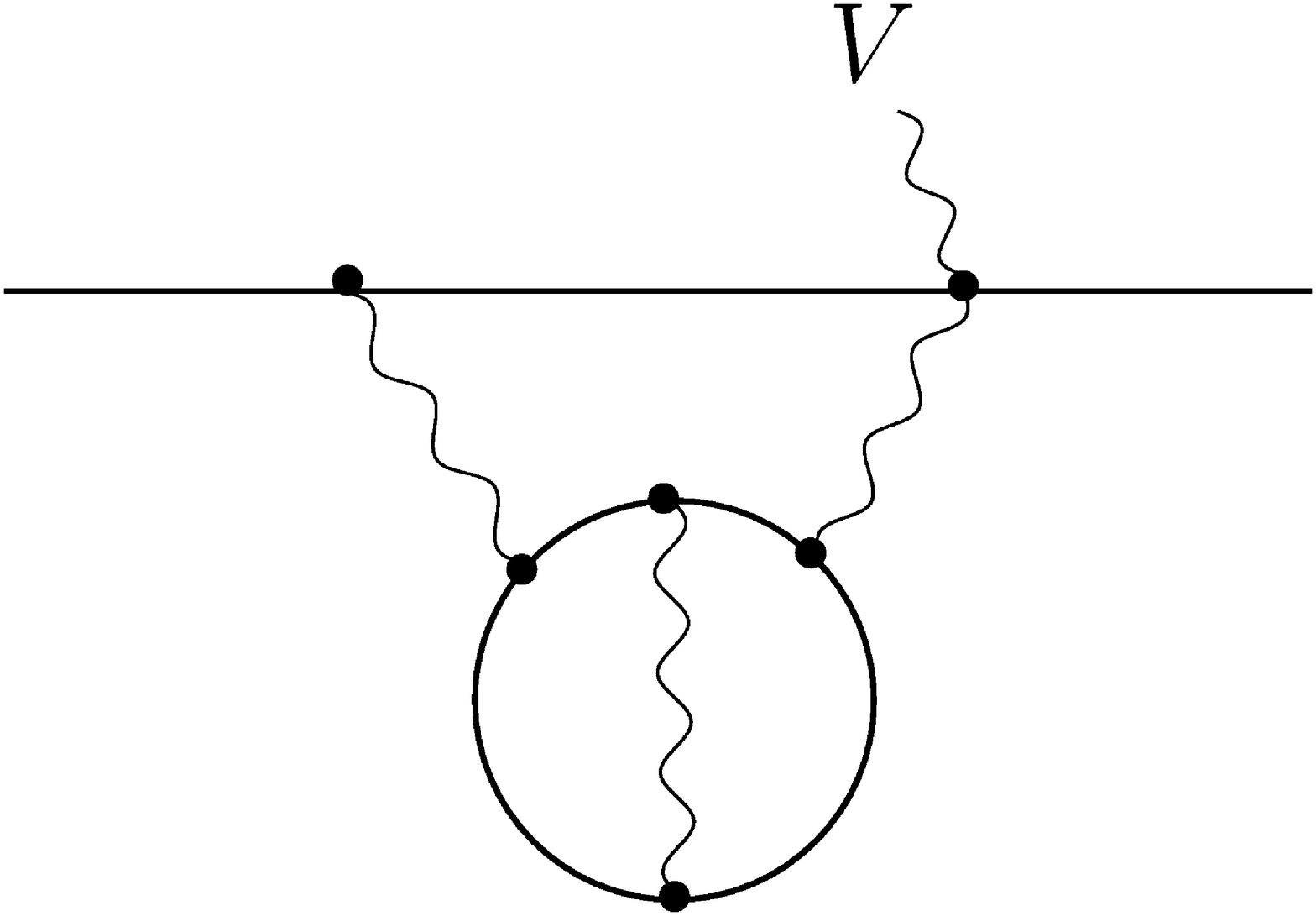}}
\settowidth{\lb}{\usebox{\boxb}}
\begin{eqnarray*}
\parbox{\la}{\usebox{\boxa}} \quad+\quad \parbox{\lb}{\usebox{\boxb}}
 \quad \sim \quad 0.
\end{eqnarray*}
The following diagram, however, remains:
\sbox{\boxa}{\includegraphics[width=4cm]{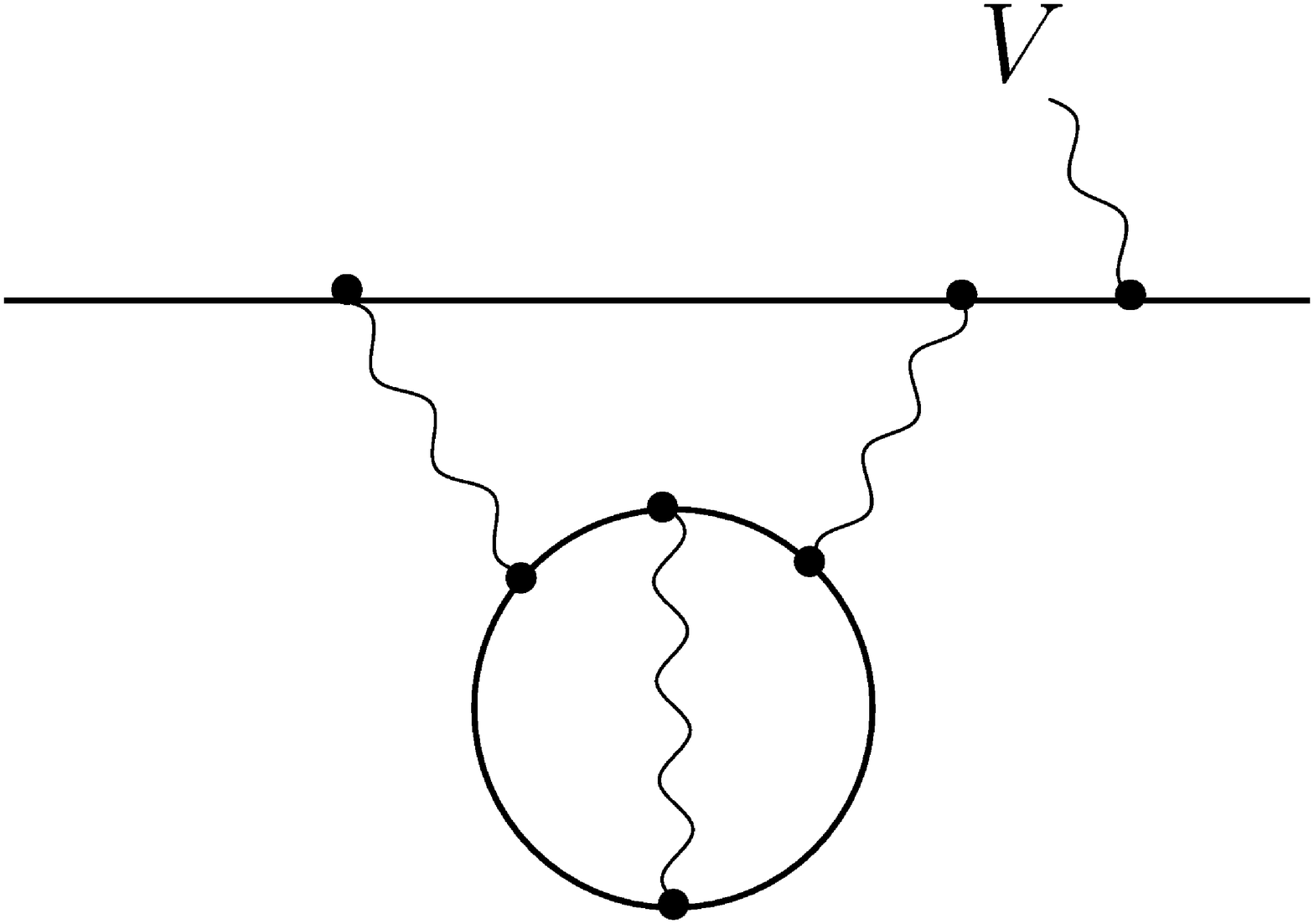}}
\settowidth{\la}{\usebox{\boxa}}
\begin{eqnarray}
 \parbox{\la}{\usebox{\boxa}}. \label{eq1100}
\end{eqnarray}
This diagram does not have a counterpart to cancel. Therefore,
summing up all diagrams obtained by inserting one external auxiliary
field in the diagram (\ref{eq1101}), all diagrams
cancel out each other except for the diagram (\ref{eq1100}).
If we again perform the partial integration, we find the remaining diagram
(\ref{eq1100}) is equivalent to
\sbox{\boxa}{\includegraphics[width=4cm]{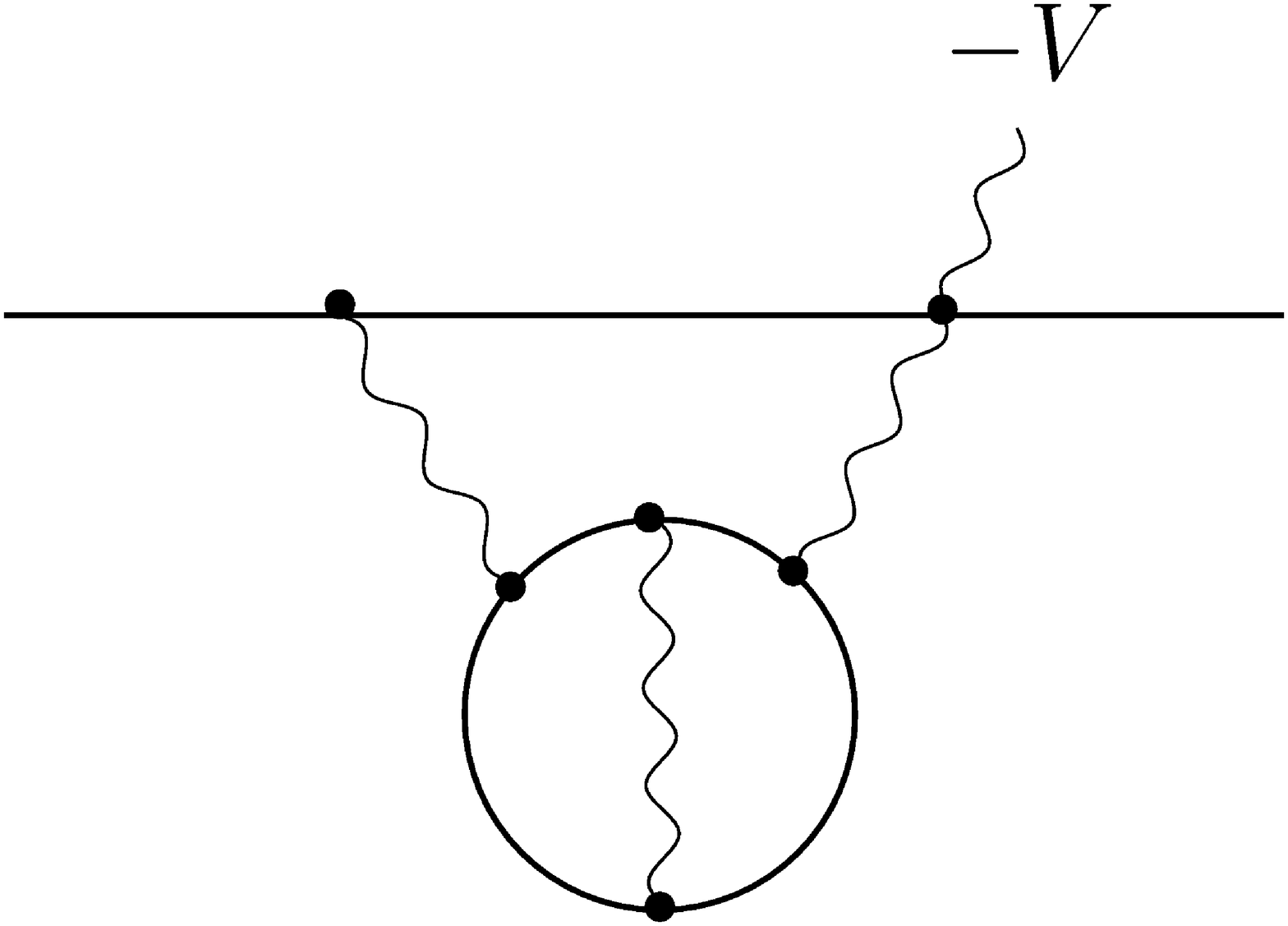}}
\settowidth{\la}{\usebox{\boxa}}
\begin{eqnarray*}
 \parbox{\la}{\usebox{\boxa}}
\end{eqnarray*}
up to terms with covariant derivatives acting on external superfileds.

When we insert two external auxiliary fields in (\ref{eq1101}), we fix
the first external auxiliary field and consider all insertion of the second external auxiliary
field. Then we find that only one diagram remains, in which the second
external field is inserted in the external line of $\tilde{\Phi}$. We
now consider all insertion of the first external auxiliary field and
again find that only one diagram remains. The remaining diagram is as
follows:
\sbox{\boxa}{\includegraphics[width=4cm]{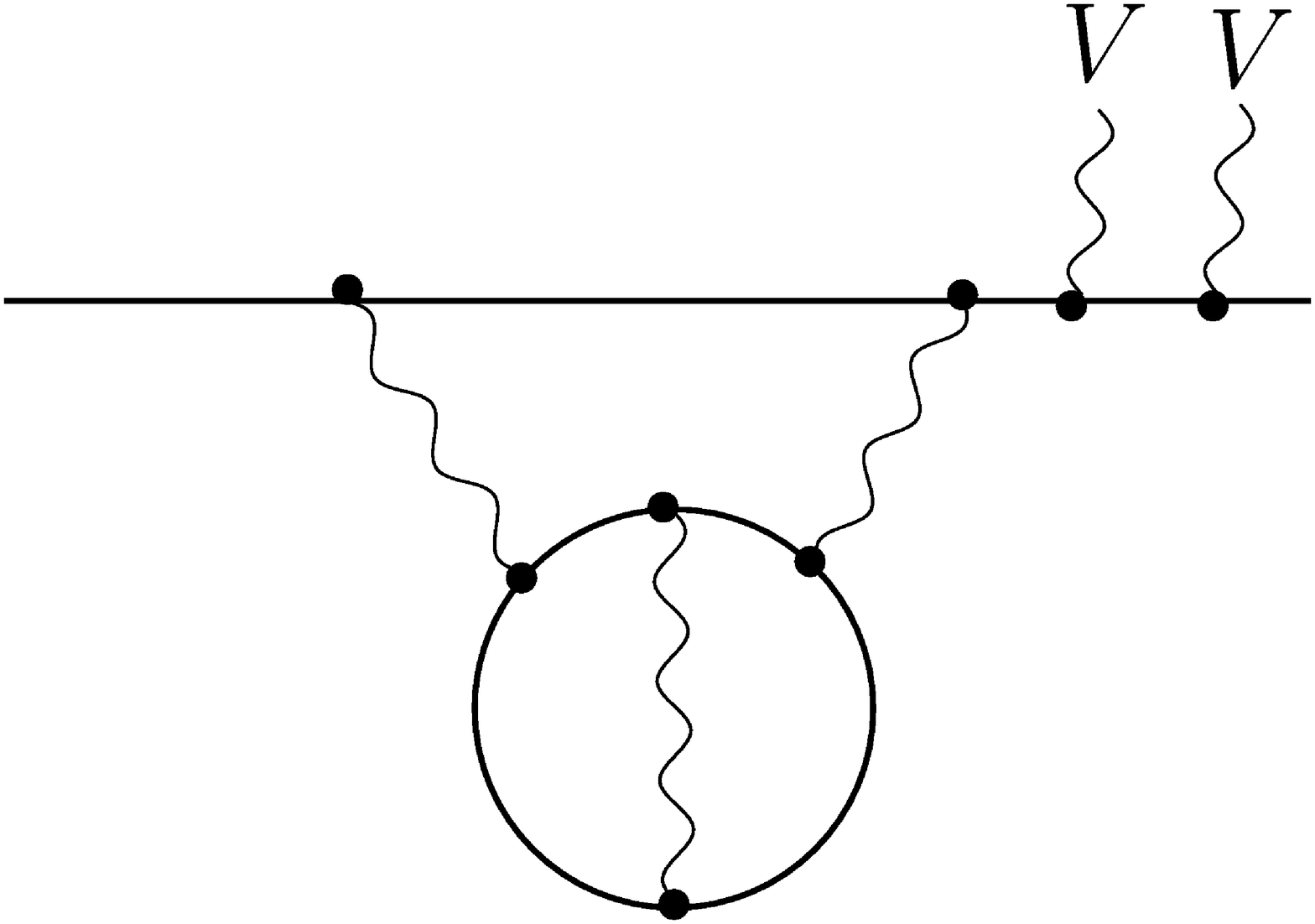}}
\settowidth{\la}{\usebox{\boxa}}
\begin{eqnarray*}
 \parbox{\la}{\usebox{\boxa}}.
\end{eqnarray*}
Notice here that if there is at
least one external auxiliary field inserted in the internal chiral loop,
we can move it along the chiral loop and obtain a vanishing
result. Performing partial integration, we find the above remaining
diagram is equivalent to
\sbox{\boxa}{\includegraphics[width=4cm]{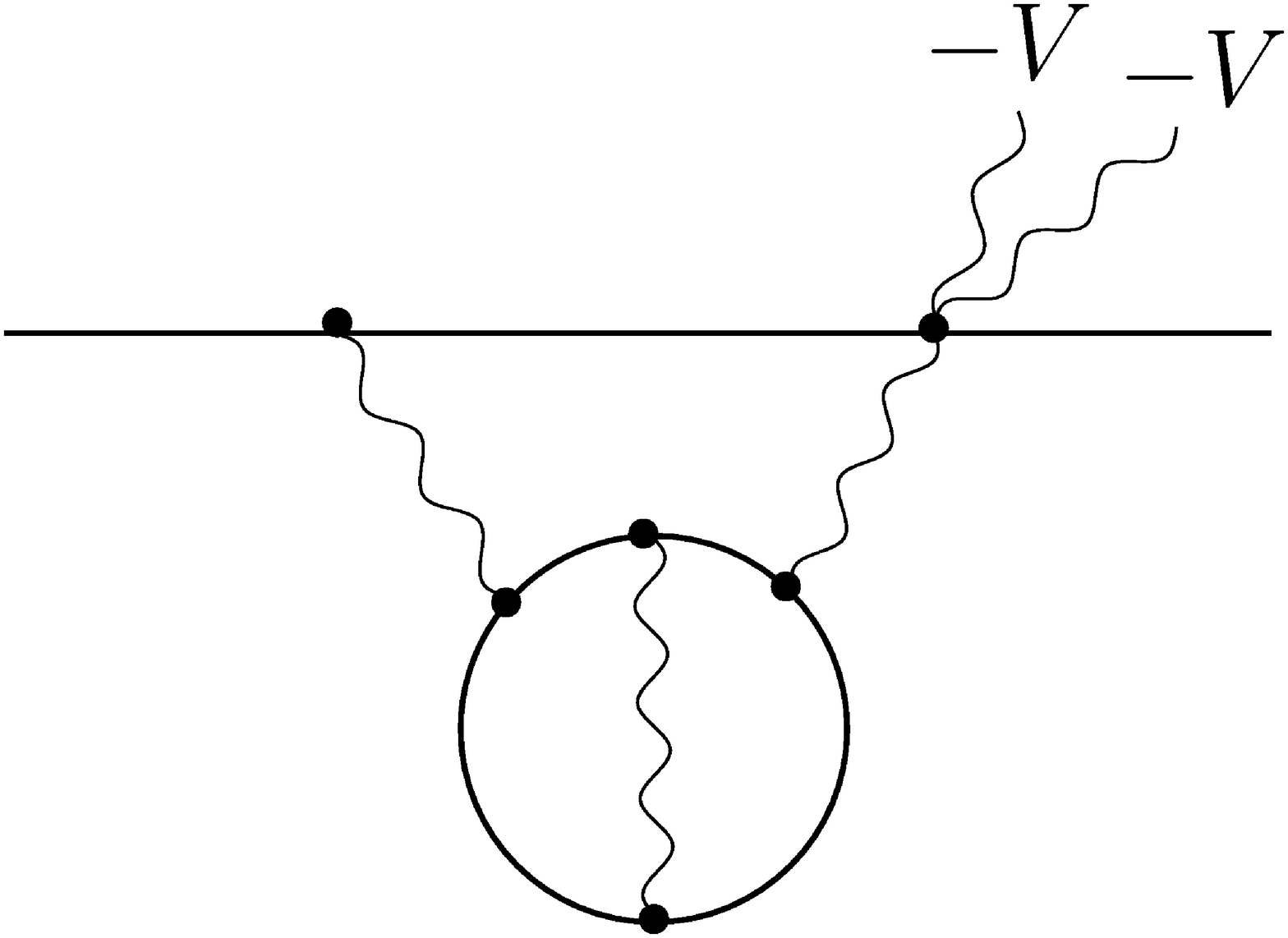}}
\settowidth{\la}{\usebox{\boxa}}
\begin{eqnarray*}
 \parbox{\la}{\usebox{\boxa}}
\end{eqnarray*}
up to terms with covariant derivatives acting on external superfields.

In the same way, when we insert $n$ external auxiliary fields in the diagram
(\ref{eq1101}), considering all insertion, all diagrams cancel
each other except for one diagram:
\sbox{\boxa}{\includegraphics[width=4cm]{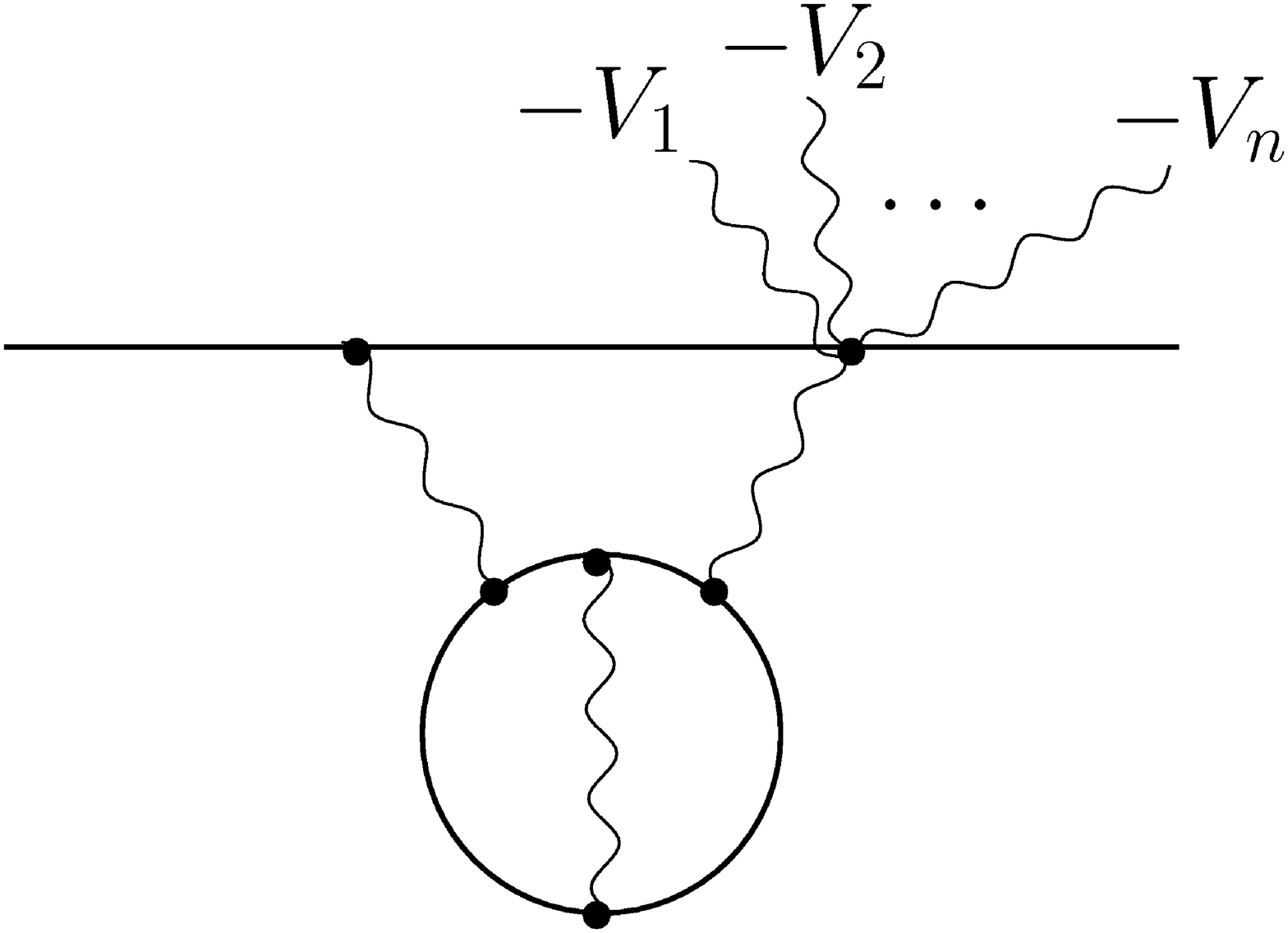}}
\settowidth{\la}{\usebox{\boxa}}
\begin{eqnarray}
 \parbox{\la}{\usebox{\boxa}}. \label{eq1102}
\end{eqnarray}
In order to obtain an amplitude from this diagram, we have to perform all integration over
grassmann coordinates of remaining vertices as well as internal
momenta. In doing so, we neglect terms with covariant derivatives acting
on external superfields because such terms have no divergence and we are
only interested in divergent terms. Particularly, we neglect terms with
$D_\alpha, \bar{D}_\alpha$ acting on external auxiliary
fields. Therefore, in order to study divergent terms,
we can rewrite (\ref{eq1102}) as
\sbox{\boxb}{\includegraphics[width=3cm]{47.eps}}
\settowidth{\lb}{\usebox{\boxb}}
\begin{eqnarray*}
\parbox{\la}{\usebox{\boxa}} \qquad \sim \qquad \left(-V_1\right)\left(-V_2\right)\cdots \left(-V_n\right) \times\left[ \parbox{\lb}{\usebox{\boxb}} \right]
\end{eqnarray*}
where ``$\sim$'' means both sides are equivalent up to terms with covariant
derivatives acting on external superfields. In the right-hand side, $n$ external auxiliary fields are just multiplied by the diagram
(\ref{eq1101}) which has no external auxiliary fields.

In general, any one-particle irreducible diagram with $n$ external
auxiliary fields and 
one pair of external $\tilde{\Phi},\tilde{\Phi}^\dagger,$ can be
obtained by insertion of $n$ external auxiliary fields in a diagram of the form
\sbox{\boxa}{\includegraphics[width=3cm]{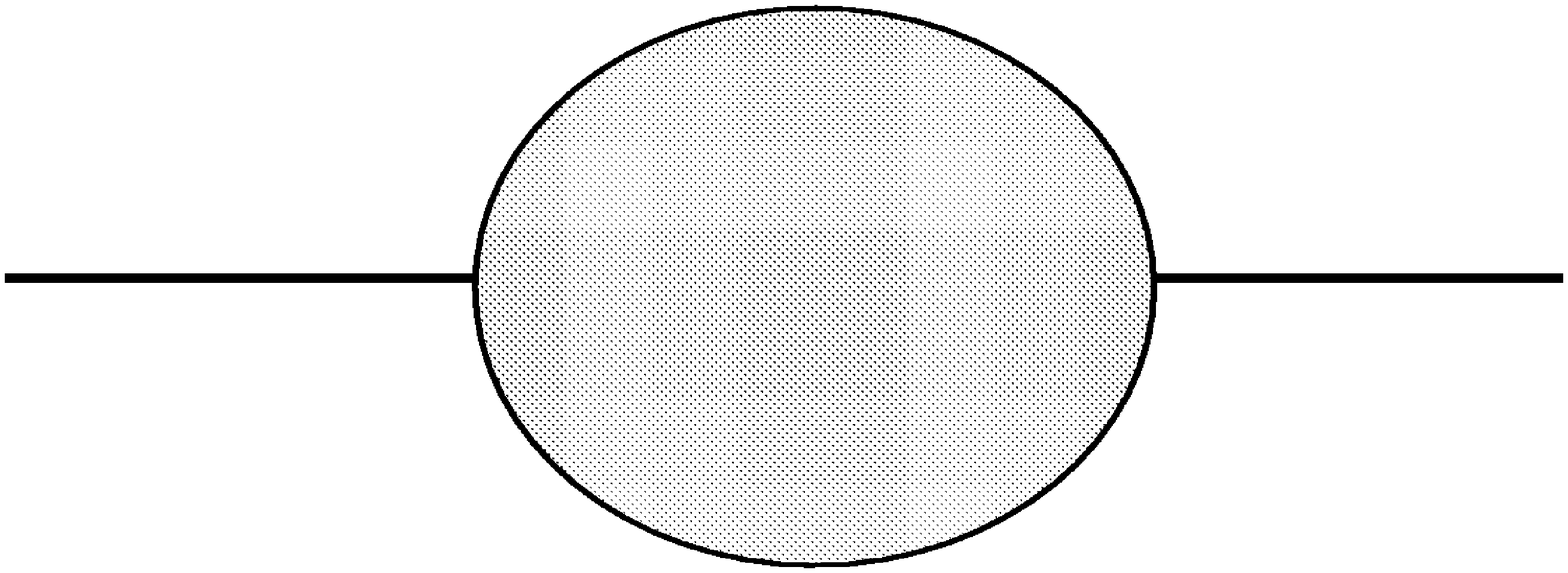}}
\settowidth{\la}{\usebox{\boxa}}
\begin{eqnarray}
\left[\parbox{\la}{\usebox{\boxa}}\right]_{\rm 1PI}. \label{eq1104}
\end{eqnarray}
If we choose one diagram of the above form and consider all insertion of
$n$ external auxiliary fields in it, the result is equivalent to
multiplication of the chosen diagram by $V_1V_2\cdots V_n$, neglecting terms
with covariant derivatives acting on external auxiliary fields. Namely,
\sbox{\boxb}{\includegraphics[width=4cm]{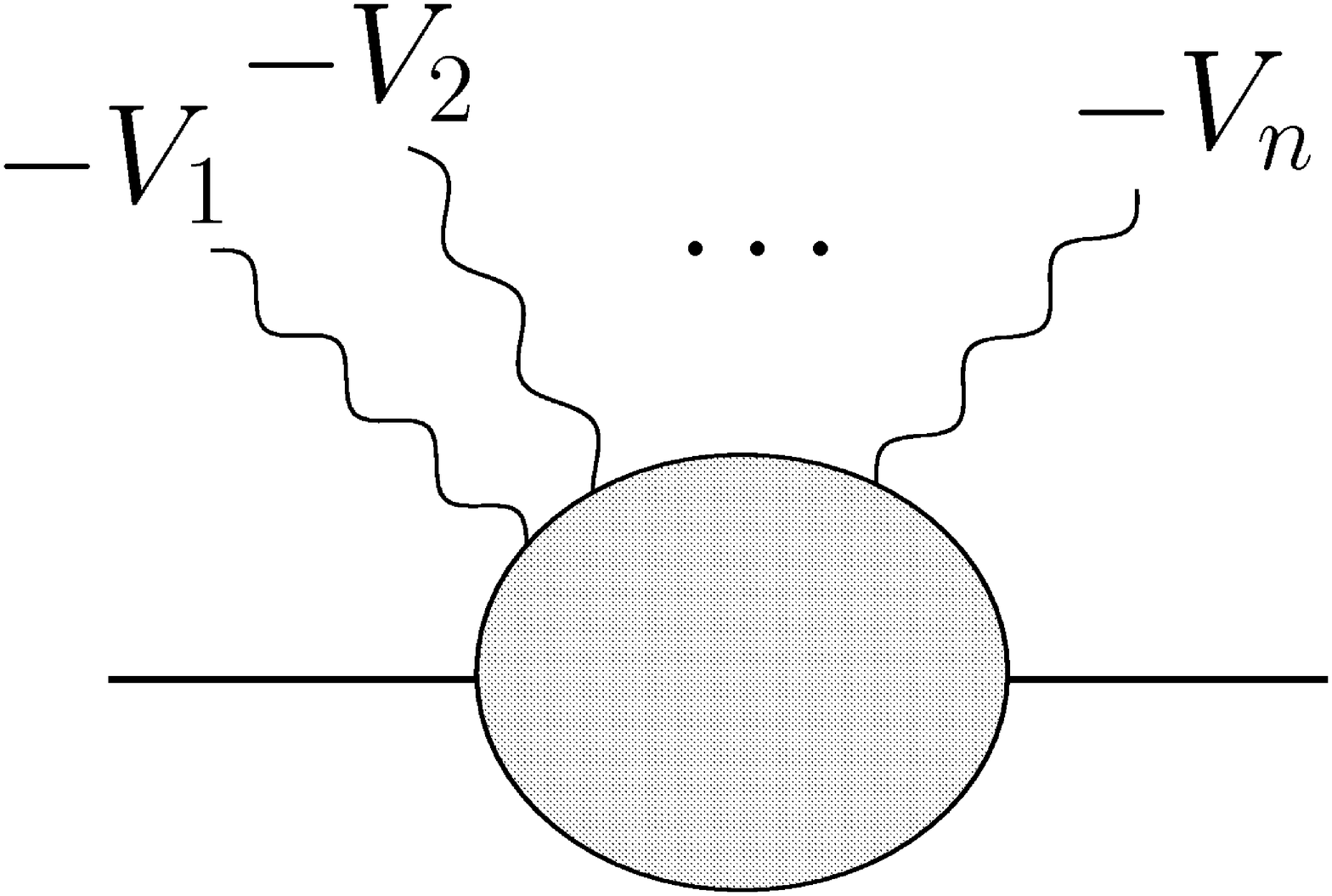}}
\settowidth{\lb}{\usebox{\boxb}}
\begin{eqnarray}
\left[\parbox{\lb}{\usebox{\boxb}}\right]_{\rm 1PI} \qquad \sim \qquad
 \left(-V_1\right)\left(-V_2\right)\cdots \left(-V_n\right) \times \left[\parbox{\la}{\usebox{\boxa}}\right]_{\rm
 1PI}. \label{eq1103}
\end{eqnarray}
We now find that all divergences included in the left-hand side of
(\ref{eq1103}) can be eliminated by a
renormalization of the wavefunction of $\tilde{\Phi}$. Indeed, the
equation (\ref{eq1103}) implies that all logarithmic divergences included in diagrams
of the form
\begin{eqnarray*}
\left[\parbox{\lb}{\usebox{\boxb}}\right]_{\rm 1PI}
\end{eqnarray*}
are canceled by a counter term of the form
\begin{eqnarray*}
 \tilde{\Phi}^{j\dagger}e^{-V}\tilde{\Phi}^j = \sum_{n=0}^\infty\frac{1}{n!}\tilde{\Phi}^{j\dagger}\left(-V\right)^n\tilde{\Phi}^j,
\end{eqnarray*} 
namely by a renormalization of the wave function of the dynamical field.

Note that the
equation (\ref{eq1103})
is satisfied for any internal diagram in shaded
circle. Therefore, it is also satisfied at each order of $1/N$. Then
divergences are eliminated at each order of $1/N$ by the renormalization.

\vspace{5ex}

\subsection{Beta function of the coupling constant}

We have shown that all divergences in the $1/N$-expansion can be
eliminated by the renormalizations of the coupling constant and the
wavefunction of the twisted chiral superfield. In this subsection, we
evaluate the beta function of the coupling constant $g_R$.

In section 2, we defined $g_R$ by
\begin{eqnarray}
 \frac{\mu}{g_R^2} := \frac{1}{g^2} -
  \frac{\Lambda}{2\pi^2}+\frac{\mu}{2\pi^2}, \label{eq1106}
\end{eqnarray}
so that the linear divergence from one-loop
diagram of the dynamical field (\ref{eq1105}) is eliminated. In the
above definition, $\mu$ is a renormalization scale and $\Lambda$ is a
momentum cut-off.
In subsection 5.2, we showed that there is no more divergence from
diagrams of the form
\sbox{\boxa}{\includegraphics[width=3cm]{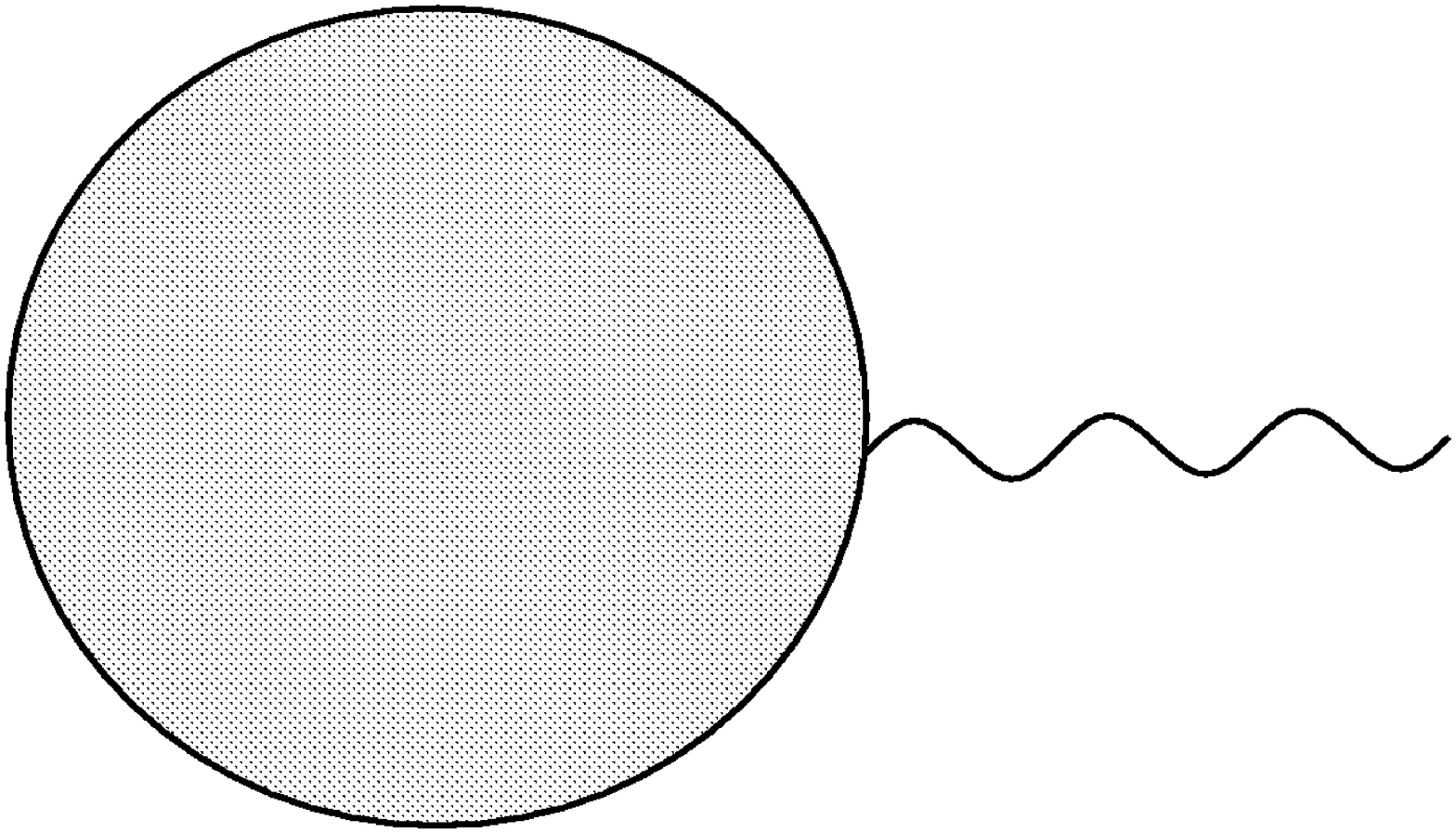}}
\settowidth{\la}{\usebox{\boxa}}
\begin{eqnarray*}
\left[\parbox{\la}{\usebox{\boxa}}\right]_{\rm 1PI}.
\end{eqnarray*}
Therefore we need no more renormalization of the coupling constant.
Then we can treat $g_R$ defined by
(\ref{eq1106}) as a renormalized coupling constant correct in all order of $1/N$-expansion.

\begin{figure}[h]
\begin{center}
\includegraphics[width=7cm]{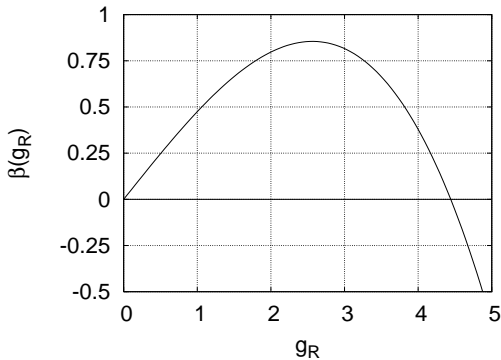}
\caption{Beta function of the coupling constant} \label{fig2}
\end{center}
\end{figure}

We now evaluate the beta function of $g_R$. Defferentiating both sides
of (\ref{eq1106}) by $\mu$, we obtain
\begin{eqnarray*}
 \frac{1}{g_R^2} - \frac{2\mu}{g_R^3}\cdot\frac{dg_{R}}{d\mu} = \frac{1}{2\pi^2},
\end{eqnarray*}
where we should note that $g$ and $\Lambda$ is independent of $\mu$ but
$g_R$ depends on $\mu$. If we define $\beta\left(g_R\right) :=
\mu\frac{dg_R}{d\mu}$, we find
\begin{eqnarray*}
 \beta\left(g_R\right) = \frac{1}{2}g_R - \frac{1}{4\pi^2}g_R^3.
\end{eqnarray*}
This beta function is shown in Figure \ref{fig2}. This vanishes when
$g_R = 0, \sqrt{2}\pi$. We find that this
theory has one ultraviolet fixed point at $g_R=\sqrt{2}\pi$. 

\vspace{5ex}

\section{Conclusions}

In this paper, we have studied a three dimensional $CP^{N-1}$ model in the method
of $1/N$-expansion. This model has ${\mathcal N}=2$
supersymmetry, $U(1)$ gauge symmetry, and global $SU(N)$ symmetry.
For the $1/N$-expansion, it is useful to use the Lagrangian with the
auxiliary field $V$. Using the super Feynman rules, we have derived the
superpropagator of auxiliary field induced by quantum effects of dynamical
field. Then we have proved that all divergences in amplitudes
can be eliminated in each order of $1/N$ by renormalizations of the coupling constant and the
wave function of the dynamical field. We have also shown that there is no
contribution to the beta function except in the leading order of $1/N$.
This model have been shown to have a non-trivial ultraviolet fixed point. 
These arguments are valid in all orders of $1/N$-expansion.

\vspace{5ex}

\section*{Acknowledgements}

We are grateful to Professor E. R. Nissimov and S. J. Pacheva for
calling our attention to \cite{Prof.Nissimov, Prof.Nissimov2}. This work was supported in part by Grants-in-Aid for
Scientific Research ($\sharp$16340075)

\newpage

{\Large \noindent {\bf Appendix A \hspace{3mm} ${\mathcal N}=2$ SUSY in Three
Dimensions}}
\vspace{3ex}

The smallest supersymmetry algebra in three dimensions has
one Majorana (real) spinor of supercharges.
It has two real degrees of freedom.
So ${\mathcal N}=2$ supersymmetry in three dimensions has 
one Dirac (complex) spinor of supercharges. 
It has four real degrees of freedom. Therefore the dimensional reduction of the
${\mathcal N}=1$ supersymmetry in four dimensions gives the ${\mathcal
N}=2$
supersymmetry in three dimensions.

The superspace has coordinates $x^\mu$, $\theta^\alpha$ and 
$\bar{\theta}^{\alpha}$ where $\mu = 0,1,2$ and
$\alpha=1,2$. Here $\theta^\alpha$ is a two-component Dirac spinor
and $\bar{\theta}^\alpha$ is the complex conjugate of
$\theta^\alpha$. 
\vspace{5ex}

\noindent {\bf A-1 \hspace{1mm} Gamma Matrices and Dirac Spinor}

\vspace{1ex}

We use the metric $\eta_{\mu\nu}= {\rm diag}(+,-,-)$ and gamma matrices
\begin{eqnarray*}
 \gamma^0 = \left[
	     \begin{array}{cc}
	      0 & -i\\
	      i & 0 \\
	     \end{array}
	    \right] \quad,\quad \gamma^1 = \left[
	    \begin{array}{cc}
	     0 & i\\
	     i & 0\\
	    \end{array}
            \right] \quad,\quad \gamma^2 = \left[
	    \begin{array}{cc}
	     i & 0\\
	     0 & -i\\
	    \end{array}
	    \right] .
\end{eqnarray*}
These matrices satisfy the anticommutation relations
$\left\{\gamma^\mu,\gamma^\nu\right\} = 2\eta^{\mu\nu}$ and the
identity
\begin{eqnarray*}
 \gamma^\mu\gamma^\nu = \eta^{\mu\nu} +
  i\epsilon^{\mu\nu\rho}\gamma_\rho
\end{eqnarray*}
where $\epsilon^{\mu\nu\rho}$ is a totally antisymmetric tensor 
so that $\epsilon^{012} = +1$.

Spinors with upper and lower indices are related through the
antisymmetric tensor $C$:
\begin{eqnarray*}
 C^{\alpha\beta} = C_{\alpha\beta} = \left[
			   \begin{array}{cc}
			    0 & -i\\
			    i & 0\\
			   \end{array}
       \right]_{\alpha\beta} \quad &,& \quad
 C^{\alpha\beta}C_{\beta\gamma} = \delta^{\alpha}_{\gamma} \\[2mm]
 \psi_{\alpha} = C_{\alpha\beta}\psi^\beta \quad , \quad \psi^{\alpha}
  &=& C^{\alpha\beta}\psi_\beta
\end{eqnarray*}
We use the following summation convention:
\begin{eqnarray*}
 \psi\chi &:=& \psi_{\alpha}\chi^{\alpha} = -\chi^\alpha\psi_\alpha
= \chi_{\alpha}\psi^\alpha = \chi\psi \\
 \bar{\psi}\bar{\chi} &:=& \bar{\psi}_\alpha \bar{\chi}^{\alpha}
= -\bar{\chi}^{\alpha}\bar{\psi}_\alpha =
\bar{\chi}_\alpha\bar{\psi}^\alpha
= \bar{\chi}\bar{\psi}\\
 \bar{\psi}\chi &:=& \bar{\psi}_\alpha\chi^\alpha =
  -\chi^\alpha\bar{\psi}_\alpha
= \chi_\alpha\bar{\psi}^\alpha = \chi\bar{\psi}
\end{eqnarray*}
The gamma matrices have the following index structure:
\begin{eqnarray*}
&\left(\gamma^\mu\right)^{\alpha}_{\hspace{.3em}\beta}&\\[2mm]
&\bar{\psi}\gamma^\mu\chi := \bar{\psi}_{\alpha}
  \left(\gamma^\mu\right)^{\alpha}_{\hspace{.3em}\beta}\chi^{\beta}&
\end{eqnarray*}
This $\gamma^\mu$ satisfies the identity
\begin{eqnarray*}
 \left(\gamma^\mu\right)_\alpha^{\hspace{.3em}\beta} :=
  C_{\alpha\gamma}C^{\beta\delta}\left(\gamma^\mu\right)^\gamma_{\hspace{.3em}\delta}
  =
  \left[\gamma^0\gamma^\mu\left(\gamma^0\right)^T\right]_\alpha^{\hspace{.32em}\beta}
  =
  \left[-\gamma^0\gamma^\mu\gamma^0\right]_\alpha^{\hspace{.32em}\beta}
  = \left[\left(\gamma^\mu\right)^T\right]_\alpha^{\hspace{.32em}\beta}
  = \left(\gamma^\mu\right)^\beta_{\hspace{.3em}\alpha}
\end{eqnarray*}
so we find
\begin{eqnarray*}
 \bar{\psi}\gamma^\mu\chi =
  \bar{\psi}_{\alpha}\left(\gamma^\mu\right)^\alpha_{\hspace{.3em}\beta}
  \chi^\beta =
  -\chi^\beta\left(\gamma^\mu\right)^\alpha_{\hspace{.3em}\beta}\bar{\psi}_{\alpha}
  =
  -\chi^\beta\left(\gamma^\mu\right)_{\beta}^{\hspace{.32em}\alpha}\bar{\psi}_{\alpha}
  =
  -\chi_\beta\left(\gamma^\mu\right)^{\beta}_{\hspace{.32em}\alpha}\bar{\psi}^{\alpha}
  = - \chi \gamma^\mu \bar{\psi}.
\end{eqnarray*}
Notice that $\bar{\psi}_\alpha := C_{\alpha\beta}\bar{\psi}^\beta \neq
\overline{C_{\alpha\beta}\psi^\beta}$ because $C_{\alpha\beta}^* =
 - C_{\alpha\beta} = C_{\beta\alpha}$. And we find the identities:
\begin{eqnarray*}
 \left(\psi\chi\right)^\dagger &=&
  \left(C_{\alpha\beta}\psi^\beta\chi^\alpha\right)^\dagger
 = \left(C_{\alpha\beta}\right)^*\bar{\chi}^\alpha\bar{\psi}^\beta =
 C_{\beta\alpha}\bar{\chi}^{\alpha}\bar{\psi}^\beta =
 \bar{\chi}\bar{\psi} = \bar{\psi}\bar{\chi}\\
 \left(\bar{\psi}\chi\right)^\dagger &=&
  \left(C_{\alpha\beta}\bar{\psi}^\beta\chi^\alpha\right)^\dagger =
  \left(C_{\alpha\beta}\right)^*\bar{\chi}^\alpha\psi^\beta =
  C_{\beta\alpha}\bar{\chi}^\alpha\psi^\beta  =\bar{\chi}\psi =
  \psi\bar{\chi}\\
 \left(\bar{\psi}\gamma^\mu\chi\right)^\dagger &=&
  \psi_\alpha\left(\bar{\gamma^\mu\chi}\right)^\alpha = \psi_\alpha
  \left(\gamma^{\mu*}\bar{\chi}\right)^\alpha = - \psi \gamma^\mu
  \bar{\chi} = \bar{\chi}\gamma^\mu \psi.
\end{eqnarray*}

\vspace{5mm}

\noindent {\bf A-2 \hspace{1mm} SUSY Algebra and Covariant Derivatives}

\vspace{1ex}

A supersymmetry transformation in the superspace
\begin{eqnarray*}
 x^\mu \to x' = x^\mu + \frac{i}{2}\left(\bar{\xi}\gamma^\mu\theta -
			       \bar{\theta}\gamma^\mu\xi\right) \quad,
 \quad \theta^\alpha \to \theta'^\alpha = \theta^\alpha + \xi^\alpha \quad , \quad
 \bar{\theta}^\alpha \to \bar{\theta}'^\alpha = \bar{\theta} + \bar{\xi}^\alpha
\end{eqnarray*}
is generated by the differential operators
\begin{eqnarray*}
 Q^{\alpha} := i\left[\frac{\partial}{\partial\theta_\alpha}
+ \frac{i}{2}\left(\Slash{\partial}\bar{\theta}\right)^\alpha\right]
 \quad , \quad
 \bar{Q}^{\alpha} := -i\left[\frac{\partial}{\partial\bar{\theta}_\alpha}
 + \frac{i}{2}\left(\Slash{\partial}\theta\right)^\alpha\right],
\end{eqnarray*}
namely
\begin{eqnarray*}
 e^{i\left(\xi Q
      -\bar{\xi}\bar{Q}\right)}F\left(x,\theta,\bar{\theta}\right)
     =  F\left(x',\theta',\bar{\theta'}\right).
\end{eqnarray*}
Supercharges $Q^\alpha$ and $\bar{Q}^\alpha$ satisfy the following
anticommutation relations
\begin{eqnarray*}
 \left\{Q^\alpha,\bar{Q}_{\beta}\right\} &=&
  -i\Slash{\partial}^{\alpha}_{\hspace{.25em}\beta}\\
  \left\{Q^{\alpha},Q_\beta\right\} &=&
  \left\{\bar{Q}^\alpha,\bar{Q}_\beta\right\} = 0.
\end{eqnarray*}
We define the covariant derivatives:
\begin{eqnarray*}
 D_\alpha := -\frac{\partial}{\partial \theta^\alpha} +
  \frac{i}{2}\left(\bar{\theta}\Slash{\partial}\right)_\alpha \quad ,
  \quad \bar{D}_\alpha := - \frac{\partial}{\partial\bar{\theta}^\alpha}
  + \frac{i}{2}\left(\theta\Slash{\partial}\right)_\alpha,
\end{eqnarray*}
and we find
\begin{eqnarray*}
 \left\{D^\alpha, \bar{D}_\beta\right\} &=&
  i\Slash{\partial}^{\alpha}_{\hspace{.25em}\beta}\\
  \left\{D^\alpha,D_\beta\right\} &=& \left\{\bar{D}^\alpha,
   \bar{D}_\beta\right\} = 0.
\end{eqnarray*}
With these definitions, supercharges $Q^\alpha, \bar{Q}^\alpha$ and
covariant derivatives $D^\alpha, \bar{D}^\alpha$ anticommute.

\vspace{5ex}

\noindent {\bf A-3 \hspace{1mm} Chiral and Vector Superfield}

\vspace{5ex}

Since $\bar{D}_\alpha$ and $Q^\alpha, \bar{Q}^\alpha$
anticommute, the ``chirality'' constraint
\begin{eqnarray*}
 \bar{D}_\alpha \Phi\left(x,\theta,\bar{\theta}\right) = 0
\end{eqnarray*}
is consistent with supersymmety transformations. The expressions for
$D_\alpha$ and $\bar{D}_\alpha$ in terms of $y^\mu:=x^\mu +
\frac{i}{2}\bar{\theta}\gamma^\mu\theta, \hspace{.125em}\theta^\alpha, \bar{\theta}^\alpha$ are
\begin{eqnarray*}
 D_\alpha = - \frac{\partial}{\partial \theta^\alpha} +
  i\left(\bar{\theta}\gamma^\mu\right)_\alpha\frac{\partial}{\partial
  y^\mu}
\quad , \quad
\bar{D}_\alpha = - \frac{\partial}{\partial\bar{\theta}^\alpha}.
\end{eqnarray*}
We can therefore expand $\Phi$ in powers of $\theta$:
\begin{eqnarray*}
 \Phi\!\left(x,\theta,\bar{\theta}\right) &=& \phi\!\left(y\right) + \theta\psi\!\left(y\right) +
  \frac{1}{2}\theta^2 F\!\left(y\right)\\
 &=& \phi\!\left(x\right) + \theta\psi\!\left(x\right)
  + \frac{1}{2}\theta^2 F\!\left(x\right)
  + \frac{i}{2}\left(\bar{\theta}\Slash{\partial}\theta\right)\!\phi\!\left(x\right)
  -
  \frac{i}{4}\theta^2\!\!\left[\hspace{.125em}\bar{\theta}\Slash{\partial}\psi\!\left(x\right)\right]
 - \frac{1}{16}\theta^2\bar{\theta}^2\partial^2\!\phi\!\left(x\right)
\end{eqnarray*}
The superfield $\Phi^\dagger$ satisfies the constraint
$D_\alpha\Phi^\dagger = 0$.

Note that there are no
chiral spinors in three dimension. Although we call $\Phi$ ``chiral''
superfield, $\psi$ is a Dirac spinor.

A vector superfield $V$ satisfies the constraint
\begin{eqnarray*}
 V^\dagger = V
\end{eqnarray*}
and has the expansion
\begin{eqnarray*}
 V\!\left(x,\theta,\bar{\theta}\right) &=& C\!\left(x\right) + \left[\theta\eta\!\left(x\right) + \bar{\theta}\bar{\eta}\!\left(x\right)\right] +
  \frac{1}{2}\left[\theta^2 \! f\!\left(x\right) +
  \bar{\theta}^2 \! f^*\!\left(x\right) \!\right] +
  \hspace{.125em}\bar{\theta}\Slash{v}\!\left(x\right)\theta\hspace{.125em} 
+ \hspace{.125em}M\!\left(x\right)\bar{\theta}\theta \\
&& \hspace{1.25em}+
  \frac{1}{2}\theta^2\hspace{.1em}\bar{\theta}\!\left[\lambda\!\left(x\right)
						    - i\Slash{\partial}\psi\right] +
	      \frac{1}{2}\bar{\theta}^2\theta\!\left[\bar{\lambda}\!\left(x\right) +
					 i\Slash{\partial}\bar{\psi}\right] +
  \frac{1}{4}\theta^2\bar{\theta}^2 \! \left[D\!\left(x\right) + \frac{1}{4}\partial^2C\!\left(x\right)\right],
\end{eqnarray*}
where $C, v^\mu, M$ and $ D$ are real.

\vspace{10ex}

{\Large\noindent {\bf Appendix B \hspace{3mm} Calculation of the
Effective Potential}}
\vspace{3ex}

We here show the last equality in (\ref{eq15000}). Namely,

\vspace{2ex}
{\bf \noindent Proposition B-1}
\begin{eqnarray}
&&-i\int^\Lambda \!\!\frac{d^3k}{\left(2\pi\right)^3}{\rm
  ln}\left(-k^2+M_c^2+D_c^2\right) +
  i\int^\Lambda\!\!\frac{d^3k}{\left(2\pi\right)^3}\hspace{.125em}{\rm tr}\hspace{.125em}{\rm
  ln}\left(\Slash{k}-M_c\right)\nonumber\\[1mm]
&& \hspace{12.5em}=\hspace{.25em} -\frac{1}{6\pi}\left|M_c^2 +
      D_c\right|^{\frac{3}{2}} + \frac{1}{6\pi}\left|M_c\right|^3 + \frac{\Lambda}{2\pi^2}D_c\hspace{1em}\label{eq35001}
\end{eqnarray}
\vspace{3ex}

In order to show this, we first note that
\begin{eqnarray*}
  \int^\Lambda\!\!\frac{d^3k}{\left(2\pi\right)^3}\hspace{.125em}{\rm tr}\hspace{.125em}{\rm
  ln}\left(\Slash{k}-M_c\right) &=&  \frac{1}{2}\left[\int^\Lambda\!\!\frac{d^3k}{\left(2\pi\right)^3}\hspace{.125em}{\rm tr}\hspace{.125em}{\rm
  ln}\left(\Slash{k}-M_c\right) +  \int^\Lambda\!\!\frac{d^3k}{\left(2\pi\right)^3}\hspace{.125em}{\rm tr}\hspace{.125em}{\rm
  ln}\left(-\Slash{k}-M_c\right)\right]\\
&=& \frac{1}{2}\int^\Lambda\!\!\frac{d^3k}{\left(2\pi\right)^3}\hspace{.125em}{\rm tr}\hspace{.125em}{\rm
  ln}\left\{\left(\Slash{k}-M_c\right)\left(-\Slash{k}-M_c\right)\right\}\\
&=&
 \frac{1}{2}\int^\Lambda\!\!\frac{d^3k}{\left(2\pi\right)^3}\hspace{.125em}{\rm tr}\hspace{.125em}{\rm ln}\left(-k^2+M_c^2\right)\\
 &=& \int^\Lambda\!\!\frac{d^3k}{\left(2\pi\right)^3}\hspace{.125em}{\rm ln}\left(-k^2+M_c^2\right).
\end{eqnarray*}
Then we perform the Wick rotation in the left-hand side:
\begin{eqnarray*}
 ({\rm LHS}) =  \int^\Lambda \!\frac{d^3k_E}{\left(2\pi\right)^3}{\rm
  ln}\left(k^2_E+M_c^2+D_c^2\right) -
\int^\Lambda\!\!\frac{d^3k_E}{\left(2\pi\right)^3}\hspace{.125em}{\rm ln}\left(k^2_E+M_c^2\right).
\end{eqnarray*}
We now combine two integrals as follows:
\begin{eqnarray}
\int_{M_c^2}^{M_c^2+D_c}\!\!dm^2\int^\Lambda\!\!\frac{d^3k_E}{\left(2\pi\right)^3}\hspace{.125em}\frac{1}{k_E^2+m^2}
 &=&
 \int_{M_c^2}^{M_c^2+D_c}\!\!dm^2\frac{1}{\left(2\pi\right)^3}\int^\Lambda_0\!\!dK\hspace{.125em}\frac{4\pi
 K^2}{K^2+m^2} \nonumber \\[1mm]
 &=&
  \int_{M_c^2}^{M_c^2+D_c}\!\!dm^2\frac{1}{2\pi^2}\int^\Lambda_0\!\!dK\hspace{.125em}\left(1 - \frac{m^2}{K^2+m^2}\right) \nonumber \\[1mm]
&=&   \int_{M_c^2}^{M_c^2+D_c}\!\!dm^2\frac{1}{2\pi^2}\left(\Lambda -
						       \int^\Lambda_0\!\!dK\hspace{.125em}\frac{m^2}{K^2+m^2}\right) \nonumber \\[1mm]
&=& \frac{\Lambda}{2\pi^2}D_c -
\frac{1}{2\pi^2}\int^{M_c^2+D_c}_{M_c^2}\!\!dm^2\int_0^\Lambda\!\!dK\frac{m^2}{K^2+m^2}. \label{eq35000}
\end{eqnarray}
Notice that the first term is linearly divergent while the second term
has no divergence. Therefore we take the limit $\Lambda\to\infty$ at the
second term:
\begin{eqnarray*}
 \int_0^\Lambda\!\!dK\frac{m^2}{K^2+m^2} \longrightarrow
  \int_0^\infty\!\!dK\frac{m^2}{K^2+m^2} &=&
  \frac{m^2}{2}\int_{-\infty}^{\infty}\frac{1}{K^2+m^2}\\
 &=& \frac{m^2}{2}\cdot\frac{\pi}{\left|m\right|}.
\end{eqnarray*}
In the last equality, we perform a contour integral. Then we can
prove the statement as follows:
\begin{eqnarray*}
 ({\rm LHS}) &=& \frac{\Lambda}{2\pi^2}D_c -
  \frac{1}{4\pi}\int_{M_c^2}^{M_c^2+D_c}\!\!dm^2\hspace{.25em}\left|m\right|\\[1mm]
&=& \frac{\Lambda}{2\pi^2}D_c -
\frac{1}{4\pi}\int_{M_c^2}^{M_c^2+D_c}\!\!dm^2\hspace{.25em}\sqrt{m^2}\\[1mm]
&=& \frac{\Lambda}{2\pi^2}D_c -
\frac{1}{6\pi}\left(\left|M_c^2+D_c\right|^{\frac{3}{2}}-
	       \left|M_c\right|^3\right)\\[5mm]
&=& ({\rm RHS}).
\end{eqnarray*}

We can also evaluate the left-hand side without the Wick rotation. Namely,
\begin{eqnarray*}
({\rm LHS}) &=& -i\int^\Lambda
  \!\!\frac{d^3k}{\left(2\pi\right)^3}\ln\left(-k^2+M_c^2+D_c\right) +
  i\int^\Lambda\!\!\frac{d^3k}{\left(2\pi\right)^3}\ln\left(-k^2+M_c^2\right)\\
&=& \int^{M_c^2+D_c}_{M_c^2}\!\!dm^2\int^\Lambda\!\!\frac{d^3k}{\left(2\pi\right)^3}\frac{i}{k^2-m^2}.
\end{eqnarray*}
If we take the limit  $\Lambda\to \infty$ and perform the contour
integral over $k^0$, then we obtain
\begin{eqnarray*}
({\rm LHS}) = \int^{M_c^2+D_c}_{M_c^2}\!\!dm^2\int\!\!\frac{d^2k}{\left(2\pi\right)^2}\frac{1}{2\sqrt{\vec{k}^2+m^2}}
  &=&
  \int\!\!\frac{d^2k}{\left(2\pi\right)^2}\sqrt{\vec{k}^2+M_c^2+D_c}
					  - \int\!\!\frac{d^2k}{\left(2\pi\right)^2}\sqrt{\vec{k}^2+M_c^2}.
\end{eqnarray*}
The first term in the right-hand side is a zero-point energy of $\phi$,
and the second is that of $\psi$. These include a linear divergence
proportional to $D$.

\vspace{10ex}

{\Large \noindent {\bf Appendix C \hspace{3mm} Fierz transformations}}

\vspace{3ex}

In this appendix, we show some useful fomulae for spinor calculations. We
can first show
\begin{eqnarray}
 \left(\theta\xi\right)\left(\theta\chi\right) =
  -\frac{1}{2}\theta^2\left(\xi\chi\right) \label{eq41}
\end{eqnarray}
through a straightforward calculations. Substituting $\bar{\theta}$ for both
$\xi$ and $\chi$, we find
\begin{eqnarray*}
 \left(\theta\bar{\theta}\right)^2 = -\frac{1}{2}\theta^2\bar{\theta}^2
\end{eqnarray*}
On the other hand, if we substitute
$\gamma^\mu\bar{\theta}$ for $\xi$ and $\partial_\mu\!\psi$ for $\chi$
in (\ref{eq41}), we obtain the equation
\begin{eqnarray*}
\left(\theta\gamma^\mu\bar{\theta}\right)\left(\theta\partial_\mu\!\psi\right)
 =
 -\frac{1}{2}\theta^2\hspace{.125em}C_{\alpha\beta}\left(\gamma^\mu\bar{\theta}\right)^\beta\!\left(\partial_\mu\!\psi\right)^\alpha
 = \frac{1}{2}\theta^2\!\left(\bar{\theta}\Slash{\partial}\psi\right)
\end{eqnarray*}
Substituting $\gamma^\mu\bar{\theta}$ for $\xi$ and $\gamma^\nu\bar{\theta}$
for $\chi$ in the equation (\ref{eq41}), we can show
$\left(\theta\gamma^\mu\bar{\theta}\right)\left(\theta\gamma^\nu\bar{\theta}\right)
  =\frac{1}{2}\theta^2\!\left(\bar{\theta}\gamma^\mu\gamma^\nu\bar{\theta}\right)
$. Using the equation $\gamma^\mu\gamma^\nu = \eta^{\mu\nu} +
i\epsilon^{\mu\nu\rho}\gamma_\rho$ and the fact that
$\epsilon^{\nu\mu\rho} = - \epsilon^{\mu\nu\rho}$, we find
\begin{eqnarray*}
  \left(\theta\gamma^\mu\bar{\theta}\right)\left(\theta\gamma^\nu\bar{\theta}\right)
  = \frac{1}{2}\theta^2\!\bar{\theta}^2\eta^{\mu\nu}.
\end{eqnarray*}
 At last, if we substitute
$\left(i\Slash{\partial}+m\right)\bar{\theta}$ for both $\xi$ and
$\chi$ in (\ref{eq41}), we obtain
\begin{eqnarray*}
 \left[\theta\left(i\Slash{\partial} + m\right)\bar{\theta}\right]^2 &=&
  -\frac{1}{2}C_{\alpha\beta}\left[\left(i\Slash{\partial}+m\right)\bar{\theta}\right]^\beta\!\left[\left(i\Slash{\partial}+m\right)\bar{\theta}\right]^\alpha = -\frac{1}{2}\left[\bar{\theta}\left(-i\Slash{\partial} + m\right)\left(i\Slash{\partial}+m\right)\bar{\theta}\right]\\
 &=& \frac{1}{2}\bar{\theta}^2\left(-\partial^2-m^2\right).
\end{eqnarray*}
We use this equation in section 3.2 to construct the superpropagator of
the dynamical field.

In the following, we will show the Fierz transformation for general
two-by-two complex matrices.
In general, any two-by-two complex matrix $\Gamma$ can be expanded by the
following four matrices:
\begin{eqnarray*}
 \Gamma^\mu := \gamma^\mu \hspace{.5em} \left(\mu=0,1,2\right) \quad ,
  \quad \Gamma^3 := i{\bf 1},
\end{eqnarray*}
such that
$
 \Gamma =  c_A \Gamma^A
$
where the capital index $A$ runs over $0,1,2,3$.
Note the relation
\begin{eqnarray*}
 {\rm tr}\left[\Gamma_A\Gamma_B\right] = 2\eta_{AB}
\end{eqnarray*}
where $\eta_{AB}
:= {\rm diag}\!\left(+---\right)$ and $\Gamma_A := \eta_{AB}\Gamma^B$. Then we find $c_A = \frac{1}{2}{\rm
tr}\left[\Gamma_A\Gamma\right]$ and therefore
\begin{eqnarray}
 \Gamma = \frac{1}{2}{\rm tr}\left[\Gamma_A\Gamma\right]\Gamma^A. \label{eq40}
\end{eqnarray}
If we take an another two-by-two matrix $\Gamma'$,
the product of the matrix elements $\Gamma_{ab}\Gamma'_{cd}$ can be
treated as the $(a,b)$-element of a two-by-two matrix fixing
the indices $b,c$. Then we use the above relation (\ref{eq40}):
\begin{eqnarray*}
 \Gamma_{ab}\Gamma'_{cd} =
  \frac{1}{2}\sum_{e,f}\left[\left(\Gamma_A\right)_{ef}\Gamma_{fb}\Gamma'_{ce}\right]
  \left(\Gamma^A\right)_{ad} =
  \frac{1}{2}\left(\Gamma'\Gamma_A\Gamma\right)_{cb}\left(\Gamma^{A}\right)_{ad}
  = \frac{1}{4}{\rm tr}\!\left(\Gamma_B\Gamma'\Gamma_A\Gamma\right)\left(\Gamma^B\right)_{cb}\left(\Gamma^A\right)_{ad}.
\end{eqnarray*}
In the last equality, we again used the relation (\ref{eq40}).

\vspace{10ex}

{\Large \noindent {\bf Appendix D \hspace{3mm} Superpropagator with twisted covariant derivatives}}
\vspace{3ex}

In this appendix, we will explicitly show the calculation to rewrite
the superpropagator of the dynamical field (\ref{eq20}) into
(\ref{eq38}) which is written in terms of the twisted covariant
derivatives. We first show a proposition.

\vspace{3ex}

\noindent {\bf Proposition D-1}
\begin{eqnarray}
 H^2\left(\theta-\theta'\right)^2 =
  -4\exp\left\{-\frac{1}{2}\left[\bar{\theta}\!\left(i\Slash{\partial}-m\right)\!\left(\theta-\theta'\right)\right]\right\}\\ \nonumber
\end{eqnarray}
The proof is straightforward. Noting that $H_\alpha = -
\frac{\partial}{\partial\theta} +
\frac{1}{2}\left[\bar{\theta}\!\left(i\Slash{\partial}-m\right)\right]_\alpha$,
we see
\begin{eqnarray*}
 H^2\left(\theta-\theta'\right)^2 =
  \left\{-\frac{\partial}{\partial\theta^\alpha}\frac{\partial}{\partial\theta_\alpha} + \left[\bar{\theta}\!\left(i\Slash{\partial}-m\right)\right]_\alpha\!\frac{\partial}{\partial\theta_\alpha} + \frac{1}{4}\bar{\theta}^2\!\left(\partial^2 + m^2\right)\right\}\left(\theta-\theta'\right)^2,
\end{eqnarray*}
where we should note that
$C_{\alpha\beta}\frac{\partial}{\partial\theta^\beta} =
-\frac{\partial}{\partial\theta_\alpha}$. We can easily show
\begin{eqnarray*}
 -\frac{\partial}{\partial\theta^\alpha}\frac{\partial}{\partial\theta_\alpha}\left(\theta-\theta'\right)^2 = -4 \quad , \quad \left[\bar{\theta}\!\left(i\Slash{\partial}-m\right)\right]_\alpha\!\frac{\partial}{\partial\theta_\alpha}\left(\theta-\theta'\right)^2 = 2\hspace{.25em}\bar{\theta}\!\left(i\Slash{\partial}-m\right)\!\left(\theta-\theta'\right),
\end{eqnarray*}
and then we find
\begin{eqnarray*}
 H^2\left(\theta-\theta'\right)^2 = -4\left\{ 1 -
     \frac{1}{2}\hspace{.25em}\bar{\theta}\!\left(i\Slash{\partial}-m\right)\!\left(\theta-\theta'\right)
     -
     \frac{1}{16}\bar{\theta}^2\!\left(\theta-\theta'\right)^2\!\left(\partial^2+m^2\right)\right\}.
\end{eqnarray*}
Noting that
$\left[-\frac{1}{2}\bar{\theta}\!\left(i\Slash{\partial}-m\right)\!\left(\theta-\theta'\right)\right]^2
=
-\frac{1}{8}\bar{\theta}^2\!\left(\theta-\theta'\right)^2\!\left(\partial^2+m^2\right)$,
we can prove the statement. 

Using this, we can show the equation (\ref{eq120}) in page \pageref{eq120}.
\vspace{5ex}

\noindent {\bf Proposition D-2}
\begin{eqnarray*}
 \frac{1}{4}\bar{E}^2 H^2\delta^{(4)}\!\left(\theta-\theta'\right) =
  \exp\left[-\bar{\theta}'\!\left(i\Slash{\partial}-m\right)\theta + \frac{1}{2}\bar{\theta}\!\left(i\Slash{\partial}-m\right)\theta
	+ \frac{1}{2}\bar{\theta}'\!\left(i\Slash{\partial}-m\right)\theta'\right]\\
\end{eqnarray*}
In the above equation, we define $\delta^{(4)}\!\left(\theta-\theta'\right) = \frac{1}{4}\left(\theta-\theta'\right)^2\!\left(\bar{\theta}-\bar{\theta}'\right)^2$.
The proof is again straightforward but needs a large amount of
calculation. We first use proposition 3 and obtain
\begin{eqnarray*}
 \bar{E}^2 H^2\delta^{(4)}\!\left(\theta-\theta'\right) &=& -
  \bar{E}^2\left\{\exp\!\left[-\frac{1}{2}\bar{\theta}\!\left(i\Slash{\partial}-m\right)\!\left(\theta-\theta'\right)\right]\cdot\left(\bar{\theta}-\bar{\theta}'\right)^2\right\}.
\end{eqnarray*}
Since we can expand $\bar{E}^2$ as
\begin{eqnarray*}
 \bar{H^2} =
  -\frac{\partial}{\partial\bar{\theta}^\alpha}\frac{\partial}{\partial\bar{\theta}_\alpha} + \left[\theta\!\left(i\Slash{\partial}+m\right)\right]_\alpha\frac{\partial}{\partial\bar{\theta}_\alpha} + \frac{1}{4}\theta^2\!\left(\partial^2+m^2\right),
\end{eqnarray*}
we find
\begin{eqnarray}
  \bar{E}^2 H^2\delta^{(4)}\!\left(\theta-\theta'\right) &=& -
  \exp\!\left[-\frac{1}{2}\bar{\theta}\!\left(i\Slash{\partial}-m\right)\!\left(\theta-\theta'\right)\right]\bar{E}^2\!\left(\bar{\theta}-\bar{\theta}'\right)\nonumber
  \\[2mm]
&&\hspace{1.25em} +
\left\{\frac{\partial}{\partial\bar{\theta}^\alpha}\frac{\partial}{\partial\bar{\theta}_\alpha}
 \exp\!\left[-\frac{1}{2}\bar{\theta}\!\left(i\Slash{\partial}-m\right)\!\left(\theta-\theta'\right)\right]\right\}\left(\bar{\theta}-\bar{\theta}'\right)^2 \nonumber \\[2mm]
  &&\hspace{2.5em}+
  2\left\{\frac{\partial}{\partial\bar{\theta}^\alpha}\exp\!\left[-\frac{1}{2}\bar{\theta}\!\left(i\Slash{\partial}-m\right)\!\left(\theta-\theta'\right)\right]\right\}\frac{\partial}{\partial\bar{\theta}_\alpha}\left(\bar{\theta}-\bar{\theta}'\right)^2\nonumber
  \\[2mm]
&&\hspace{3.75em}-
\left[\bar{\theta}\!\left(i\Slash{\partial}+m\right)\right]_\alpha\left\{\frac{\partial}{\partial\bar{\theta}_\alpha}
 \exp\!\left[-\frac{1}{2}\bar{\theta}\!\left(i\Slash{\partial}-m\right)\!\left(\theta-\theta'\right)\right]\right\}\left(\bar{\theta}-\bar{\theta}'\right)^2 \hspace{2.5em} \label{eq62}
\end{eqnarray}
where we use the Leibniz rule for the derivative with respect to
$\bar{\theta}$. We can easily show
\begin{eqnarray*}
\bar{E}^2\left(\bar{\theta}-\bar{\theta}'\right)^2 &=& -4 -
 2\left(\bar{\theta}-\bar{\theta}'\right)\!\left(i\Slash{\partial}-m\right)\!\theta
 - \frac{1}{4}\theta^2\left(\bar{\theta}-\bar{\theta}'\right)^2\!\left(\partial^2+m^2\right)\\[1mm]
 \frac{\partial}{\partial\bar{\theta}^\alpha}\frac{\partial}{\partial\bar{\theta}_\alpha}
  \exp\!\left[-\frac{1}{2}\bar{\theta}\!\left(i\Slash{\partial}-m\right)\!\left(\theta-\theta'\right)\right]
  &=&
  -\frac{1}{4}\left(\theta-\theta'\right)^2\!\left(\partial^2+m^2\right)\exp\!\left[-\frac{1}{2}\bar{\theta}\!\left(i\Slash{\partial}-m\right)\!\left(\theta-\theta'\right)\right]\\[1mm]
\frac{\partial}{\partial\bar{\theta}^\alpha}\exp\!\left[-\frac{1}{2}\bar{\theta}\!\left(i\Slash{\partial}-m\right)\!\left(\theta-\theta'\right)\right]
 &=&
 -\frac{1}{2}\left[\left(\theta-\theta'\right)\!\left(i\Slash{\partial}+m\right)\right]_\alpha \exp\!\left[-\frac{1}{2}\bar{\theta}\!\left(i\Slash{\partial}-m\right)\!\left(\theta-\theta'\right)\right].
\end{eqnarray*} 
Therefore the equation
(\ref{eq62}) becomes
\begin{eqnarray*}
 \bar{E}^2 H^2\delta^{(4)}\!\left(\theta-\theta'\right) &=&
  \exp\!\left[-\frac{1}{2}\bar{\theta}\!\left(i\Slash{\partial}-m\right)\left(\theta-\theta'\right)\right]\\[1mm]
&&\hspace{1.25em}\cdot \hspace{.25em}4\left\{1
	 +
	 \frac{1}{2}\left(\bar{\theta}-\bar{\theta}'\right)\!\left(i\Slash{\partial}-m\right)\!\left(2\theta-\theta'\right) - \frac{1}{16}\left(\bar{\theta}-\bar{\theta}'\right)^2\!\left(2\theta-\theta'\right)^2\!\left(\partial^2 + m^2\right)\right\}.
\end{eqnarray*}
We can rewrite this equation as
\begin{eqnarray*}
  \bar{E}^2 H^2\delta^{(4)}\!\left(\theta-\theta'\right) &=&
  4\exp\!\left[-\frac{1}{2}\bar{\theta}\!\left(i\Slash{\partial}-m\right)\left(\theta-\theta'\right)\right]\exp\left[+\frac{1}{2}\left(\bar{\theta}-\bar{\theta}'\right)\!\left(i\Slash{\partial}-m\right)\left(2\theta-\theta'\right)\right].
\end{eqnarray*}
Then, at last, we can prove the statement.

\vspace{10ex}

{\Large \noindent {\bf Appendix E \hspace{3mm} Useful formulae with
twisted covarinat derivatives}}
\vspace{3ex}

In this appendix, we will show some useful formulae for loop calculations involving
twisted covariant derivatives, which is shown only by using the
anticommutation relation of them. 

\noindent{\bf Proposition E-1}
\begin{eqnarray*}
\left\{
\begin{array}{l}
 \left[\bar{E}^2, H_\alpha\right] =
  2\left[\bar{E}\!\left(i\Slash{\partial}-m\right)\right]_\alpha \quad
  ,  \quad \left[\bar{E}^2, H^\alpha\right] = -2\left[\left(i\Slash{\partial}+m\right)\!\bar{E}\right]^\alpha\\[2mm]
 \left[\bar{H}^2, E_\alpha\right] =
  2\left[\bar{H}\!\left(i\Slash{\partial}+m\right)\right]_\alpha \quad
  ,  \quad \left[\bar{H}^2, E^\alpha\right] = -2\left[\left(i\Slash{\partial}-m\right)\!\bar{H}\right]^\alpha\\
\end{array}
\right.
\end{eqnarray*}
We can easily find the anticommutation relations $\left\{H^\alpha, \bar{E}_\beta\right\} =
\left(i\Slash{\partial} + m\right)^\alpha_{\hspace{1.5mm}\beta}$ and
$\left\{E^\alpha, \bar{H}_\beta\right\} =
\left(i\Slash{\partial}-m\right)^\alpha_{\hspace{1.5mm}\beta}$. Using
these, this proposition can be shown through a direct calculation. 

By using this proposition, we can show the following important formula.
\vspace{5ex}

\noindent {\bf Proposition E-2}
\begin{eqnarray}
\left\{ 
\begin{array}{l}
 \left[\bar{E}^2, H^2\right] = 4\left(-\partial^2 - m^2\right) -
  4\hspace{.125em}H\!\left(i\Slash{\partial}+m\right)\!\bar{E}\\[2mm]
 \left[E^2, \bar{H}^2\right] = 4\left(-\partial^2 - m^2\right) -
  4\hspace{.125em}\bar{H}\!\left(i\Slash{\partial}+m\right)\!E
\end{array}
\right. \label{eq50}
\end{eqnarray}
We first show the upper equation. 
The second term $H\!\left(i\Slash{\partial}+m\right)\!\bar{E}$ in the
right-hand side means $H_\alpha
\left(i\Slash{\partial} +
m\right)^\alpha_{\hspace{1.5mm}\beta}H^\beta$. When we rewrite the left-hand side as
$H_\alpha\left[\bar{E}^2, H^\alpha\right] + \left[\bar{E}^2,
H_\alpha\right]H^\alpha$ and use the proposition 1, the left-hand side becomes
\begin{eqnarray*}
 -2\hspace{.125em}H\!\left(i\Slash{\partial}+m\right)\!\bar{E} +
  2\hspace{.125em}\bar{E}\!\left(i\Slash{\partial}-m\right)\!H =
  -4\hspace{.125em}H\!\left(i\Slash{\partial}+m\right)\!\bar{E} +
  2\left(i\Slash{\partial} -
    m\right)^\alpha_{\hspace{.3em}\beta}\cdot\left\{\bar{E}_\alpha, H^\beta\right\}.
\end{eqnarray*}
Then using the anticommutation relations of $\bar{E}_\alpha, H^\beta$, we
can prove the statement. The lower equation of (\ref{eq50}) is proved in
the same way. 

\vspace{10ex}

{\Large \noindent {\bf Appendix F \hspace{3mm} $I(p^2)^{-1} = \frac{\arctan\sqrt{\frac{-p^2}{4m^2}}}{\sqrt{-p^2}}$}}
\vspace{3ex}

We here explicitly show the equation
\begin{eqnarray*}
 I(p^2)^{-1} = \frac{\arctan\sqrt{\frac{-p^2}{4m^2}}}{\sqrt{-p^2}}.
\end{eqnarray*}
The definition of $I(p^2)^{-1}$ is
\begin{eqnarray*}
 I(p^2)^{-1} := \frac{4\pi}{i}\int\!\!\frac{d^3q}{\left(2\pi\right)^3}\frac{1}{\left(p+q\right)^2-m^2+i\epsilon}\frac{1}{q^2-m^2+i\epsilon}.
\end{eqnarray*}
We first introduce a Feynman parameter:
\begin{eqnarray*}
 I(p^2)^{-1} &=&
  \frac{4\pi}{i}\int_0^1\!\!dx\int\!\!\frac{d^3k}{\left(2\pi\right)^3}\hspace{.125em}\frac{1}{\left[\left(1-x\right)\left(q^2-m^2\right)
  + x\left\{\left(q+p\right)^2-m^2\right\}\right]^2}\\[1mm]
&=& \frac{4\pi}{i}\int_0^1\!\!dx
\int\!\!\frac{d^3k}{\left(2\pi\right)^3}\hspace{.125em}\frac{1}{\left[\left(k-xp\right)^2-\Delta\!\left(p^2;x\right)+i\epsilon\right]^2},
\end{eqnarray*}
where $\Delta\!\left(p^2;x\right):=-x\left(1-x\right)p^2+m^2$. Then we
shift the integration variable $k$ as $k\to k+xp$ and perform the Wick rotation:
\begin{eqnarray*}
 I(p^2)^{-1} &=& \frac{4\pi}{i}\int_0^1\!\!dx
  \int\!\!\frac{d^3k}{\left(2\pi\right)^3}\hspace{.125em}\frac{1}{\left[k^2-\Delta\!\left(p^2;x\right)+i\epsilon\right]^2}\\[1mm]
&=& 4\pi\int_0^1\!\!dx\int\!\!
\frac{d^3k_E}{\left(2\pi\right)^3}\hspace{.125em}\frac{1}{\left[k_E^2+\Delta\!\left(p^2;x\right)\right]^2}\\[1mm]
&=&
\frac{2}{\pi}\int_0^1\!\!dx\int_0^\infty\!\!dK\frac{K^2}{\left[K^2+\Delta\!\left(p^2;x\right)\right]^2}.
\end{eqnarray*}
We can easily perform this integral by changing the integration variable as
$K=\sqrt{\Delta\!\left(p^2;x\right)}\tan\theta$. The result is
\begin{eqnarray*}
I(p^2)^{-1} &=&
\frac{2}{\pi}\int_0^1\!\!dx\frac{\pi}{4\sqrt{\Delta\!\left(p^2;x\right)}}
= \frac{1}{2}\int_0^1\!\!dx\frac{1}{\sqrt{-p^2x\left(1-x\right)+m^2}}.
\end{eqnarray*}
Then we shift the variable $x$ as $x \to x+\frac{1}{2}$:
\begin{eqnarray*}
 I(p^2)^{-1} &=&
  \frac{1}{2}\int_{-\frac{1}{2}}^{\frac{1}{2}}\!dx\hspace{.125em}\frac{1}{\sqrt{-p^2\left(\frac{1}{2}+x\right)\left(\frac{1}{2}-x\right)+m^2}}\\[1mm]
&=&
\frac{1}{2}\int_{-\frac{1}{2}}^{\frac{1}{2}}\!dx\hspace{.125em}\frac{1}{\sqrt{-p^2\left(\frac{1}{4}-x^2\right)+m^2}}\\[1mm]
&=&
\frac{1}{\sqrt{-p^2}}\int_0^{\frac{1}{2}}\!dx\hspace{.125em}\frac{1}{\sqrt{\frac{p^2-4m^2}{4p^2}-x^2}}\\[1mm]
&=&\frac{1}{\sqrt{-p^2}}\int_0^{\sqrt{\frac{p^2}{p^2-4m^2}}}\!d\tilde{x}\frac{1}{\sqrt{1-\tilde{x}^2}},
\end{eqnarray*}
where we define $\tilde{x}:= \sqrt{\frac{4p^2}{p^2-4m^2}}x$ and change
the integration variable. Then we find
\begin{eqnarray*}
 I(p^2)^{-1} = \frac{1}{\sqrt{-p^2}}\arcsin\sqrt{\frac{p^2}{p^2-4m^2}}.
\end{eqnarray*}
Moreover, we can show that $\arcsin\sqrt{\frac{p^2}{p^2-4m^2}} =
\arctan\sqrt{-\frac{p^2}{4m^2}}$.  The proof is as follows. Suppose
$y=\arctan \sqrt{\frac{p^2}{p^2-4m^2}}$. Then $\frac{p^2}{p^2-4m^2}$
equals to $\sin^2 y$, namely $1-\cos^2 y$. Therefore, we find $\cos^2 y
= 1-\frac{p^2}{p^2-4m^2} = \frac{-4m^2}{p^2-4m^2}$. Then we can show
$\tan^2 y = \frac{1}{\cos^2 y}-1 = \frac{p^2-4m^2}{-4m^2}-1 =
\frac{p^2}{-4m^2}$. We have shown that $y=\arctan
\sqrt{-\frac{p^2}{4m^2}}$. Therefore we find that
\begin{eqnarray*}
 I(p^2)^{-1} = \frac{\arctan\sqrt{-\frac{p^2}{4m^2}}}{\sqrt{-p^2}}.
\end{eqnarray*}

\vspace{10ex}
{\Large \noindent {\bf Appendix G \hspace{3mm} D-algebra}}
\vspace{3ex}

We here study the algebra of supercovariant derivatives
$D_\alpha,\bar{D}_\alpha$ especially in momentum space. Recall that we can obtain
$D_\alpha,\bar{D}_\alpha$ from twisted covariant derivatives
$E_\alpha,\bar{E}_\alpha,H_\alpha,\bar{H}_\alpha$ imposing
$m=0$. Therefore, from Proposition E-1 and E-2, we find in momentum
space that
\begin{eqnarray*}
 \left[\bar{D}^2, D^\alpha\right] =
  -2\left(\Slash{p}\bar{D}\right)^\alpha
  \quad,\quad\left[\bar{D}^2, D_\alpha\right] = 2\left(\bar{D}\Slash{p}\right)_\alpha,
\end{eqnarray*}
and
\begin{eqnarray}
 \left[\bar{D}^2,D^2\right] = 4p^2
  -4D\!\left(\Slash{p}\right)\!\bar{D}. \label{eq40000}
\end{eqnarray}
Moreover, we can show the following proposition:
\vspace{5ex}

{\bf \noindent Proposition G-1}
\begin{eqnarray}
 \bar{D}^2\!D^2 -2\bar{D}\Slash{p}D = D\bar{D}^2\!D \quad,\quad
  D^2\!\bar{D}^2 - 2D\Slash{p}\bar{D} = D\bar{D}^2\!D,\label{eq40004}
\end{eqnarray}
where $D\bar{D}^2\!D := D_\alpha\bar{D}^2\!D^\alpha$.

\vspace{5ex}
The proof is straightforward. We can rewrite $\bar{D}^2\!D^2$ as
\begin{eqnarray*}
\bar{D}^2\!D^2 &=& -\bar{D}_\alpha D_\beta \bar{D}^\alpha D^\beta +
\bar{D}_\alpha\left\{\bar{D}^\alpha,D_\beta\right\}D^\beta\\
 &=&
-\bar{D}_\alpha D_\beta \bar{D}^\alpha D^\beta +
\bar{D}\Slash{p}D \\
&=& D_\beta\bar{D}^2D^\beta
 -\left(\Slash{p}\right)_{\beta\alpha}\bar{D}^\alpha D^\beta
 +\bar{D}\Slash{p}D\\
&=& D\bar{D}^2\!D  -\left(\Slash{p}\right)_{\alpha\beta}\bar{D}^\alpha D^\beta
 +\bar{D}\Slash{p}D\\
&=& D\bar{D}^2\!D  +\left(\Slash{p}\right)^{\alpha}_{\hspace{.3em}\beta}\bar{D}_\alpha D^\beta
 +\bar{D}\Slash{p}D\\
&=& D\bar{D}^2\!D + 2\bar{D}\Slash{p}D.
\end{eqnarray*}
Therefore the left equation in (\ref{eq40004}) is proved. The right
equation can be shown in the similar way.

We now define the following projection operators:
\begin{eqnarray}
 P_1 := \frac{D^2\!\bar{D}^2}{4p^2} \quad,\quad P_2:=
  \frac{\bar{D}^2\!D^2}{4p^2}. \label{eq40001}
\end{eqnarray}
Suppose $\Phi$ is an arbitrary chiral superfield. Then we can show
$P_1\Phi = 0$ and
\begin{eqnarray*}
P_2\Phi = \left(P_1 + \frac{\left[\bar{D^2},D^2\right]}{4p^2}\right)\Phi
 = \Phi.
\end{eqnarray*}
In the last equality, we use the equation (\ref{eq40000}). In the same
way, we can show that $P_1\Phi^\dagger = \Phi^\dagger$ and
$P_2\Phi^\dagger = 0$ for arbitrary anti-chiral superfield
$\Phi^\dagger$. Therefore, $P_1$ and $P_2$ is projection operators to
anti-chiral and chiral superfield respectively.

In addition to (\ref{eq40001}), we can define the following four
Lorentz invariant operators from $D_\alpha$ and $\bar{D}_\alpha$:
\begin{eqnarray*}
 P_+ := -\frac{iD^2}{2\sqrt{-p^2}} \quad, \quad P_-:=
  -\frac{i\bar{D}^2}{2\sqrt{-p^2}} \quad, \quad P_T:=
  -\frac{D\bar{D}^2\!D}{2p^2} \quad, \quad P_D := -\frac{i\bar{D}D}{\sqrt{-p^2}}.
\end{eqnarray*}
Since $D_\alpha D^2 = \bar{D}_\alpha\bar{D}^2 = 0$,  any other
differential operator composed of $D_\alpha,\bar{D}_\alpha$ can be written as a linear combination of these
six operators.
Note that $D\Slash{p}\bar{D}$ can be written as a linear combination of
$P_2$ and $P_T$ using Proposition G-1.

Moreover, we can show the following useful formula:
\vspace{5ex}

{\bf \noindent Proposition G-2}
\begin{eqnarray*}
 P_1+P_2+P_T=1
\end{eqnarray*}
The proof is straightforward. Recalling the definition of projection
operators, we can show that
\begin{eqnarray*}
 P_1+P_2+P_T &=& \frac{1}{4p^2}\left(D^2\!\bar{D}^2 +
				\bar{D}^2\!D^2 - 2D\bar{D}^2\!D\right)\\
&=& \frac{1}{4p^2}\left\{\left(D^2\bar{D}^2-D\bar{D}^2\!D\right) +
		   \left(\bar{D}^2\!D^2-D\bar{D}^2\!D\right)\right\}\\
 &=& \frac{1}{2p^2}\left(D\Slash{p}\bar{D} + \bar{D}\Slash{p}D\right).
\end{eqnarray*}
In the last equality, we use Proposition G-1. We moreover rewrite this
as
\begin{eqnarray*}
 P_1+P_2+P_T &=&
  \frac{1}{2p^2}\left(D_\alpha\Slash{p}^\alpha_{\hspace{.3em}\beta}\bar{D}^\beta
		+
		\bar{D}_\alpha\Slash{p}^\alpha_{\hspace{.3em}\beta}D^\beta\right)\\
&=&   \frac{1}{2p^2}\left(D^\alpha\Slash{p}_\alpha^{\hspace{.3em}\beta}\bar{D}_\beta
		+
		\bar{D}_\alpha\Slash{p}^\alpha_{\hspace{.3em}\beta}D^\beta\right)\\
&=&   \frac{1}{2p^2}\left(D^\alpha\Slash{p}^{\beta}_{\hspace{.3em}\alpha}\bar{D}_\beta
		+
		\bar{D}_\alpha\Slash{p}^\alpha_{\hspace{.3em}\beta}D^\beta\right)\\
&=&
 \frac{1}{2p^2}\Slash{p}^{\beta}_{\hspace{.3em}\alpha}\left\{D^\alpha,\bar{D}_\beta\right\}\\
&=&
 \frac{1}{2p^2}\hspace{.125em}\Slash{p}^\beta_{\hspace{.3em}\alpha}\hspace{.125em}\Slash{p}^{\alpha}_{\hspace{.3em}\beta}\\
&=& \frac{1}{2p^2}{\rm tr}\left({\Slash{p}^2}\right)\\[3mm]
&=& 1.
\end{eqnarray*}
Then the statement has been proved.

We can indicate the multiplication rules of these projection operators
as follows:
\begin{table}[h]
\begin{center}
\begin{tabular}{|c||c|c|c|c|c|c|}\hline
left $\backslash$ right & $P_1$ & $P_2$ & $P_+$ & $P_-$ & $P_T$ & $P_D$
 \\ \hline\hline
$P_1$ & $P_1$ &  & $P_+$ &  &  &  \\ \hline
$P_2$ &  & $P_2$ &  & $P_-$ &  &  \\ \hline
$P_+$ &  & $P_+$ &  & $P_1$ &  &  \\ \hline
$P_-$ & $P_-$ &  & $P_2$ &  &  &  \\ \hline
$P_T$ &  &  &  &  & $P_T$ & $P_D$ \\ \hline
$P_D$ &  &  &  &  & $P_D$ & $P_T$ \\ \hline
\end{tabular}
\caption{The multiplicative property of oparators}
\end{center}
\end{table}

\noindent In particular, 
\begin{eqnarray*}
 P_+P_2 = \frac{i}{8\left(-p^2\right)^{\frac{3}{2}}}D^2\bar{D}^2D^2 =
  \frac{i}{8\left(-p^2\right)^{\frac{3}{2}}}D^2\left[\bar{D}^2,D^2\right]
  = -\frac{iD^2}{2\sqrt{-p^2}} = P_+
\end{eqnarray*}
and similarly,
\begin{eqnarray*}
 P_1P_+ = \frac{i}{8\left(-p^2\right)^{\frac{3}{2}}}D^2\bar{D}^2D^2 = P_+.
\end{eqnarray*}
We can also show that
\begin{eqnarray*}
 P_-P_1 =
  \frac{i}{8\left(-p^2\right)^{\frac{3}{2}}}\bar{D}^2D^2\bar{D}^2 =
  \frac{i}{8\left(-p^2\right)^{\frac{3}{2}}}\left[\bar{D}^2,D^2\right]\bar{D}^2
  = -\frac{i\bar{D}^2}{2\sqrt{-p^2}} = P_-
\end{eqnarray*}
and
\begin{eqnarray*}
 P_2P_- =
  \frac{i}{8\left(-p^2\right)^{\frac{3}{2}}}\bar{D}^2D^2\bar{D}^2 = P_-.
\end{eqnarray*}
Noting that
\begin{eqnarray*}
 P_T=-\frac{D\bar{D}^2\!D}{2p^2} = -\frac{\bar{D}D^2\!\bar{D}}{2p^2},
\end{eqnarray*}
it leads to a vanishing result to multiply $P_T$ by $P_1,P_2,P_+$ or
$P_-$. Similarly, since
\begin{eqnarray*}
 P_D=-\frac{i\bar{D}D}{\sqrt{-p^2}} = -\frac{iD\bar{D}}{\sqrt{-p^2}},
\end{eqnarray*}
we obtain a vanishing result if we multiply $P_D$ by $P_1, P_2, P_+,$ or $P_-$.

We can show $P_T^2=P_T$ as follows. First, noting that
\begin{eqnarray*}
 D\bar{D}^2\!D = D^2\bar{D}^2 - 2D\Slash{p}\bar{D} = \bar{D}^2D^2 - 2\bar{D}\Slash{p}D,
\end{eqnarray*}
we see that
\begin{eqnarray}
 P_T^2 &=&
  \frac{1}{p^4}\left(D\Slash{p}\bar{D}\right)\left(\bar{D}\Slash{p}D\right)
  = \frac{1}{p^4}D_\alpha\bar{D}^\beta\bar{D}_\gamma
  D^\delta\Slash{p}^\alpha_{\hspace{.3em}\beta}\Slash{p}^{\gamma}_{\hspace{.3em}\delta}. \label{EQ38}
\end{eqnarray}
Then we use the Fierz identity:
\begin{eqnarray*}
 \Slash{p}^\alpha_{\hspace{.3em}\beta}\Slash{p}^\gamma_{\hspace{.3em}\delta}
  &=& \frac{1}{4}{\rm
  tr}\left[\Gamma_B\Slash{p}\hspace{.25em}\Gamma_A\Slash{p}\right]\Gamma^{B\gamma}_{\hspace{1em}\beta}\Gamma^{A\alpha}_{\hspace{1em}\delta}=
  \frac{1}{4}{\rm
  tr}\left[\Slash{p}^2\right]\delta^\gamma_{\hspace{.3em}\beta}\delta^{\alpha}_{\hspace{.3em}\delta}
  + \frac{1}{4}{\rm
  tr}\left[\gamma_\mu\Slash{p}\gamma_\nu\Slash{p}\right]\gamma^{\mu\gamma}_{\hspace{1em}\beta}\gamma^{\nu\alpha}_{\hspace{1em}\delta}\\
&=&
 \frac{1}{2}p^2\delta^\gamma_{\hspace{.3em}\beta}\delta^{\alpha}_{\hspace{.3em}\delta} - \frac{1}{2}\left(\eta_{\mu\nu}p^2-2p_\mu p_\nu\right)\gamma^{\mu\gamma}_{\hspace{1em}\beta}\gamma^{\nu\alpha}_{\hspace{1em}\delta}\\
&=&
 \frac{1}{2}p^2\delta^\gamma_{\hspace{.3em}\beta}\delta^{\alpha}_{\hspace{.3em}\delta} - \frac{1}{2}p^2\gamma^{\mu\gamma}_{\hspace{1em}\beta}\gamma^{\hspace{.3em}\alpha}_{\mu\hspace{.5em}\delta} + \Slash{p}^\gamma_{\hspace{.3em}\beta}\Slash{p}^{\alpha}_{\hspace{.3em}\delta}.
\end{eqnarray*}
When we substitute this for (\ref{EQ38}) and  note that
$\bar{D}\gamma^\mu \bar{D} = 0$, we can show that
\begin{eqnarray*}
 P_T^2 &=& \frac{1}{2p^2}D_\alpha\bar{D}^\beta\bar{D}_\gamma
  D^\delta\delta^\gamma_{\hspace{.3em}\beta}\delta^{\alpha}_{\hspace{.3em}\delta}
  = \frac{1}{2p^2}D_\alpha\bar{D}^\beta\bar{D}_\beta
  D^\alpha = -\frac{1}{2p^2}D_\alpha\bar{D}_\beta\bar{D}^\beta
  D^\alpha = - \frac{1}{2p^2}D\bar{D}^2\!D = P_T.
\end{eqnarray*}

We can similarly show that $P_D^2=P_D$. Note that
\begin{eqnarray*}
 P_D^2 &=& \frac{1}{p^2}\left(D\bar{D}\right)\left(\bar{D}D\right) =
  \frac{1}{p^2}D_\alpha\bar{D}^\beta\bar{D}_\gamma D^\delta
  \delta^\alpha_{\hspace{.3em}\beta}\delta^\gamma_{\hspace{.3em}\delta}.
\end{eqnarray*}
By using the Fierz identity, we see
\begin{eqnarray*}
 \delta^\alpha_{\hspace{.3em}\beta}\delta^\gamma_{\hspace{.3em}\delta}
  &=& \frac{1}{4}{\rm
  tr}\left[\Gamma_A\Gamma_B\right]\Gamma^{B\gamma}_{\hspace{1em}\beta}\Gamma^{A\alpha}_{\hspace{1em}\delta}
  =
  \frac{1}{2}\Gamma^{A\gamma}_{\hspace{1em}\beta}\Gamma_{A\hspace{.3em}\delta}^{\hspace{.3em}\alpha}
  =
  \frac{1}{2}\left(\gamma^{\mu\gamma}_{\hspace{1em}\beta}\gamma_{\mu\hspace{.3em}\delta}^{\hspace{.3em}\alpha}
	     + \delta^\gamma_{\hspace{.3em}\beta}\delta^{\alpha}_{\hspace{.3em}\delta}\right).
\end{eqnarray*}
We can therefore show that
\begin{eqnarray*}
 P_D^2 &=& \frac{1}{2p^2}D_\alpha\bar{D}^\beta\bar{D}_\gamma D^\delta
  \delta^\gamma_{\hspace{.3em}\beta}\delta^{\alpha}_{\hspace{.3em}\delta}
  = \frac{1}{2p^2}D_\alpha\bar{D}^\beta\bar{D}_\beta D^\alpha= -
  \frac{1}{2p^2}D\bar{D}^2\!D = P_T.
\end{eqnarray*}

We will now show that $P_TP_D = P_D$. First note that
\begin{eqnarray}
 P_TP_D &=&
  -\frac{i}{2\left(-p^2\right)^{\frac{3}{2}}}\left(D\bar{D}^2\!D\right)\left(\bar{D}D\right) =   -\frac{i}{2\left(-p^2\right)^{\frac{3}{2}}}\left(D_\alpha\left[\bar{D}^2,D^\alpha\right]\right)\left(\bar{D}D\right)\nonumber \\
&=&
 \frac{i}{\left(-p^2\right)^{\frac{3}{2}}}\left(D\Slash{p}\bar{D}\right)\left(\bar{D}D\right)
 =
 \frac{i}{\left(-p^2\right)^{\frac{3}{2}}}D_\alpha\bar{D}^\beta\bar{D}_\gamma
 D^\delta \Slash{p}^\alpha_{\hspace{.3em}\beta}\delta^\gamma_{\hspace{.3em}\delta}.\label{EQ39}
\end{eqnarray}
Then use the Fierz identity as follows:
\begin{eqnarray*}
 \Slash{p}^\alpha_{\hspace{.3em}\beta}\delta^\gamma_{\hspace{.3em}\delta}
  &=& \frac{1}{4}{\rm tr}\left[\Gamma_A\hspace{.125em}\Slash{p}\hspace{.25em}\Gamma_B\right]\Gamma^{B\gamma}_{\hspace{1em}\beta}\Gamma^{A\alpha}_{\hspace{1em}\delta}.
\end{eqnarray*}
Subsituting this for (\ref{EQ39}), we see that
the factor $\bar{D}^\beta\Gamma^{B\gamma}_{\hspace{4mm}\beta}\bar{D}_{\gamma}$
vanishes if
$\Gamma^B=\gamma^\mu$. The reason for this is $\bar{D}\gamma^\mu
\bar{D}=0$. Non-zero contribution, therefore, occurs if and only only if
$\Gamma^B=i{\bf 1}$, namely,
\begin{eqnarray*}
 P_TP_D &=&
  \frac{i}{4\left(-p^2\right)^{\frac{3}{2}}}D_\alpha\bar{D}^\beta\bar{D}_\gamma
  D^\delta {\rm
  tr}\left[\Slash{p}\gamma_\mu\right]\delta^{\gamma}_{\hspace{.3em}\beta}\gamma^{\mu\alpha}_{\hspace{1em}\delta}\\
&=&  \frac{i}{2\left(-p^2\right)^{\frac{3}{2}}}D_\alpha\bar{D}^\beta\bar{D}_\gamma
  D^\delta
  \delta^{\gamma}_{\hspace{.3em}\beta}\Slash{p}^{\alpha}_{\hspace{.3em}\delta}\\
&=&  \frac{i}{2\left(-p^2\right)^{\frac{3}{2}}}D_\alpha\bar{D}^\beta\bar{D}_\beta
  D^\delta
  \Slash{p}^{\alpha}_{\hspace{.3em}\delta}\\
&=& -\frac{i}{2\left(-p^2\right)^{\frac{3}{2}}}D_\alpha\bar{D}^2
  \left(\Slash{p}D\right)^\alpha\\
&=& -\frac{i}{2\left(-p^2\right)^{\frac{3}{2}}}\bar{D}^2
  \left(D\Slash{p}D\right) -
  \frac{i}{2\left(-p^2\right)^{\frac{3}{2}}}\left[D_\alpha,\bar{D}^2\right]\left(\Slash{p}D\right)^\alpha\\
&=&
 \frac{i}{\left(-p^2\right)^\frac{3}{2}}\left(\bar{D\Slash{p}}^2D\right)\\[1mm]
&=& -\frac{i\bar{D}D}{\sqrt{-p^2}}\\[2mm]
&=& P_D.
\end{eqnarray*}
In the same way, we can show that
\begin{eqnarray*}
 P_DP_T =P_D.
\end{eqnarray*}

\vspace{10ex}
{\Large \noindent {\bf Appendix H \hspace{3mm} Propagators of auxiliary
component fields}}
\vspace{3ex}

We here derive the expression (\ref{v}) and (\ref{M}) from the
superpropagator of the auxiliary field
\begin{eqnarray}
\left<V(-p,\theta',\bar{\theta}')V(p,\theta,\bar{\theta})\right>_0 = \frac{4\pi
 i}{N}I(p^2)\cdot\nabla_V^{-1}\delta^{(4)}\!\left(\theta-\theta'\right),\label{EQ40}
\end{eqnarray}
where
\begin{eqnarray}
 \nabla_V^{-1} &=& \frac{1}{p^2-4m^2}\cdot\frac{D\bar{D}^2D-4mD\bar{D}}{p^2} +
 \frac{\alpha}{2p^4}\left(D^2\bar{D}^2 + \bar{D}^2D^2\right).\label{EQ48}
\end{eqnarray}
We will first expand the left-hand side of (\ref{EQ40}) in
components. Choosing the Wess-Zumino gauge, the auxiliary field $V$ can be written as
\begin{eqnarray*}
V = \bar{\theta}\Slash{v}\theta + M\bar{\theta}\theta +
 \frac{1}{2}\bar{\theta}^2\theta\lambda +
 \frac{1}{2}\theta^2\bar{\theta}\bar{\lambda} + \frac{1}{4}\theta^2\bar{\theta}^2D.
\end{eqnarray*}
Then we see that the propagators of $M$ and $v^\mu$ can be obtained by
 taking terms proportional to $\bar{\theta}'\theta'\bar{\theta}\theta$ in
(\ref{EQ40}), namely,
\begin{eqnarray}
 \left.\left<V(-p,\theta',\bar{\theta}')V(p,\theta,\bar{\theta})\right>_0\right|_{O(\bar{\theta}'\theta'\bar{\theta}\theta)}
  &=&
  \left<\bar{\theta}'\!\left[\Slash{v}(-p)+M(-p)\right]\!\theta'\hspace{.25em}\bar{\theta}\!\left[\Slash{v}(p)+M(p)\right]\!\theta\right>_0.\label{EQ49}
\end{eqnarray}

In the right-hand side of (\ref{EQ40}), we can show that
\begin{eqnarray}
 \left.D^2\bar{D}^2\delta^{(4)}\!\left(\theta-\theta'\right)\right|_{O(\bar{\theta}'\theta'\bar{\theta}\theta)}
  &=&
  \left[4\exp\left(-\frac{1}{2}\bar{\theta}\Slash{p}\theta-\frac{1}{2}\bar{\theta}'\Slash{p}\theta'+\bar{\theta}\Slash{p}\theta'\right)\right]_{O(\bar{\theta}'\theta'\bar{\theta}\theta)}\nonumber
  \\
&=&
 \left[4\cdot\frac{1}{2!}\left(-\frac{1}{2}\bar{\theta}\Slash{p}\theta-\frac{1}{2}\bar{\theta}'\Slash{p}\theta'+\bar{\theta}\Slash{p}\theta'\right)^2\right]_{O(\bar{\theta}'\theta'\bar{\theta}\theta)}\nonumber \\[1mm]
&=& \left(\bar{\theta}\Slash{p}\theta\right)\left(\bar{\theta}'\Slash{p}\theta'\right).\label{EQ45}
\end{eqnarray}
Similarly, we see
\begin{eqnarray}
  \left.\bar{D}^2D^2\delta^{(4)}\!\left(\theta-\theta'\right)\right|_{O(\bar{\theta}'\theta'\bar{\theta}\theta)}
  &=&\left(\bar{\theta}\Slash{p}\theta\right)\left(\bar{\theta}'\Slash{p}\theta'\right).\label{EQ46}
\end{eqnarray}
We can also show that
\begin{eqnarray}
 \left.D\!\left(\Slash{p}+2m\right)\!\bar{D}\hspace{.125em}\delta^{(4)}\!\left(\theta-\theta'\right)\right|_{O(\bar{\theta}'\theta'\bar{\theta}\theta)}
  &=&
  \left.\frac{1}{2}\left(\bar{\theta}\Slash{p}\right)_\alpha\left(\Slash{p}+2m\right)^\alpha_{\hspace{.3em}\beta}\frac{\partial}{\partial\bar{\theta}_\beta}\delta^{(4)}\!\left(\theta-\theta'\right)\right|_{O(\bar{\theta}'\theta'\bar{\theta}\theta)}\nonumber
  \\
&&\hspace{1.25em}+\left.\left(-\frac{\partial}{\partial\theta^\alpha}\right)\left(\Slash{p}+2m\right)^\alpha_{\hspace{1.5mm}\beta}\left(-\frac{1}{2}\left(\Slash{p}\theta\right)^\beta\right)\delta^{(4)}\!\left(\theta-\theta'\right)\right|_{O(\bar{\theta}'\theta'\bar{\theta}\theta)}\nonumber
 \\[2mm]
&=&
\left.\frac{1}{2}\left(\bar{\theta}\Slash{p}\right)_\alpha\left(\Slash{p}+2m\right)^\alpha_{\hspace{.3em}\beta}\frac{\partial}{\partial\bar{\theta}_\beta}\delta^{(4)}\!\left(\theta-\theta'\right)\right|_{O(\bar{\theta}'\theta'\bar{\theta}\theta)}\nonumber
\\
&&\hspace{1.25em}-\left.\frac{1}{2}\left(\Slash{p}+2m\right)^\alpha_{\hspace{.3em}\beta}\left(\Slash{p}\theta\right)^\beta\frac{\partial}{\partial\theta^\alpha}\delta^{(4)}\!\left(\theta-\theta'\right)\right|_{O(\bar{\theta}'\theta'\bar{\theta}\theta)}\nonumber
 \\
&&\hspace{2.5em}+\left.\frac{1}{2}\left(\Slash{p}+2m\right)^\alpha_{\hspace{.3em}\beta}\Slash{p}^\beta_{\hspace{.3em}\alpha}\delta^{(4)}\!\left(\theta-\theta'\right)\right|_{O(\bar{\theta}'\theta'\bar{\theta}\theta)}\nonumber
 \\[2mm]
&=&
\left.\left(\bar{\theta}\Slash{p}\right)_\alpha\left(\Slash{p}+2m\right)^\alpha_{\hspace{.3em}\beta}\left(\bar{\theta}-\bar{\theta}'\right)^\beta\left(\theta-\theta'\right)^2\right|_{O(\bar{\theta}'\theta'\bar{\theta}\theta)}\nonumber
\\
&&\hspace{1.25em}+\left.\left(\Slash{p}+2m\right)^\alpha_{\hspace{.3em}\beta}\left(\Slash{p}\theta\right)^\beta\left(\theta-\theta'\right)_\alpha\left(\bar{\theta}-\bar{\theta}'\right)\right|_{O(\bar{\theta}'\theta'\bar{\theta}\theta)}\nonumber
 \\
&&\hspace{2.5em}+\left.\frac{1}{8}{\rm
		 tr}\left[p^2+2m\Slash{p}\right]\left(\theta-\theta'\right)^2\left(\bar{\theta}-\bar{\theta}'\right)^2\right|_{O(\bar{\theta}'\theta'\bar{\theta}\theta)}\nonumber \\[2mm]
&=&
\frac{1}{2}\hspace{.125em}\bar{\theta}\!\left(p^2+2m\Slash{p}\right)\!\bar{\theta}'\left(\theta\theta'\right)
- \frac{1}{2}\hspace{.125em}\theta\!\left(p^2 -
				    2m\Slash{p}\right)\!\theta'\left(\bar{\theta}\bar{\theta}'\right) + p^2\left(\theta\theta'\right)\left(\bar{\theta}\bar{\theta}'\right)\nonumber\\[2mm]
&=& p^2\left(\theta\theta'\right)\left(\bar{\theta}\bar{\theta}'\right)
 +
 m\left(\bar{\theta}\Slash{p}\bar{\theta}'\right)\left(\theta\theta'\right) + m\left(\theta\Slash{p}\theta'\right)\left(\bar{\theta}\bar{\theta}'\right).\label{EQ44}
\end{eqnarray}
By using the Fierz identity, we obtain
\begin{eqnarray*}
 \left(\theta\theta'\right)\left(\bar{\theta}\bar{\theta}'\right) &=&
  -\frac{1}{4}{\rm
  tr}\left[\Gamma_A\Gamma_B\right]\left(\theta\Gamma^A\bar{\theta}\right)\left(\bar{\theta}'\Gamma^B\theta'\right) = -\frac{1}{2}\left(\theta\gamma^\mu\bar{\theta}\right)\left(\bar{\theta}'\gamma_\mu\theta'\right) - \frac{1}{2}\left(\theta\bar{\theta}\right)\left(\bar{\theta}'\theta'\right),\\
\left(\bar{\theta}\Slash{p}\bar{\theta}'\right)\left(\theta\theta'\right)
 &=& -\frac{1}{4}{\rm
 tr}\left[\Gamma_A\hspace{.125em}\Slash{p}\hspace{.25em}\Gamma_B\right]\left(\bar{\theta}\Gamma^A\theta\right)\left(\theta'\Gamma^B\bar{\theta}'\right)\\
&=& -\frac{1}{4}{\rm
 tr}\left(\gamma_\mu\Slash{p}\gamma_\nu\right)\left(\bar{\theta}\gamma^\mu\theta\right)\left(\theta'\gamma^\nu\bar{\theta}'\right)
 - \frac{1}{4}{\rm
 tr}\left(\gamma_\mu\Slash{p}\right)\left(\bar{\theta}\gamma^\mu\theta\right)\left(\theta'\bar{\theta}'\right)
 - \frac{1}{4}{\rm
 tr}\left(\Slash{p}\gamma_\nu\right)\left(\bar{\theta}\theta\right)\left(\theta'\gamma^\nu\bar{\theta}'\right)\\
&=&
 -\frac{i}{2}\epsilon_{\mu\rho\nu}p^\rho\left(\bar{\theta}\gamma^\mu\theta\right)\left(\theta'\gamma^\nu\bar{\theta}'\right)
 -
 \frac{1}{2}\left(\bar{\theta}\Slash{p}\theta\right)\left(\theta'\bar{\theta}'\right)
 -
 \frac{1}{2}\left(\bar{\theta}\theta\right)\left(\theta'\Slash{p}\bar{\theta}'\right),\\
\left(\theta\Slash{p}\theta'\right)\left(\bar{\theta}\bar{\theta}'\right)
 &=& -\frac{i}{2}\epsilon_{\mu\rho\nu}p^\rho\left(\bar{\theta}\gamma^\mu\theta\right)\left(\theta'\gamma^\nu\bar{\theta}'\right)
 +
 \frac{1}{2}\left(\bar{\theta}\Slash{p}\theta\right)\left(\theta'\bar{\theta}'\right)
 +
 \frac{1}{2}\left(\bar{\theta}\theta\right)\left(\theta'\Slash{p}\bar{\theta}'\right).
\end{eqnarray*}
Then the equation (\ref{EQ44}) becomes
\begin{eqnarray}
  \left.D\!\left(\Slash{p}+2m\right)\!\bar{D}\hspace{.125em}\delta^{(4)}\!\left(\theta-\theta'\right)\right|_{O(\bar{\theta}'\theta'\bar{\theta}\theta)}
  &=&
  -\frac{p^2}{2}\left[\left(\theta\gamma^\mu\bar{\theta}\right)\left(\bar{\theta}'\gamma_\mu\theta'\right)
		+
		\left(\theta\bar{\theta}\right)\left(\bar{\theta}'\theta'\right)\right] - im\epsilon_{\mu\rho\nu}p^\rho\left(\bar{\theta}\gamma^\mu\theta\right)\left(\theta'\gamma^\nu\bar{\theta}'\right)\nonumber \\
 &=&  \frac{p^2}{2}\left[\left(\bar{\theta}\gamma^\mu\theta\right)\left(\bar{\theta}'\gamma_\mu\theta'\right)
		-
		\left(\bar{\theta}\theta\right)\left(\bar{\theta}'\theta'\right)\right] - im\epsilon_{\mu\nu\rho}p^\rho\left(\bar{\theta}\gamma^\mu\theta\right)\left(\bar{\theta}'\gamma^\nu\theta'\right).\hspace{2.5em}\label{EQ47}
\end{eqnarray}

From (\ref{EQ45}), (\ref{EQ46}), (\ref{EQ47}), and (\ref{EQ48}),
 we can evaluate terms propotional to $\bar{\theta}'\theta'\bar{\theta}\theta$ in
 (\ref{EQ40}) as follows:
\begin{eqnarray*}
&&\left.\left<V(-p,\theta',\bar{\theta}')V(p,\theta,\bar{\theta})\right>_0\right|_{O(\bar{\theta}'\theta'\bar{\theta}\theta)}\\[3mm]
 &=& \frac{4\pi
  i}{N}I(p^2)\left[\frac{2}{p^2-4m^2}\left\{\frac{\left(\bar{\theta}'\Slash{p}\theta'\right)\left(\bar{\theta}\Slash{p}\theta\right)}{2p^2}-\frac{\frac{p^2}{2}\left[\left(\bar{\theta}'\gamma^\mu\theta'\right)\left(\bar{\theta}\gamma_\mu\theta\right)-\left(\bar{\theta}'\theta'\right)\left(\bar{\theta}\theta\right)\right]-im\epsilon_{\mu\nu\rho}p^\rho\left(\bar{\theta}'\gamma^\mu\theta'\right)\left(\bar{\theta}\gamma^\nu\theta\right)}{p^2}\right\}\right.\\[1mm]
&&\hspace{32.5em}\left.-\frac{\alpha}{p^4}\left(\bar{\theta}'\Slash{p}\theta'\right)\left(\bar{\theta}\Slash{p}\theta\right)\right]\\[3mm]
&=&
 \left(\bar{\theta}'\gamma^\mu\theta'\right)\left(\bar{\theta}\gamma^\nu\theta\right)\times\frac{4\pi
 i}{N}I(p^2)\left\{\frac{1}{p^2-4m^2}\left(-\eta_{\mu\nu}+\frac{p_\mu
				      p_\nu}{p^2}\left(1-\frac{\alpha}{p^2}\left(p^2-4m^2\right)\right) + \frac{2mi}{p^2}\epsilon_{\mu\nu\rho}p^\rho\right)\right\}\\[1mm]
&&\hspace{2.5em}+\left(\bar{\theta'}\theta'\right)\left(\bar{\theta}\theta\right)\times\frac{4\pi
 i}{N}I(p^2)\frac{1}{p^2-4m^2}.
\end{eqnarray*}
Comparing this to (\ref{EQ49}), we can show that
\begin{eqnarray*}
 \left<v^\mu(-p)v^\nu(p)\right>_0 &=& \frac{4\pi
 i}{N}I(p^2)\left\{\frac{1}{p^2-4m^2}\left(-\eta_{\mu\nu}+\frac{p_\mu
				      p_\nu}{p^2}\left(1-\frac{\alpha}{p^2}\left(p^2-4m^2\right)\right) + \frac{2mi}{p^2}\epsilon_{\mu\nu\rho}p^\rho\right)\right\},\\
\left<M(-p)M(p)\right>_0 &=& \frac{4\pi
 i}{N}I(p^2)\frac{1}{p^2-4m^2},\\[2mm]
\left<v^\mu(-p)M(p)\right>_0 &=& 0.
\end{eqnarray*}
These coincide with the results in \cite{paper1}.

\end{document}